\documentclass[useAMS, usenatbib, fleqn]{mnras}

\usepackage{times, graphicx, aas_macros, amsmath, url, natbib, acronym, rotating, flushend, etoolbox, bm, hyperref}
\usepackage[referable]{threeparttablex}
\usepackage[noabbrev]{cleveref}
\usepackage[T1]{fontenc}

\hypersetup{colorlinks=True, linkcolor=blue, citecolor=blue, filecolor=blue, urlcolor=blue}

\makeatletter

\patchcmd{\NAT@citex}
  {\@citea\NAT@hyper@{%
     \NAT@nmfmt{\NAT@nm}%
     \hyper@natlinkbreak{\NAT@aysep\NAT@spacechar}{\@citeb\@extra@b@citeb}%
     \NAT@date}}
  {\@citea\NAT@nmfmt{\NAT@nm}%
   \NAT@aysep\NAT@spacechar\NAT@hyper@{\NAT@date}}{}{}

\patchcmd{\NAT@citex}
  {\@citea\NAT@hyper@{%
     \NAT@nmfmt{\NAT@nm}%
     \hyper@natlinkbreak{\NAT@spacechar\NAT@@open\if*#1*\else#1\NAT@spacechar\fi}%
       {\@citeb\@extra@b@citeb}%
     \NAT@date}}
  {\@citea\NAT@nmfmt{\NAT@nm}%
   \NAT@spacechar\NAT@@open\if*#1*\else#1\NAT@spacechar\fi\NAT@hyper@{\NAT@date}}
  {}{}
  
\makeatother

\acrodef{qso}[QSO]{quasar}
\acrodef{agn}[AGN]{active galactic nuclei}
\acrodef{sn}[SN]{supernova}
\acrodef{smbh}[SMBH]{super massive black hole}
\acrodef{igm}[IGM]{intergalactic medium}
\acrodef{ism}[ISM]{interstellar medium}
\acrodef{cgm}[CGM]{circumgalactic medium}
\acrodef{sfr}[SFR]{star formation rate}
\acrodef{sed}[SED]{spectral energy distribution}
\acrodef{aal}[AAL]{associated absorption line}
\acrodef{bal}[BAL]{broad absorption line}
\acrodef{cie}[CIE]{collisional ionization equilibrium}
\acrodef{blr}[BLR]{broad-line region}
\acrodef{nlr}[NLR]{narrow-line region}
\acrodef{hvc}[HVC]{high velocity cloud}
\acrodef{ivc}[IVC]{intermediate velocity cloud}
\acrodef{cmb}[CMB]{cosmic microwave background}
\acrodef{whim}[WHIM]{warm-hot intergalactic medium}

\acrodef{hst}[\emph{HST}]{Hubble Space Telescope}
\acrodef{cos}[COS]{Cosmic Origins Spectrograph}
\acrodef{fos}[FOS]{Faint Object Spectrograph}
\acrodef{stis}[STIS]{Space Telescope Imaging Spectrograph}
\acrodef{acs}[ACS]{Advanced Camera for Surveys}
\acrodef{vlt}[VLT]{Very Large Telescope}
\acrodef{vimos}[VIMOS]{\ac{vlt} Visible Multi-Object Spectrograph}
\acrodef{deimos}[DEIMOS]{Deep Imaging Multi-Object Spectrograph}
\acrodef{gmos}[GMOS]{Gemini Multi-Object Spectrograph}
\acrodef{cfht}[CFHT]{Canada-France-Hawaii Telescope}
\acrodef{wfccd}[WFCCD]{Wide Field Reimaging CCD Camera}
\acrodef{imacs}[IMACS]{Inamori-Magellan Areal Camera \& Spectrograph}
\acrodef{ldss3}[LDSS3]{Low Dispersion Survey Spectrograph 3}
\acrodef{fuse}[FUSE]{Far-Ultraviolet Spectroscopic Explorer}
\acrodef{ghrs}[GHRS]{Goddard High Resolution Spectrograph}

\acrodef{eagle}[\textsc{Eagle}]{Evolution and Assembly of GaLaxies and their Environments}
\acrodef{gimic}[\textsc{gimic}]{Galaxies-Intergalactic Medium Calculation}
\acrodef{owls}[\textsc{owls}]{OverWhelmingly Large Simulations}
\acrodef{sph}[SPH]{smoothed particle hydrodynamics}

\acrodef{sdss}[SDSS]{Sloan Digital Sky Survey}
\acrodef{2dfgrs}[2dFGRS]{2dF Galaxy Redshift Survey}
\acrodef{gama}[GAMA]{Galaxy and Mass Assembly}
\acrodef{vvds}[VVDS]{\ac{vlt} \ac{vimos} Deep Survey}
\acrodef{vipers}[VIPERS]{\ac{vimos} Public Extragalactic Redshift Survey}
\acrodef{gdds}[GDDS]{Gemini Deep Deep Survey}
\acrodef{cfhtls}[CFHTLS]{\ac{cfht} Legacy Survey}

\acrodef{stsci}[STScI]{Space Telescope Science Institute}
\acrodef{lco}[LCO]{Las Campanas Observatory}
\acrodef{opus}[OPUS]{Operational Pipeline Unified System}
\acrodef{mast}[MAST]{Mikulski Archive for Space Telescopes}

\acrodef{fwhm}[FWHM]{full width at half maximum}
\acrodef{snr}[SNR]{signal-to-noise ratio}
\acrodef{uv}[UV]{ultraviolet}
\acrodef{nuv}[NUV]{near ultraviolet}
\acrodef{fuv}[FUV]{far ultraviolet}
\acrodef{euv}[EUV]{extreme ultraviolet}
\acrodef{los}[LOS]{line-of-sight}
\acrodef{ota}[OTA]{optical tube assembly}
\acrodef{lsf}[LSF]{line-spread function}
\acrodef{psf}[PSF]{point-spread function}
\acrodef{psa}[PSA]{primary science aperture}
\acrodef{boa}[BOA]{bright object aperture}
\acrodef{mcp}[MCP]{microchannel plate}
\acrodef{mama}[MAMA]{multi-anode microchannel array}
\acrodef{hxdl}[HXDL]{helical cross delay line}
\acrodef{go}[GO]{general observer}
\acrodef{gto}[GTO]{guaranteed time observations}

\acrodef{iraf}[\textsc{iraf}]{Image Reduction and Analysis Facility}
\newcommand{\calcos}{\textsc{Calcos}}
\newcommand{\calfos}{\textsc{Calfos}}
\newcommand{\vpfit}{\textsc{vpfit}}

\newcommand{\esorex}{\textsc{esorex}}
\newcommand{\vipgi}{\textsc{vipgi}}
\newcommand{\scamp}{\textsc{Scamp}}
\newcommand{\sextractor}{\textsc{SExtractor}}

\newcommand{\gadget}{\textsc{gadget}}
\newcommand{\specwizard}{\textsc{SpecWizard}}

\crefname{figure}{fig.}{figs.}
\Crefname{figure}{Fig.}{Figs.}

\title[The O$\,$\textsc{vi}--galaxy cross-correlation at $z < 1$]{On the connection
between the metal-enriched intergalactic medium and galaxies: an
O$\,$\textsc{vi}--galaxy cross-correlation study at $\bmath{z < 1}$}
\author[Charles W. Finn et al.]{Charles W. Finn\hyperlink{a1}{$^1$}$^,$\hyperlink{a2}{$^2$}\thanks{E-mail: \href{mailto:c.w.finn2301@gmail.com}{c.w.finn2301@gmail.com}}, Simon L. Morris\hyperlink{a1}{$^1$}, Nicolas Tejos\hyperlink{a3}{$^3$}, Neil H. M. Crighton\hyperlink{a4}{$^4$}, \newauthor Robert Perry\hyperlink{a2}{$^2$}, Michele Fumagalli\hyperlink{a1}{$^1$}$^,$\hyperlink{a2}{$^2$}, Rich Bielby\hyperlink{a1}{$^1$}, Tom Theuns\hyperlink{a2}{$^2$}$^,$, Joop Schaye\hyperlink{a5}{$^5$}, \newauthor Tom Shanks\hyperlink{a1}{$^1$}, Jochen Liske\hyperlink{a6}{$^6$}, Madusha L.P. Gunawardhana\hyperlink{a1}{$^1$}$^,$\hyperlink{a2}{$^2$} and Stephanie Bartle\hyperlink{a1}{$^1$} \\
\hypertarget{a1}{$^1$}Department of Physics, Durham University, South Road, Durham, DH1 3LE, UK \\
\hypertarget{a2}{$^2$}Institute for Computational Cosmology, Department of Physics, Durham University, South Road, Durham, DH1 3LE, UK \\
\hypertarget{a3}{$^3$}University of California Observatories-Lick Observatory, University of California, Santa Cruz, CA 95064, USA \\
\hypertarget{a4}{$^4$}Centre for Astrophysics and Supercomputing, Swinburne University of Technology, PO Box 218, Victoria 3122, Australia \\
\hypertarget{a6}{$^5$}Leiden Observatory, Leiden University, PO Box 9513, NL-2300 RA Leiden, the Netherlands \\
\hypertarget{a7}{$^6$}Hamburger Sternwarte, Universit{\"a}t Hamburg, Gojenbergsweg 112, 21029 Hamburg, Germany}

\begin{document}
\date{Draft version}
\pagerange{\pageref{firstpage}--\pageref{lastpage}} \pubyear{2015}
\maketitle
\label{firstpage}

\begin{abstract}
We present new results on the auto- and cross-correlation functions of galaxies and \ion{O}{6} absorbers in a $\sim 18~\textrm{Gpc}^3$ comoving volume at $z < 1$. We use a sample of 51 296 galaxies and 140 \ion{O}{6} absorbers in the column density range $13 \lesssim \log N \lesssim 15$ to measure two-point correlation functions in the two dimensions transverse and orthogonal to the line-of-sight $\xi(r_{\perp}, r_{\parallel})$. We furthermore infer the corresponding `real-space' correlation functions, $\xi(r)$, by projecting $\xi(r_{\perp}, r_{\parallel})$ along $r_{\parallel}$, and assuming a power-law form, $\xi(r) = (r / r_0)^{-\gamma}$. Comparing the results from the absorber-galaxy cross-correlation function, $\xi_{\textrm{ag}}$, the galaxy auto-correlation function, $\xi_{\textrm{gg}}$, and the absorber auto-correlation function, $\xi_{\textrm{aa}}$, we constrain the statistical connection between galaxies and the metal-enriched intergalactic medium as a function of star-formation activity. We also compare these results to predictions from the \ac{eagle} cosmological hydrodynamical simulation and find a reasonable agreement. We find that: (i) \ion{O}{6} absorbers show very little velocity dispersion with respect to galaxies on $\sim$ Mpc scales, likely $\lesssim$ 100 \kms; (ii) \ion{O}{6} absorbers and galaxies may not linearly trace the same underlying distribution of matter in general. In particular, our results demonstrate that \ion{O}{6} absorbers are less clustered, and potentially more extended around galaxies than galaxies are around themselves; (iii) On $\gtrsim 100$ kpc scales, the likelihood of finding \ion{O}{6} absorbers around star-forming galaxies is similar to the likelihood of finding \ion{O}{6} absorbers around non star-forming galaxies; and (iv) \ion{O}{6} absorbers are either not ubiquitous to galaxies in our sample, or their distribution around them is patchy on scales $\gtrsim 100$ kpc (or both), at least for the column densities at which most are currently detected.
\end{abstract}


\section{Introduction}
\label{sec:introduction}
The connection between the \ac{igm} and galaxies is fundamental to our understanding of the formation and evolution of galaxies and the large-scale structure of the Universe. This is because there exists a continuous interplay between galaxies and the plasma around them, which fuels the formation of stars and the hierarchical assembly of cosmic structures. In the theoretical $\Lambda$ cold dark matter ($\Lambda$CDM) paradigm, the two main physical processes that drive this assembly are: (i) the accretion of intergalactic matter in `hot' and `cold' modes \cite[e.g.][]{1977MNRAS.179..541R,1978MNRAS.183..341W,1991ApJ...379...52W,2005MNRAS.363....2K,2006MNRAS.368....2D,2009ApJ...703..785D,2011MNRAS.414.2458V}; and (ii) winds emanating from galaxies generated mostly by \ac{sn} explosions and \ac{agn} \cite[e.g.][]{2005MNRAS.356.1191B,2005ARA&A..43..769V,2006MNRAS.370..645B,2008MNRAS.388..587L,2013MNRAS.429.1922C}. These winds are also thought to be responsible for enriching the \ac{igm} with metals \cite[e.g.][]{2001ApJ...559L...1S,2012Natur.492...79S}. Observational studies are producing results largely consistent with this picture, but better constraints are nevertheless required if we are to understand these processes in detail.

To achieve a thorough understanding of the formation of galaxies and cosmic structure, we must correctly describe the behaviour and evolution of the baryonic matter in the Universe. For this we require hydrodynamical simulations following the evolution of baryons and dark matter together within a cosmological volume \cite[e.g.][]{2009MNRAS.399.1773C,2010MNRAS.408.2051D,2014MNRAS.444.1518V,2015MNRAS.446..521S}. Large volumes are fundamentally important, since the simulations must be able to reproduce the statistics of the present day galaxy population. Unfortunately, due to the computational cost, there is a fundamental reliance on uncertain `sub-grid' prescriptions to capture the relevant physics on scales smaller than the resolution limit \cite[e.g.][]{2010MNRAS.402.1536S,2012MNRAS.423.1726S}. To glean trust-worthy physical insight from these simulations, we must therefore place tight constraints on the sub-grid physics. Observations of the gaseous environments around galaxies play a major role in this goal, as the simulations are not typically calibrated to match these observations. They therefore provide an important test that is independent of any `fine tuning'. In particular, observations of the metal-enriched components of the \ac{igm} are expected to provide especially robust constraints, since their distribution and dynamics are found to be sensitive to details of the typically implemented subgrid feedback prescriptions \cite[e.g.][]{2011MNRAS.415..353W,2013MNRAS.430.1548H,2013MNRAS.432...89F,2015MNRAS.448..895S}.

Unfortunately, despite being the main reservoir of baryons at all epochs, the extremely low density of the \ac{igm} makes it difficult to observe. The best method at present is through the analysis of absorption lines in \ac{qso} spectra. These lines appear due to the scattering of \ac{uv} photons by intervening gas along the \ac{los}. The resulting characterisation of the \ac{igm} is therefore often limited to being one-dimensional. Nevertheless, by combining information from multiple \ac{los}, we are able to construct a statistical picture of the distribution and dynamics of gas in the Universe \cite[e.g.][]{2009ApJ...701.1219C,2014MNRAS.437.2017T}.

Observations of the \ac{igm} at low redshifts ($z < 1$) have improved dramatically over the last few years with the advent of the \ac{cos} on the \ac{hst} \cite[][]{2012ApJ...744...60G}. With a sensitivity more than 10 times that of its predecessor, \ac{cos} has provided observations of hundreds of \acp{qso} in the \ac{fuv}. Observations at these wavelengths are fundamentally important, as they allow for a mapping of the \ion{H}{1} and metal content of the \ac{igm} to $z = 0$. The capabilities of \ac{cos} have been exploited extensively to probe both cool ($T \sim 10^4~\rm{K}$) gas, traced mostly by the \lya\ forest, and warmer ($\sim 10^5 - 10^6~\rm{K}$) gas, traced by broad \lya, \ion{O}{6} and \ion{Ne}{8} absorption \cite[e.g.][]{2011ApJ...743..180S,2011Sci...334..948T,2013ApJ...770..138L,2013ApJ...767...49M,2014MNRAS.445.2061L,2014ApJ...796...49S,2014ApJ...791..128S,2015MNRAS.446.2444H,2016ApJ...817..111D}. These ions probably trace up to $\sim 60$\% of all the baryons, with only $\sim 10$\% in the luminous constituents of the Universe (stars and galaxies), and the rest in an even hotter plasma at $T > 10^6~\rm{K}$ \cite[][]{2004ApJ...616..643F}.

To date, much of the work on the low-redshift IGM in relation to galaxies has taken a `galaxy-centric' approach, with a primary focus on the properties of the so-called \ac{cgm}. A number of successful programs have been designed with this goal in mind, notably the `\ac{cos}-Halos' survey \cite[][]{2013ApJ...777...59T}, and various programs by the \ac{cos} \ac{gto} team \cite[e.g.][]{2013ApJ...763..148S,2013ApJ...765...27K}. These studies implicitly assume a one-to-one correspondence between absorption systems and the closest observed galaxy, which is problematic due to the incomplete sampling of galaxies in any galaxy survey. Despite this shortcoming, it is clear from these studies that there is a nearly ubiquitous presence of cool ($T \approx 10^4-10^5$ K) metal-enriched gas surrounding galaxies to impact parameters of $\sim 150$ kpc \cite[see e.g.][]{2011ApJ...740...91P,2013ApJS..204...17W}. Ionization models suggest this cool \ac{cgm}, combined with an additional hotter component traced by collisionally ionized \ion{O}{6}, can account for at least half of the baryons expected from Big Bang nucleosynthesis that were originally unaccounted for \cite[][]{1998ApJ...503..518F,2010ApJ...708L..14M,2012ApJ...759...23S,2014ApJ...792....8W}, although see \cite{2013MNRAS.434.1063O} and \cite{2015MNRAS.446.3078V} for important caveats. Nevertheless, 30 -- 40\% of the baryons may still be unaccounted for, residing in the so-called \ac{whim} predicted by cosmological hydrodynamical simulations \cite[e.g.][]{1999ApJ...514....1C,2001ApJ...552..473D}. An unambiguous detection of the \ac{whim} is needed if we are to validate the predictions of these simulations \cite[see][for a recent attempt at addressing this problem]{2016MNRAS.455.2662T}.

In this paper, we address the connection between the metal-enriched \ac{igm} and galaxies at $z < 1$ via an analysis of the two-point cross- and auto-correlation functions of galaxies and \ion{O}{6} absorbers. The advantage of this approach is two-fold: (i) we do not rely on associating a particular intergalactic absorber with a particular galaxy (or set of galaxies), which in many instances is ambiguous; and (ii) we are robust to galaxy/absorber completeness variations, since we are measuring a clustering excess as a function of scale relative to a random expectation that takes into account the relevant selection functions \cite[][]{2014MNRAS.437.2017T}. Using these measurements, we are therefore able to investigate the distribution and dynamics of the metal-enriched \ac{igm} around galaxies on both the \ac{cgm} scale ($\lesssim 300$ kpc), and to much larger scales ($\gg 1$ Mpc). We use the \ion{O}{6} $\lambda\lambda1031,1037$ doublet largely because the component transitions have high oscillator strengths, and possess rest-frame wavelengths that make them accessible in the redshift range $0.1 \lesssim z \lesssim 0.7$ with current \ac{fuv} instrumentation. \ion{O}{6} absorbers are thus a convenient tracer of the metal-enriched gas in the \ac{igm}. In addition, they are thought to trace both cool, photoionized plasmas in the temperature range $10^4 \lesssim T \lesssim 10^5~\textrm{K}$, and hotter, collisionally ionized gas at temperatures $10^5 \lesssim T \lesssim 10^7~\textrm{K}$ \cite[e.g.][]{2001ApJ...563..724T,2005ApJ...624..555D,2008ApJ...679..194D,2008ApJS..179...37T,2011MNRAS.413..190T,2012MNRAS.420..829O,2014ApJS..212....8S,2014ApJ...791..128S}, the latter of which is commonly referred to as the \ac{whim} \cite[e.g.][]{1999ApJ...514....1C,2001ApJ...552..473D,2004ApJ...616..643F}. They may also form in more complicated scenarios, e.g. in conductive or turbulent interfaces between gaseous components at multiple temperatures \cite[e.g.][]{1990ApJ...355..501B,2010ApJ...719..523K}. Furthermore, the ionization fractions of \ion{O}{6} absorbers may be high in environments close to star-forming or post-starburst galaxies, or any galaxy where there has been recent or ongoing \ac{agn} activity due to non-equilibrium effects and long recombination time-scales \cite[e.g.][]{2013MNRAS.434.1063O,2013MNRAS.434.1043O,2015MNRAS.446.3078V}. This makes \ion{O}{6} absorbers effective tracers of metal enriched gas in environments like these, even for low metallicities. We bear in mind the many, potentially complex formation scenarios for \ion{O}{6} absorbers in the interpretation of our results.

This paper is structured as follows. In \Cref{sec:observations}, we describe the observational data sets used in this work. In \Cref{sec:igm_analysis,sec:galaxy_analysis}, we describe the data analysis relating to \ac{igm} absorption systems and galaxies respectively. In \Cref{sec:eagle}, we describe the creation of a set of comparison data drawn from the \ac{eagle} cosmological hydrodynamical simulation. In \Cref{sec:correlation_analysis}, we describe the mathematical formalisms used to compute the auto- and cross-correlation functions of galaxies and absorbers, and describe the creation of the random samples that are crucial to this analysis. In \Cref{sec:results}, we present the results of our correlation function analysis. In \Cref{sec:discussion}, we present a discussion of these results and a comparison to the literature. In \Cref{sec:summary_conclusions}, we summarise our findings and outline the main conclusions of this work.

All distances are in comoving coordinates unless otherwise stated. We assume a $\Lambda$CDM cosmology, with parameters set to the best-fit values determined from the 2013 analysis of data from the \emph{Planck} satellite \cite[][]{2014A&A...571A..16P}.

\section{Observations}
\label{sec:observations}
The observational data consists of 50 independent fields with small angular coverage in each of which we have at least one QSO spectrum obtained from \ac{hst}/\ac{cos} and a large number of spectra, or spectroscopic measurements of galaxies at $z \lesssim 1$. There are 60 QSOs in the sample. Out of 50 fields, 6 are inherited from our previous work on the \ion{H}{1}-galaxy cross correlation \cite[][]{2014MNRAS.437.2017T}. We have greatly expanded on this original sample by incorporating a large number of publicly available data sets.  We describe in detail the sample of galaxies and \acp{qso} and summarise the data reduction procedures in the following subsections.

	\subsection{QSOs}
	\label{sec:qsos}
	We have used \ac{hst}/\ac{cos} and \ac{fos} spectroscopy of 60 \acp{qso} to characterise the diffuse \ac{igm} through analysis of intervening \ion{H}{1} and metal absorption systems. Of these, 7 have had \ac{cos} spectroscopy obtained and presented by our collaboration in previous work \cite[][]{2013MNRAS.433..178C,2014MNRAS.437.2017T,2014MNRAS.440.3317F}. Details of the \ac{qso} observations are summarised in \Cref{tab:qso_sample}.

    \begin{table*}
    \caption{The \ac{qso} sample}
    \begin{threeparttable}
    \begin{tabular}{l c c c c c c c l}
        \hline
        QSO name &
        \multicolumn{2}{c}{G130M} &
        \multicolumn{2}{c}{G160M} &
        \multicolumn{2}{c}{NUV\tnotex{qso_sample:3}} &
        Programme ID(s) &
        P.I.(s) \\
        &
        $t_{\textrm{exp}}$ (ks)\tnotex{qso_sample:1} &
        $\textrm{S} / \textrm{N}$\tnotex{qso_sample:2} &
        $t_{\textrm{exp}}$ (ks)\tnotex{qso_sample:1} &
        $\textrm{S} / \textrm{N}$\tnotex{qso_sample:2} &
        $t_{\textrm{exp}}$ (ks)\tnotex{qso_sample:1} &
        $\textrm{S} / \textrm{N}$\tnotex{qso_sample:2}
        &
        & \\
        \hline
        PG $0003{+}158$              & 10.4 & 22 & 10.9 & 19 & --   & -- & 12038                   & Green           \\
        PG $0026{+}129$              & 1.9  & 18 & --   & -- & --   & -- & 12569                   & Veilleux        \\
        HE $0056{-}3622$             & 7.8  & 25 & 5.7  & 15 & --   & -- & 12604                   & Fox             \\
        LBQS $0107{-}0235$A          & 28.1 & 9  & 44.3 & 8  & 35.6 & 30 & 11585, 6592, 6100, 5320 & Crighton, Foltz \\
        LBQS $0107{-}0235$B          & 21.2 & 9  & 21.2 & 7  & 13.0 & 30 & 11585, 6592, 6100, 5320 & Crighton, Foltz \\
        LBQS $0107{-}0232$           & --   & -- & 83.5 & 7  & 32.8 & 18 & 11585, 6592, 6100       & Crighton, Foltz \\
        B$0117{-}2837$               & 5.2  & 24 & 8.5  & 19 & --   & -- & 12204                   & Thom            \\
        Ton S210                     & 5.0  & 41 & 5.5  & 26 & --   & -- & 12204                   & Thom            \\
        PG $0157{+}001$              & 1.8  & 16 & --   & -- & --   & -- & 12569                   & Veilleux        \\
        FBQS J$0209{-}0438$          & 14.0 & 12 & 28.1 & 10 & 14.4 & 12 & 12264                   & Morris          \\
        HE $0226{-}4110$             & 8.8  & 34 & 7.8  & 24 & --   & -- & 11541                   & Green           \\
        PKS $0405{-}123$             & 24.2 & 59 & 11.1 & 30 & --   & -- & 11508, 11541            & Noll, Green     \\
        RBS 542                      & 23.2 & 61 & 15.2 & 35 & --   & -- & 11686                   & Arav            \\
        PKS $0558{-}504$             & 1.8  & 19 & 0.7  & 10 & --   & -- & 11692                   & Howk            \\
        SDSS J$080908.13{+}461925.6$ & 5.7  & 15 & 5.0  & 13 & --   & -- & 12248                   & Tumlinson       \\
        PG $0832{+}251$              & 8.8  & 14 & 6.8  & 12 & --   & -- & 12025                   & Green           \\
        PG $0844{+}349$              & 1.9  & 18 & --   & -- & --   & -- & 12569                   & Veilleux        \\
        Mrk 106                      & 9.3  & 28 & 7.6  & 18 & --   & -- & 12029                   & Green           \\
        RXS J$09565{-}0452$          & 7.7  & 16 & --   & -- & --   & -- & 12275                   & Wakker          \\
        PG $0953{+}414$              & 6.2  & 38 & 5.6  & 26 & --   & -- & 12038                   & Green           \\
        PG $1001{+}291$              & 7.1  & 21 & 6.8  & 17 & --   & -- & 12038                   & Green           \\
        HE $1003{+}0149$             & 14.0 & 9  & 22.3 & 9  & --   & -- & 12264                   & Morris          \\
        FBQS J$1010{+}3003$          & 12.8 & 17 & 10.8 & 10 & --   & -- & 12025                   & Green           \\
        Ton 1187                     & 2.0  & 16 & --   & -- & --   & -- & 12275                   & Wakker          \\
        PG $1011{-}040$              & 6.7  & 29 & 4.7  & 18 & --   & -- & 11524                   & Green           \\
        LBQS $1019{+}0147$           & 2.2  & 6  & 2.9  & 5  & --   & -- & 11598                   & Tumlinson       \\
        1ES $1028{+}511$             & 20.0 & 20 & 14.6 & 13 & --   & -- & 12025                   & Green           \\
        1SAX J$1032.3{+}5051$        & 13.5 & 12 & 11.3 & 6  & --   & -- & 12025                   & Green           \\
        PG $1048{+}342$              & 7.8  & 23 & 11.0 & 16 & --   & -- & 12024                   & Green           \\
        PG $1049{-}005$              & 3.6  & 14 & 2.8  & 12 & --   & -- & 12248                   & Tumlinson       \\
        HS $1102{+}3441$             & 11.3 & 17 & 11.3 & 13 & --   & -- & 11541                   & Green           \\
        SBS $1108{+}560$             & 10.6 & 4  & 8.9  & 14 & --   & -- & 12025                   & Green           \\
        PG $1115{+}407$              & 5.1  & 23 & 5.7  & 15 & --   & -- & 11519                   & Green           \\
        PG $1116{+}215$              & 6.1  & 39 & 5.5  & 28 & --   & -- & 12038                   & Green           \\
        PG $1121{+}422$              & 5.0  & 21 & 5.8  & 13 & --   & -- & 12604                   & Fox             \\
        SBS $1122{+}594$             & 9.9  & 14 & 10.5 & 12 & --   & -- & 11520                   & Green           \\
        Ton 580                      & 4.9  & 21 & 5.6  & 16 & --   & -- & 11519                   & Green           \\
        3C 263                       & 15.4 & 34 & 18.0 & 23 & --   & -- & 11541                   & Green           \\
        PG $1216{+}069$              & 5.1  & 24 & 5.6  & 16 & --   & -- & 12025                   & Green           \\
        3C 273                       & 4.0  & 73 & --   & -- & --   & -- & 12038                   & Green           \\
        HE $1128{+}0131$             & 13.2 & 44 & 11.0 & 36 & --   & -- & 11686                   & Arav            \\
        PG $1229{+}204$              & 1.9  & 17 & --   & -- & --   & -- & 12569                   & Veilleux        \\
        PG $1259{+}593$              & 12.0 & 32 & 11.2 & 24 & --   & -- & 11541                   & Green           \\
        PKS $1302{-}102$             & 7.4  & 26 & 6.9  & 20 & --   & -- & 12038                   & Green           \\
        PG $1307{+}085$              & 1.8  & 18 & --   & -- & --   & -- & 12569                   & Veilleux        \\
        SDSS J$135726.27{+}043541.4$ & 14.0 & 9  & 28.1 & 7  & 14.4 & 11 & 12264                   & Morris          \\
        PG $1424{+}240$              & 6.4  & 21 & 7.9  & 21 & --   & -- & 12612                   & Stocke          \\
        PG $1435{-}067$              & 1.9  & 14 & --   & -- & --   & -- & 12569                   & Veilleux        \\
        LBQS $1435{-}0134$           & 22.3 & 23 & 34.2 & 16 & --   & -- & 11741                   & Tripp           \\
        Mrk 478                      & 1.9  & 18 & --   & -- & --   & -- & 12569                   & Veilleux        \\
        Ton 236                      & 8.3  & 18 & 9.4  & 15 & --   & -- & 12038                   & Green           \\
        1ES $1553{+}113$             & 10.8 & 33 & 12.3 & 26 & --   & -- & 11520, 12025            & Green           \\
        Mrk 877                      & 1.8  & 18 & --   & -- & --   & -- & 12569                   & Veilleux        \\
        PKS $2005{-}489$             & 2.5  & 24 & 1.9  & 15 & --   & -- & 11520                   & Green           \\
        Mrk 1513                     & 6.9  & 32 & 4.8  & 20 & --   & -- & 11524                   & Green           \\
        PHL 1811                     & 3.9  & 36 & 3.1  & 24 & --   & -- & 12038                   & Green           \\
        PKS $2155{-}304$             & 4.6  & 45 & --   & -- & --   & -- & 12038                   & Green           \\
        FBQS J$2218{+}0052$          & --   & -- & --   & -- & 20.2 & 10 & 12264                   & Morris          \\
        MR $2251{-}178$              & 5.6  & 38 & 5.4  & 30 & --   & -- & 12029                   & Green           \\
        4C 01.61                     & 1.8  & 20 & --   & -- & --   & -- & 12569                   & Veilleux        \\
        \hline
    \end{tabular}
    \begin{tablenotes}
        \item[1] \label{qso_sample:1} Total exposure time in ks.
        \item[2] \label{qso_sample:2} Median \ac{snr} per resolution element.
        \item[3] \label{qso_sample:3} FOS gratings G270H and/or G190H for LBQS $0107{-}0235$A, LBQS $0107{-}0235$B and LBQS $0107{-}0232$, COS G230L grating otherwise.
    \end{tablenotes}
    \end{threeparttable}
    \label{tab:qso_sample}
    \end{table*}

		\subsubsection{The sample}
		\label{sec:qso_sample}
		The \acp{qso} used in this work were selected to lie in fields well surveyed for their galaxy content, and having spectroscopy with good \ac{snr} ($\gtrsim 10$). Their spectra have been obtained by a number of collaborations, including our own, for a variety of specific programmes. In \Cref{tab:qso_sample}, we list the \ac{hst} proposal ID(s) and principal investigator(s) associated with the data obtained for each \ac{qso}, and we refer the reader to the proposal abstracts for details on the associated science cases.

		To obtain high \ac{snr} observations in a reasonable amount of observing time with \ac{hst}, most of the \acp{qso} in this sample were selected on the basis of having \ac{fuv} fluxes $\gtrsim 100 \mu\textrm{J}$. This biases the \ac{qso} sample to be of high luminosity, which potentially has implications for their local environments. However, we note that the regions of the Universe along the \ac{los} to most of these \acp{qso} are effectively random, and we proceed with this assumption throughout the forthcoming analysis.

		\subsubsection{Data reduction}
		\label{sec:qso_data_reduction}
		All of the \ac{cos} data were reduced with the \calcos\ pipeline. In particular, the \ac{cos} \ac{qso} spectra obtained by our collaboration, namely, LBQS J$0107{-}0235$A, LBQS J$0107{-}0235$B, LBQS J$0107{-}0232$, FBQS J$0209{-}0438$, HE $1003{+}0149$, SDSS J$135726.27{+}043541.4$ and FBQS J$2218{+}0052$, and additionally LBQS $1019{+}0147$ and LBQS $1435{-}067$, were reduced using v2.18.5 of the pipeline in combination with \textsc{Python} routines developed by the authors,\footnote{Available at \url{https://github.com/cwfinn/COS/}} which are based loosely on IDL routines developed by the \ac{cos} GTO team.\footnote{\url{http://casa.colorado.edu/danforth/science/cos/costools.html}} For full details, see \cite{2014MNRAS.437.2017T} and \cite{2014MNRAS.440.3317F}. All of the other \ac{cos} spectra were reduced as described in \cite{2014arXiv1402.2655D}, using \calcos\ versions contemporary with their observation epoch.

		\ac{fos} data were reduced using the \calfos\ pipeline. We refer the reader to \cite{2014MNRAS.437.2017T} for full details. 

	\subsection{Galaxies}
	\label{sec:galaxies}
	The galaxy data is obtained from a number of different instruments and surveys. We include data collected by our own collaboration from the \ac{deimos}, \ac{gmos}, \ac{cfht} multi-object spectrograph and \ac{vimos} \cite[hereafter, T14 and T14-Q0107;][]{2006MNRAS.367.1261M,2014MNRAS.437.2017T}. We make use of the \ac{sdss} \cite[][]{2009ApJS..182..543A}, \ac{2dfgrs} \cite[][]{2001MNRAS.328.1039C}, \ac{gama} survey \cite[][]{2011MNRAS.413..971D}, \ac{vvds} \cite[][]{2005A&A...439..845L} and \ac{vipers} \cite[][]{2014A&A...566A.108G}. We also include data from the \ac{lco}/\ac{wfccd} galaxy survey of 20 fields surrounding UV-bright \acp{qso} \cite[hereafter, P11;][]{2011ApJS..193...28P}, galaxy data around PKS $0405{-}123$ presented in \citet[hereafter, J13]{2013MNRAS.434.1765J}; and galaxy data around HE $0226{-}4110$ and PG $1216{+}069$ presented in \citet[hereafter, C09]{2009ApJ...701.1219C}. The latter two surveys made use of the \ac{imacs} and \ac{ldss3} at \ac{lco}.

	\begin{table*}
	\setlength{\tabcolsep}{12pt}
	\caption{The galaxy sample}
	\begin{threeparttable}
	\begin{tabular}{l c c c c l}
		\hline
		Survey &
		$N_{\textrm{galaxies}}$\tnotex{galaxy_sample:1} &
		$z_{\textrm{median}}$\tnotex{galaxy_sample:2} &
		$z_{95}$\tnotex{galaxy_sample:3} &
		$m_{\textrm{limit}}$\tnotex{galaxy_sample:4} &
		Reference \\
		\hline
		SDSS      & 41 342 & 0.10 & 0.19 & $r < 17.77$                        & {\cite{2009ApJS..182..543A}} \\
		2dFGRS    & 10 643 & 0.11 & 0.22 & $b_{\textrm{J}} < 19.45$           & {\cite{2001MNRAS.328.1039C}} \\
		GAMA      & 8636   & 0.22 & 0.40 & $r < 19.8$                         & {\cite{2011MNRAS.413..971D}} \\
		VVDS      & 18 181 & 0.58 & 1.07 & $I < 22.5$                         & {\cite{2005A&A...439..845L}} \\
		VIPERS    & 24 183 & 0.70 & 1.06 & $I < 22.5$\tnotex{galaxy_sample:a} & {\cite{2014A&A...566A.108G}} \\
		T14       & 1049   & 0.43 & 0.93 & $R < 23.5$\tnotex{galaxy_sample:b} & {\cite{2014MNRAS.437.2017T}} \\
		T14-Q0107 & 962    & 0.55 & 1.07 & various\tnotex{galaxy_sample:c}    & {\cite{2014MNRAS.437.2017T}} \\
		P11       & 900    & 0.16 & 0.36 & R < 20\tnotex{galaxy_sample:d}     & {\cite{2011ApJS..193...28P}} \\
		C09       & 810    & 0.36 & 0.64 & R < 22                             & {\cite{2009ApJ...701.1219C}} \\
		J13       & 443    & 0.41 & 0.81 & R < 23                             & {\cite{2013MNRAS.434.1765J}} \\
		\hline
		\textbf{ALL} & \textbf{107 149} & \textbf{0.19} & \textbf{1.00} & -- & -- \\
		\hline
	\end{tabular}
	\begin{tablenotes}
		\item[1] \label{galaxy_sample:1} Number of galaxies with spectroscopically confirmed redshifts (not labelled `c' - see \Cref{sec:redshift_determination}).
		\item[2] \label{galaxy_sample:2} Median redshift for the survey.
		\item[3] \label{galaxy_sample:3} The 95th percentile of the redshift distribution.
		\item[4] \label{galaxy_sample:4} Magnitude limit for the survey.
		\item[a] \label{galaxy_sample:a} Colour cuts also applied.
		\item[b] \label{galaxy_sample:b} Priority given to objects with $R < 22.5$.
		\item[c] \label{galaxy_sample:c} VIMOS: $R < 23$, priority given to objects with $R < 22$. DEIMOS: $R < 24.5$, priority given to brighter objects, colour cuts also applied. GMOS: Top priority given to objects with $R < 22$, second priority given to objects with $22 < R < 23$, last priority given to objects with $23 < R < 24$. CFHT: $R < 23.5$ (indicative only).
		\item[d] \label{galaxy_sample:d} $R < 19.5$ for some fields.
	\end{tablenotes}
	\end{threeparttable}
	\label{tab:galaxy_sample}
	\end{table*}

	\begin{table*}
	\setlength{\tabcolsep}{12pt}
	\caption{QSO sight-line fields}
	\begin{threeparttable}
	\begin{tabular}{l c c c l}
		\hline
  		Field name &
    	$N_{\textrm{QSO}}$ &
    	Area (sr)\tnotex{tn3:a} &
    	$V_{\textrm{c}}$ (Gpc$^3$)\tnotex{tn3:b} &
    	Instrument/survey \\
  		\hline
  		J$0005{+}1609$ & 1 & 0.00221 & 0.181 & SDSS                      \\
    	J$0029{+}1316$ & 1 & 0.00003 & 0.002 & WFCCD                     \\
    	J$0058{-}3606$ & 1 & 0.00218 & 0.153 & 2dFGRS                    \\
    	J$0110{-}0218$ & 3 & 0.00004 & 0.049 & CFHT, VIMOS, DEIMOS, GMOS \\
    	J$0120{-}2821$ & 2 & 0.00487 & 0.623 & 2dFGRS                    \\
    	J$0159{+}0023$ & 1 & 0.00309 & 0.215 & SDSS                      \\
    	J$0209{-}0438$ & 1 & 0.00314 & 4.700 & VIPERS                    \\
    	J$0228{-}1904$ & 1 & 0.00005 & 0.024 & IMACS, LDSS3              \\
    	J$0407{-}1211$ & 1 & 0.00029 & 0.176 & WFCCD, IMACS, LDSS3       \\
    	J$0426{-}5712$ & 1 & 0.00135 & 0.041 & 2dFGRS                    \\
   	 	J$0559{-}5026$ & 1 & 0.00004 & 0.002 & WFCCD                     \\
    	J$0809{+}4619$ & 1 & 0.00487 & 0.436 & SDSS                      \\
    	J$0835{+}2459$ & 1 & 0.00487 & 0.369 & SDSS                      \\
    	J$0847{+}3445$ & 1 & 0.00487 & 0.059 & SDSS                      \\
    	J$0919{+}5521$ & 1 & 0.00487 & 0.201 & SDSS                      \\
    	J$0956{-}0453$ & 1 & 0.00332 & 0.210 & 2dFGRS                    \\
    	J$0956{+}4115$ & 2 & 0.00487 & 0.452 & SDSS                      \\
    	J$1005{+}0134$ & 1 & 0.00082 & 1.082 & SDSS, VVDS, VIMOS         \\
    	J$1007{+}2929$ & 2 & 0.01097 & 1.042 & SDSS                      \\
    	J$1013{+}3551$ & 1 & 0.00487 & 0.088 & SDSS                      \\
    	J$1014{-}0418$ & 1 & 0.00487 & 0.049 & 2dFGRS                    \\
    	J$1022{+}0132$ & 1 & 0.00004 & 0.027 & SDSS, VIMOS               \\
    	J$1031{+}5052$ & 1 & 0.00487 & 0.467 & SDSS                      \\
    	J$1051{-}0051$ & 1 & 0.00289 & 0.213 & SDSS                      \\
    	J$1058{+}3412$ & 2 & 0.01097 & 1.010 & SDSS                      \\
    	J$1118{+}5728$ & 2 & 0.01950 & 1.697 & SDSS                      \\
    	J$1119{+}2119$ & 1 & 0.00363 & 0.291 & SDSS, WFCCD               \\
    	J$1121{+}4113$ & 1 & 0.01097 & 1.060 & SDSS                      \\
    	J$1131{+}3114$ & 1 & 0.00487 & 0.502 & SDSS                      \\
    	J$1139{+}6547$ & 1 & 0.00487 & 0.378 & SDSS                      \\
    	J$1226{+}0319$ & 3 & 0.01950 & 3.638 & SDSS, IMACS, LDSS3, WFCCD \\
    	J$1232{+}2009$ & 1 & 0.00487 & 0.057 & SDSS                      \\
    	J$1301{+}5902$ & 2 & 0.00487 & 0.488 & SDSS                      \\
    	J$1305{-}1033$ & 1 & 0.00004 & 0.007 & WFCCD                     \\
    	J$1309{+}0819$ & 1 & 0.00487 & 0.309 & SDSS, WFCCD               \\
    	J$1357{+}0435$ & 1 & 0.00091 & 1.225 & SDSS, VVDS                \\
    	J$1427{+}2348$ & 1 & 0.00487 & 0.372 & SDSS                      \\
    	J$1437{-}0147$ & 1 & 0.00487 & 1.646 & GAMA                      \\
    	J$1438{-}0658$ & 1 & 0.00182 & 0.078 & 2dFGRS                    \\
    	J$1442{+}3526$ & 1 & 0.00487 & 0.088 & SDSS                      \\
    	J$1528{+}2825$ & 1 & 0.00487 & 0.382 & SDSS                      \\
    	J$1555{+}1111$ & 1 & 0.00487 & 0.577 & SDSS, WFCCD               \\
    	J$1620{+}1724$ & 1 & 0.00487 & 0.169 & SDSS                      \\
    	J$2009{-}4849$ & 1 & 0.00004 & 0.001 & WFCCD                     \\
    	J$2132{+}1008$ & 1 & 0.00296 & 0.035 & SDSS                      \\
    	J$2155{-}0922$ & 1 & 0.00003 & 0.003 & WFCCD                     \\
    	J$2158{-}3013$ & 1 & 0.00487 & 0.183 & 2dFGRS, WFCCD             \\
    	J$2218{+}0052$ & 1 & 0.00109 & 1.464 & SDSS, VVDS, VIMOS         \\
    	J$2254{-}1734$ & 1 & 0.00122 & 0.015 & 2dFGRS                    \\
    	J$2351{-}0109$ & 1 & 0.00256 & 0.201 & SDSS                      \\
  		\hline
	\end{tabular}
	\begin{tablenotes}
    	\item[a] \label{tn3:a} Field area, approximating the survey region as a rectangle.
    	\item[b] \label{tn3:b} Comoving volume covered by the survey up to the minimum of $(z_{\textrm{QSO}}, z_{95})$.
	\end{tablenotes}
	\end{threeparttable}
	\label{tab:fields}
	\end{table*}

		\subsubsection{The sample}
		\label{sec:galaxy_sample}
		The surveys that make up our galaxy sample cover regions close to all of the \ac{qso} sight-lines used to characterise the \ac{igm}. Some were conducted for the primary purpose of mapping galaxies close to a particular \ac{qso} sight-line, while others serendipitously cover regions where there are bright \acp{qso} with \ac{hst} spectroscopy. Those that fall in the latter category are the large \ac{sdss}, \ac{2dfgrs}, \ac{gama}, \ac{vvds} and \ac{vipers} surveys. For SDSS, we adopt just those galaxies in the main sample, i.e. SDSS-I/II \cite[see][for details]{2009ApJS..182..543A}. We restrict our combined galaxy sample to $4 \times 4$ square degree fields centred on each \ac{qso}.\footnote{Note that a number of surveys cover areas of sky that are smaller than this.} This means that we can sample galaxy-absorber pairs to transverse separations of $\sim 15$ comoving Mpc at the median redshift of our sample ($z_{\textrm{median}} = 0.19$), $\sim 10$ comoving Mpc at $z \sim 0.07$ and $\sim 1$ comoving Mpc at $z \sim 0.005$. We discard all objects with $z < 0.005$, regardless of their classification, on the basis that they may be stars. Some fields are made larger by virtue of there being more than one \ac{qso} that inhabits a particular $4 \times 4$ square degree region.
		
		We summarise our combined galaxy sample in \Cref{tab:galaxy_sample}. As an indication of survey depth, we list the median redshift for each survey, and the 95th percentile of the redshift distribution, which we denote $z_{95}$. This is more informative than the maximum of the redshift distribution, as many surveys show long tails to high redshift due to the presence of luminous \ac{agn}. We also list the magnitude limit for each survey, which in many cases is only indicative (see the table footnotes for more details). There are 107 149 galaxies in our combined sample, which has a median redshift of 0.19. In \Cref{tab:fields}, we summarise the \ac{qso} sight-line fields. We list the number of \acp{qso} in each field and give an indication of the area and comoving volume covered by each. For the latter, we define the edge of the volume by the minimum of $(z_{\textrm{QSO}}, z_{95})$, where $z_{\textrm{QSO}}$ denotes the maximum \ac{qso} redshift for the field.

		\subsubsection{Data reduction}
		\label{sec:galaxy_data_reduction}
		Galaxy data obtained with VIMOS pre-2011 were reduced using \vipgi\ pipeline \cite[][]{2005PASP..117.1284S}, and after this time using the \esorex\ pipeline with the exception of that taken for \ac{vipers}, which has its own dedicated pipeline \cite[][]{2014A&A...566A.108G}. For full details, see \cite{2005A&A...439..845L} and \cite{2014MNRAS.437.2017T}. Data from \ac{deimos} was reduced using the DEEP2 \ac{deimos} Data Pipeline \cite[][]{2013ApJS..208....5N}.\footnote{\url{http://astro.berkeley.edu/\~cooper/deep/spec2d/}} \ac{gmos} data was reduced using the Gemini \ac{iraf} \cite[see][for details]{2014MNRAS.437.2017T}. The reduction of the \ac{cfht} data is described in \cite{2006MNRAS.367.1261M}. SDSS, GAMA and 2dFGRS data reduction procedures are described in \cite{2002AJ....123..485S}, \cite{2013MNRAS.430.2047H} and \cite{2001MNRAS.328.1039C} respectively. Details on the reduction of the \ac{lco}/\ac{wfccd} data can be found in \cite{2011ApJS..193...28P}. The reduction of the \ac{lco}/\ac{imacs} and \ac{ldss3} data is described fully in \cite{2009ApJ...701.1219C} and \cite{2013MNRAS.434.1765J}.

\section{Analysis of the IGM data}
\label{sec:igm_analysis}
The following sections briefly describe the processes involved in creating absorption line lists from the reduced \ac{cos} and \ac{fos} data obtained and/or analysed by our collaboration. For more description, see \cite{2014MNRAS.437.2017T} and \cite{2014MNRAS.440.3317F}. For the majority of the \acp{qso}, we obtained absorption line lists directly as a result of the analysis in \cite{2014arXiv1402.2655D}, which was downloaded as a high-level science product from the \ac{mast}.\footnote{\url{http://archive.stsci.edu/prepds/igm/}} These lists were assembled using an automated line identification and fitting algorithm, with subsequent human verification. We refer the reader to \cite{2014arXiv1402.2655D} for a full description of the analysis and line list creation for these spectra. It is important to note that the absorption line lists presented by \cite{2014arXiv1402.2655D} have been updated since the analysis conducted in this paper. These new results are presented in \cite{2016ApJ...817..111D}. They are based on the analysis of a further seven \ac{agn} sight-lines, and have better detection statistics owing to improved spectrum extraction and background subtraction. Although a refreshed analysis using this new data set is desirable, the general conclusions in this paper are likely to remain valid. In particular, any spurious line detections in the older analysis of \cite{2014arXiv1402.2655D} should only act to decrease the overall statistical significance of our results, rather than changing their implications.

	\subsection{Continuum fitting}
	\label{sec:continuum_fitting}
	Before line identification and fitting, the reduced QSO spectra are normalised by an estimate of the pseudo continuum (continuum emission $+$ line emission). We estimate this using a technique similar to that described in \cite{1979ApJ...229..891Y}, \cite{1982MNRAS.198...91C} and \cite{2002ApJ...576....1A}. Each spectrum is split up into an arbitrary number of wavelength intervals, and a cubic spline fit through the set of points defined by the median flux in each interval. Pixels falling an arbitrary $n \sigma$ below the continuum are rejected, the median flux is recalculated, and the fit performed again. Here $\sigma$ is the standard deviation of the flux in each wavelength interval. We iterate over this process until the fit converges with an approximately Gaussian distribution of flux values above the continuum. The appropriate value of $n$ is found to vary from spectrum to spectrum, with values adopted in the range 1.5 to 3. From trial and error, the best value depends on \ac{snr} and location either within, or outside of the \lya\ forest.

	The continuum fitting process described above generally works well in regions of the spectra where the continuum varies smoothly. For regions where it fails, we adjust the continuum manually by hand. This is typically at the cusps of emission lines, in the Galactic \lya\ absorption trough, at the absorption edge of Lyman limit systems and at the detector edges.

	\subsection{Absorption line identification}
	\label{sec:line_identification}
	We identified absorption lines attributable to a particular ion and transition by performing a manual search through each \ac{qso} spectrum. We begin by searching for Galactic absorption lines at $z = 0$ and associated absorption lines at $z_{\textrm{QSO}}$. We then work systematically from $z_{\textrm{QSO}}$ to $z = 0$, identifying \ion{H}{1} absorbers on the basis of there being at least two clearly detected Lyman series transitions at a given redshift. We simultaneously identify any metal absorbers coincident with the redshift of these \ion{H}{1} absorbers.\footnote{Coincident here loosely means at $\Delta v \lesssim 50~\textrm{km s}^{-1}$. We are empirically motivated to search for metal absorbers at small (or zero) velocity separations from high column density H$\,$\textsc{i} absorbers ($\log N > 10^{15}~\textrm{cm}^{-2}$), but we do not make any prior assumption on the physical mechanisms that give rise to these coincidences.} Next we scan through each spectrum again, identifying any `high-ionization' doublets (namely \ion{Ne}{8}, \ion{O}{6}, \ion{N}{5}, \ion{C}{4} and \ion{Si}{4}) that may appear independently of any \ion{H}{1} absorption. Finally we assume lines in short wavelength regions of the spectra where there is no \lyb\ coverage to be attributable to \lya, and again look for coincident metal absorbers. For all identified ions we set an initial guess for the number of velocity components, and for each component a column density and Doppler broadening parameter. This process typically accounts for $> 95$\% of all absorption lines with equivalent widths at the $> 3 \sigma$ significance level.\footnote{See \cite{2012PASP..124..830K} for a detailed discussion on the significance of absorption lines in \ac{hst}/\ac{cos} spectra.}

	\subsection{Voigt profile fitting}
	\label{sec:vp_fitting}
	We fit Voigt profiles to the identified absorption-line systems with \vpfit,\footnote{\url{http://www.ast.cam.ac.uk/\~rfc/vpfit.html}} accounting for the non-Gaussian \ac{cos} \ac{lsf} at each wavelength by interpolating between the tables provided by the \ac{stsci}.\footnote{\url{http://www.stsci.edu/hst/cos/performance/spectral_resolution}} In \vpfit, a $\chi^2$ minimisation is performed to fit Voigt profiles that are first convolved with the wavelength dependent \ac{cos} \ac{lsf}.

	We begin with an initial list of guesses provided by the identification algorithm described in the previous section, and give these as input to \vpfit. All transitions of a given ion in a given system are fitted simultaneously, such that for each system, every transition of that ion shares the same redshift. Here by `system' we refer to the set of transitions belonging to a single ion at a single redshift, and we shall hereafter refer to these `systems' as `absorbers'. In general, we do not make any assumption as to whether different ions belong together at the same redshift in the same physical absorption complex \cite[although see][for a special case]{2014MNRAS.440.3317F}. Therefore the redshifts for coincident ions are free to vary, consistent with the observation that some coincident ions show small velocity offsets. Fitted profiles are visually inspected, and initial guesses tweaked in rare cases where \vpfit\ fails to converge on a sensible result. We adopt only the minimum number of velocity components needed to minimise the reduced $\chi^2$ value on the fit.

	\subsection{The absorption line catalogues}
	\label{sec:absorption_line_catalogues}
	For each \ac{qso} spectrum, we compile an absorption line list based on the identification and Voigt profile fitting procedures just described. These are lists of absorbers (in a given ion), where each has a redshift ($z$), log column density ($\log N$) and Doppler broadening parameter ($b$), together with the associated $1 \sigma$ uncertainties calculated during the fitting process. We also assign each absorber the right-ascension and declination of the \ac{qso}, such that it can be assigned a unique position in redshift space for cross-correlation. Additionally, a flagging system is employed to categorise the reliability of each absorber identification/fit. This scheme is similar to that in \cite{2014MNRAS.437.2017T} and is defined as follows:
	\begin{itemize}
		\item \emph{Secure} (`a'): systems that are detected on the basis of at least two transitions (in the same ion), with $\log N / \Delta(\log N) > 30$, and each transition having an equivalent width significant at the $> 3\sigma$ level.
		\item \emph{Probable} (`b'): \ion{H}{1} systems detected on the basis of \lya\ only (after ruling out all other possibilities and with equivalent widths significant at the $> 4\sigma$ level), or metal-line systems detected on the basis of one transition with equivalent widths significant at the $> 3\sigma$ level and with one or more accompanying Lyman series transitions. Both possibilities also with the requirement $\log N / \Delta(\log N) > 30$.
		\item \emph{Uncertain} (`c'): systems for which $\log N / \Delta(\log N) < 30$ and/or equivalent widths detected at the $< 3 \sigma$ level.
	\end{itemize}
	Absorbers in category `c' are excluded from scientific analysis. This scheme is also applied to the measurements presented in \cite{2014arXiv1402.2655D}. The scheme ensures that we only include absorbers in our sample that are both well constrained and statistically significant. The requirement that \ion{H}{1} absorbers detected on the basis of \lya\ only must have equivalent widths significant at the $> 4\sigma$ level is motivated in \cite{2014arXiv1402.2655D}, and is estimated to reduce the number of spurious detections to $\sim 3$ per spectrum.
	
	\begin{figure}
		\centering
		\includegraphics[width=8.4cm]{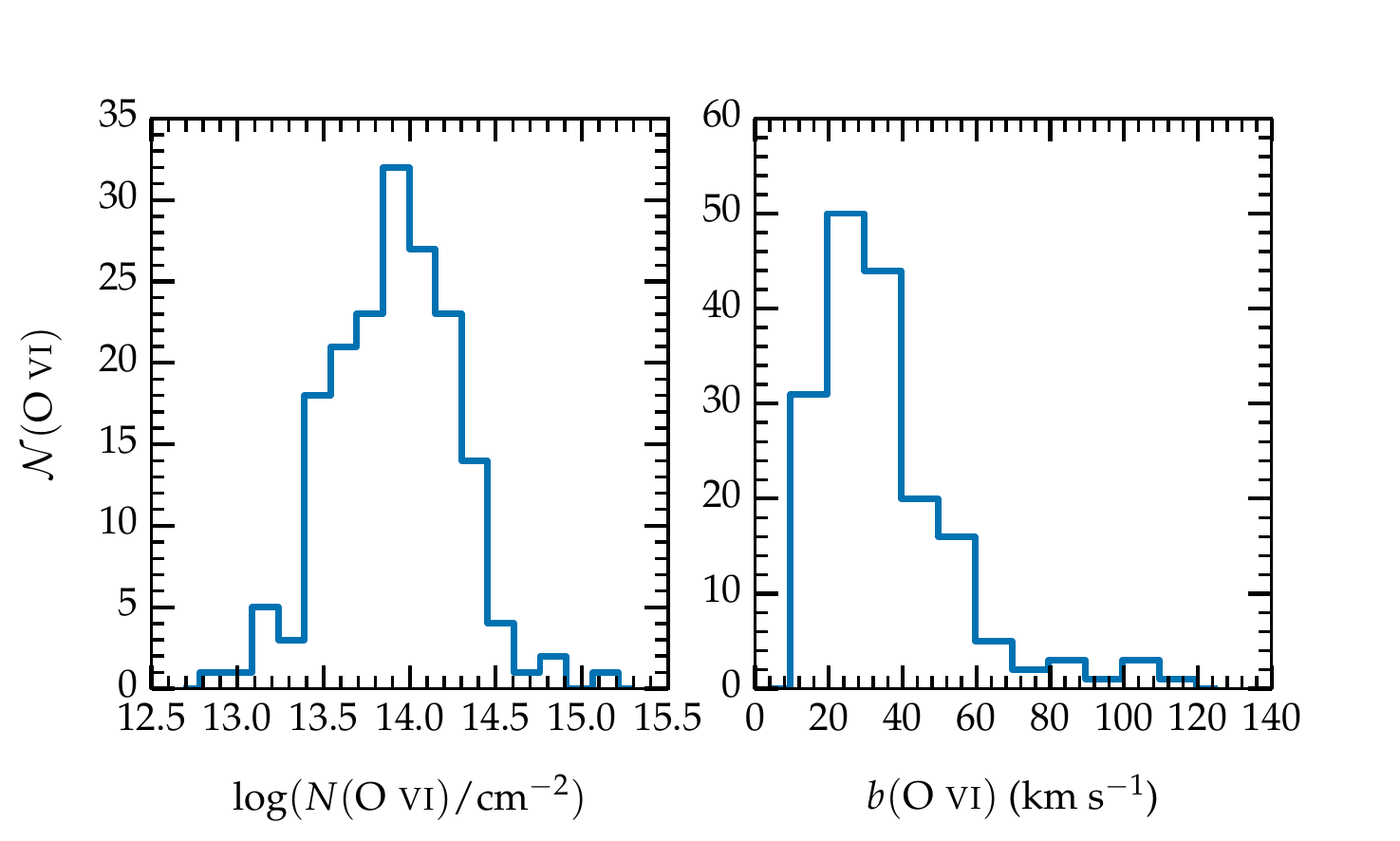}
		\caption[Statistics of $\textrm{O}\;\textsc{vi}$ absorbers in our survey]{Statistics of $\textrm{O}\;\textsc{vi}$ absorbers in our survey. The left panel shows the histogram of column densities, and the right shows the histogram of Doppler broadening parameters.}
		\label{fig:o6_histograms}
	\end{figure}

	For the analysis performed in this paper, we consider just the \ion{O}{6} samples. In \Cref{fig:o6_histograms}, we show histograms of column density and Doppler broadening parameter for our \ion{O}{6} sample. There are a total of 181 \ion{O}{6} systems that possess reliability flags `a' or `b'. These absorption systems range over a factor of 100 in column density down to our detection limit ($N(\textrm{O}\;\textsc{vi}) \approx 10^{13}~\textrm{cm}^{-2}$), in marked contrast to \ion{H}{1} absorbers that are observed to span $\sim 10$ orders of magnitude in column density. The number of \ion{O}{6} absorbers drops off fairly rapidly below $10^{13.5}~\textrm{cm}^{-2}$, and we are typically 100\% complete at $N(\textrm{O}\;\textsc{vi}) > 10^{14}~\textrm{cm}^{-2}$. Doppler broadening parameters show a long tail to high values, and a sharp cut off at $\sim 10$ \kms, which roughly corresponds to the spectral resolution of \ac{cos}. This distribution is similar to that presented in \cite{2016ApJ...817..111D}. There may be a population of very narrow \ion{O}{6} absorbers, but we are not sensitive to them here.

\section{Analysis of the galaxy data}
\label{sec:galaxy_analysis}
The following sections describe the analyses performed on the 1D extracted galaxy spectra and photometric parent samples. Much of this analysis builds on that already presented in \cite{2014MNRAS.437.2017T}. For the \ac{sdss}, \ac{2dfgrs}, \ac{vvds} and \ac{vipers} surveys, and all galaxy data presented in \cite{2009ApJ...701.1219C}, \cite{2011ApJS..193...28P} and \cite{2013MNRAS.434.1765J}, we work from the catalogued magnitudes, redshifts and spectral line measurements (where available).\footnote{Catalogues for \ac{2dfgrs}, \ac{sdss}, \ac{vvds} and \ac{vipers} galaxies are obtained from \url{http://www.2dfgrs.net}, \url{http://skyserver.sdss.org/casjobs}, \url{http://cesam.oamp.fr/vvdsproject/vvds.htm} and \url{http://vipers.inaf.it/rel-pdr1.html} respectively. Catalogues from the analysis in \cite{2009ApJ...701.1219C}, \cite{2011ApJS..193...28P} and \cite{2013MNRAS.434.1765J} were obtained from \url{http://vizier.cfa.harvard.edu/viz-bin/VizieR}.} GAMA galaxy catalogues used in this study are from phase II of the survey, and are not publicly available at the time of writing. For \ac{sdss}, we make use of the spectral line measurements presented in \cite{2004MNRAS.351.1151B} (see \Cref{sec:sf_activity} for details).

	\subsection{Redshift determination}
	\label{sec:redshift_determination}
	The majority of the galaxy redshifts in our \ac{vimos}, \ac{deimos}, \ac{gmos} and \ac{cfht} samples were obtained by cross-correlating galaxy, star and \ac{qso} templates from \ac{sdss}\footnote{\url{http://www.sdss.org/dr7/algorithms/spectemplates/}} with each observed spectrum \cite[see][for a full description]{2006MNRAS.367.1261M,2014MNRAS.437.2017T}. Each galaxy was then assigned a quality flag to indicate the reliability of the assigned redshift. The scheme is designed as follows:
	\begin{itemize}
		\item \emph{Secure} (`a'): At least three well-identified spectral features (emission or absorption lines) or two well identified emission lines.
		\item \emph{Possible} (`b'): Only one or two spectral features.
		\item \emph{Uncertain} (`c'): No clear spectral features.
	\end{itemize}
	\cite[][]{2014MNRAS.437.2017T}. Flag `c' is typically raised for spectra with low \ac{snr}, or due to an intrinsic lack of observable features at the instrumental resolution. We do not use these redshifts in any of the forthcoming analysis. For all other galaxy redshifts, we map the corresponding quality flags onto our scheme to ensure a unified definition for `secure', `possible', or `uncertain' as follows.

	In \ac{sdss}, we simply adopt all galaxies with a warning flag of 0 (indicating no warnings) as being secure (label `a'), and flag all other redshifts as `c' \cite[see][for details on \ac{sdss} flags]{2002AJ....123..485S}.

	The \ac{2dfgrs} scheme is defined in terms of absorption redshifts and emission redshifts separately. In brief, for absorption redshifts, a quality parameter $\textrm{Q}_{\textrm{a}}$ is defined in terms of a variable R, being the ratio of peak to noise in the cross-correlation with the best fitting template, as follows:
	\[ \textrm{Q}_{\textrm{a}} = \left\{
		\begin{array}{l l}
			4 & \quad \textrm{R} > 5.0, \\
			3 & \quad \textrm{R} > 4.5, \\
			2 & \quad \textrm{R} > 4.0, \\
			1 & \quad \textrm{R} > 3.5, \\
			0 & \quad \textrm{otherwise,}
		\end{array} \right.\]
	with a further requirement that $\textrm{Q}_{\textrm{a}} = 3$ and $\textrm{Q}_{\textrm{a}} = 4$ redshifts are obtained to within 600 \kms\ across four and six of the eight spectral templates respectively. For emission redshifts, the parameter $\textrm{Q}_{\textrm{e}}$ is defined as
	\[ \textrm{Q}_{\textrm{e}} = \left\{
		\begin{array}{l l}
			4 & \quad \textrm{three or more detected lines,} \\
			2 & \quad \textrm{two lines, or one strong line,} \\
			1 & \quad \textrm{one weak line,} \\
			0 & \quad \textrm{no lines.}
		\end{array} \right.\]
	The combined redshift quality flag, $\textrm{Q}_{\textrm{b}}$, is then determined as $\textrm{Q}_{\textrm{b}} = \textrm{max}(\textrm{Q}_{\textrm{a}}, \textrm{Q}_{\textrm{e}})$, unless the difference between the absorption and emission redshifts is $< 600~\textrm{km s}^{-1}$, in which case, $\textrm{Q}_{\textrm{b}} = \textrm{max}(\textrm{Q}_{\textrm{a}}, \textrm{Q}_{\textrm{e}}, 3)$, or if $\textrm{Q}_{\textrm{a}} \geq 2$ and $\textrm{Q}_{\textrm{e}} \geq 2$ and the difference between absorption and emission redshifts is $> 600~\textrm{km s}^{-1}$, in which case $\textrm{Q}_{\textrm{b}} = 1$. An overall redshift flag, $\textrm{Q}$, is determined via human verification of the automated redshift measurement, with the option of manually fitting Gaussian lines to spectral features as a means to obtain the redshift. The scheme is then as follows:
	\[ \textrm{Q} = \left\{
		\begin{array}{l l}
			5 & \quad \textrm{reliable redshift, high quality spectrum,} \\
			4 & \quad \textrm{reliable redshift,} \\
			3 & \quad \textrm{probable redshift,} \\
			2 & \quad \textrm{possible, but doubtful redshift,} \\
			1 & \quad \textrm{no redshift could be estimated.}
		\end{array} \right.\]
	\cite[see][for a detailed description]{2001MNRAS.328.1039C}. We perform the mapping $\textrm{Q} \geq 4 \rightarrow \textrm{`a'}$, $\textrm{Q} = 3 \rightarrow \textrm{`b'}$, $\textrm{Q} < 3 \rightarrow \textrm{`c'}$.

	\ac{gama} redshifts are derived using the \textsc{autoz} code \cite[][]{2014MNRAS.441.2440B} and assigned a quality parameter, nQ, in the range 1-4 based on quantitative estimates of their reliability \cite[][]{2014MNRAS.441.2440B,2015MNRAS.452.2087L}. We use the same mapping as above to translate nQ to our quality flag.

	The \ac{vvds} scheme is defined according to the following numbering scheme: (0) no redshift (no features); (1) tentative redshift (weak features, continuum shape); (2) secure redshift (several features); (3) very secure redshift (strong spectral features); (4) completely secure redshift (obvious spectral features); (9) redshift based on single secure feature. Added to this are the prefixes 1 and 2, to mean broad line \ac{agn} and secondary target respectively \cite[see][for more details]{2005A&A...439..845L}. We perform the following mapping:
		\begin{itemize}
			\item $\left\{ 4, 14, 24, 214, 3, 13, 23, 213 \right\} {\rightarrow} \textrm{`a'}$,
			\item $\left\{ 2, 12, 22, 212, 9, 19, 29, 219 \right\} {\rightarrow} \textrm{`b'}$,
			\item $\left\{ 1, 11, 21, 211, 0 \right\} {\rightarrow} \textrm{`c'}$.
		\end{itemize}

	The \ac{vipers} scheme is identical to that of \ac{vvds}, but with the addition of a decimal fraction to each flag depending on the photometric redshift from the accompanying 5-band \ac{cfhtls} photometry. If the spectroscopic redshift falls within the $1\sigma$ confidence interval on the photometric redshift, a value of 0.5 is added. If it falls within the $2\sigma$ confidence interval, a value of 0.4 is added. If it falls outside the $2\sigma$ confidence interval, a value of 0.2 is added, and when there is no photometric redshift, a value of 0.1 is added \cite[see][for more details]{2014A&A...566A.108G}. We adopt the same mapping as for \ac{vvds}, regardless of the added decimal fraction.

	Redshifts for galaxies presented in \cite{2009ApJ...701.1219C} and \cite{2011ApJS..193...28P} are only provided where they are deemed reliable. We therefore label objects having an assigned redshift with flag `a', and all other objects flag `c'.

	In \cite{2013MNRAS.434.1765J}, the redshift flagging scheme is defined (A) secure ($\geq 2$ features); (B) 1 feature; (C) observed but no features; and (N) not observed. We perform the mapping $\textrm{A} {\rightarrow} \textrm{`a'}$, $\textrm{B} {\rightarrow} \textrm{`b'}$, $\left\{\textrm{C, N}\right\} {\rightarrow} \textrm{`c'}$.

	\subsection{Global astrometry/photometry solutions}
	\label{sec:astrometry_photometry}
	In two fields, J$1005{-}0134$ and J$2218{+}0052$, \ac{vimos} observations obtained by our collaboration supplement \ac{vvds} galaxy data at small angular separations from each \ac{qso} \cite[][]{2014MNRAS.437.2017T}. We have improved the photometric and astrometric calibration for these data as follows.

	Astrometry and $R$-band photometry was originally obtained from the \ac{vimos} pre-imaging data. However, these fields overlap with the VIRMOS deep imaging survey \cite[][]{2004A&A...417..839L}, for which the astrometric and photometric calibration is superior. This photometric data set extends coverage to the $B, V$ and $I$ bands, and forms the basis for target selection in the \ac{vvds}. We therefore set about matching these data sets to ensure global astrometric and photometric consistency across these fields. We made use of the \sextractor\ \cite[][]{1996A&AS..117..393B} and \scamp\ \cite[][]{2006ASPC..351..112B} software packages to automatically map galaxy positions from the \ac{vimos} detector plane to world coordinates using sources detected in \ac{sdss} as a reference. This brought the astrometric solution to within one arcsecond of the VIRMOS deep imaging survey, which is below the typical seeing level. We then cross-matched the photometric catalogues, and calculated the mean $R$-band magnitude offset needed to bring the two into statistical agreement. We did this for each \ac{vimos} quadrant separately. The typical offset was $\sim 0.4$ magnitudes. Not all sources could be matched to those from the VIRMOS deep imaging survey due to regions of the imaging for that survey that are poorly calibrated. For these sources we keep offset \ac{vimos} $R$-band magnitudes, whereas elsewhere we assign the appropriate matched $BVRI$ photometry.

	\subsection{Spectral line measurements}
	For the \ac{vvds} and \ac{vipers} surveys, no spectral line measurements or indications of spectral type are made available. We therefore performed our own analysis where possible, as a means to estimate the star-formation activity for the galaxies in these surveys (see \cref{sec:sf_activity} for details). We also performed this analysis on the \ac{vimos}, \ac{gmos} and \ac{cfht} data collected by our collaboration \cite[][]{2006MNRAS.367.1261M,2014MNRAS.437.2017T}. Originally, the spectral types for these galaxies were determined by assigning the spectral type of the best fitting template as part of the redshift determination process. We improve upon this by measuring spectral line fluxes, as described below.

	For each galaxy spectrum, where spectral coverage, resolution and \ac{snr} allowed, we estimated the integrated fluxes and local continuum level around the [\ion{O}{2}], H$\delta$, H$\gamma$, H$\beta$, [\ion{O}{3}], H$\alpha$, [\ion{N}{2}] and [\ion{S}{2}] emission lines. For this, we used the spectral line indices defined in \Cref{tab:spectral_index}, and a direct integration over spectral pixels. The continuum level is obtained by iteratively clipping points $1.5\sigma$ below the estimated continuum in a manner similar to that described in \Cref{sec:continuum_fitting}, taking a mean of the `un-clipped' pixels either side of the line, and linearly interpolating between the points defined by these means. This makes the continuum estimate reasonably robust to underlying absorption, but it occasionally fails at the edges of some spectra where there is a loss in sensitivity, leading to poor flux-calibration, and a rapid fall-off in the continuum. The other main cause for continuum misplacement is the presence of occasional contaminating zero-orders\footnote{Crowding on the detector can often lead to zero-order spectra landing on regions inhabited by first-order spectra.} lying on top of the galaxy spectra, or regions of bad sky subtraction. Note that we do not accurately remove the stellar continuum (including the underlying stellar absorption) in our procedure. This inevitably affects the reliability of the inferred emission line fluxes (the Balmer emission lines in particular), however our approach suffices for the purposes of splitting the galaxy sample into star-forming and non star-forming populations (see \Cref{sec:sf_activity} for details).
	
	\begin{table}
	\caption{Spectral index definitions}
	\tabcolsep=0.07cm
	\begin{tabular}{c c c c}
		\hline
  		Index & Blue continuum (\AA) & Line (\AA) & Red continuum (\AA) \\
  		\hline
  		$[\textrm{O}\;\textsc{ii}]$           & $3655{-}3705$ & $3708.5{-}3748.5$ & $3750{-}3800$ \\
  		H$\delta$ 						      & $4030{-}4080$ & $4082.0{-}4122.0$ & $4125{-}4170$ \\
  		H$\gamma$ 							  & $4230{-}4270$ & $4321.5{-}4361.5$ & $4365{-}4400$ \\
  		H$\beta$ 							  & $4785{-}4820$ & $4842.5{-}4882.5$ & $5030{-}5100$ \\
  		$[\textrm{O}\;\textsc{iii}]$          & $4785{-}4820$ & $4988.0{-}5028.0$ & $5030{-}5100$ \\
  		H$\alpha + [\textrm{N}\;\textsc{ii}]$ & $6460{-}6520$ & $6544.5{-}6584.5$ & $6610{-}6670$ \\
  		$[\textrm{S}\;\textsc{ii}]$           & $6640{-}6700$ & $6713.0{-}6753.0$ & $6760{-}6810$ \\
 		\hline
	\end{tabular}
	\label{tab:spectral_index}
	\end{table}

	The line indices in \Cref{tab:spectral_index} are optimised for spectra taken with \ac{vimos} at a spectral resolution $R = 200$, appropriate for the \ac{vvds} and \ac{vipers} surveys, and for the \ac{vimos} data presented in \cite{2014MNRAS.437.2017T}. For the \ac{gmos} and \ac{cfht} data, we narrowed the line indices to reflect the higher spectral resolution obtained by these instruments \cite[see][and references therein for details]{2014MNRAS.437.2017T}. For integrated line fluxes detected above a $3\sigma$ significance threshold, we also attempt to fit Gaussian profiles, which are usually adopted in preference to the pixel measurements. We revert to the pixel measurements when the fitting routine returns a Gaussian with zero amplitude, indicating that the fit has failed. All Gaussian fits are performed with a $\chi^2$ minimisation employing the Levenberg-Marquardt algorithm. The rest-frame standard deviation of each Gaussian line is bounded between $0.5\sigma_{\textrm{LSF}}$ \AA\ and $\sqrt{10^2 + \sigma_{\textrm{LSF}}^2}$ \AA, where $\sigma_{\textrm{LSF}}$ is the standard deviation of the (assumed Gaussian) \ac{lsf}, which helps to identify broad emission lines that are likely of \ac{agn} origin, and contaminating sky lines. We fit the [\ion{O}{2}] doublet, H$\delta$, H$\gamma$ and [\ion{S}{2}] doublet lines separately. The [\ion{O}{2}] doublet is not resolved by our spectra, so we fit it as a single line. For the [\ion{S}{2}] doublet, which is marginally resolved, we tie the Gaussian standard deviations in the fit, and fix the line ratio [\ion{S}{2}] $\lambda6716$ / [\ion{S}{2}] $\lambda6731$ to the expected (but not fixed)\footnote{Assumes a gas density and temperature.} ratio of $1 / 1.4$ \cite[][]{1989agna.book.....O}. We fit the H$\beta$/[\ion{O}{3}] line complex simultaneously, tying together the Gaussian standard deviations, and fixing the [\ion{O}{3}] $\lambda4956$ / [\ion{O}{3}] $\lambda5007$ ratio to the expected value of $1 / 2.98$ \cite[][]{2000MNRAS.312..813S}. We also fit the H$\alpha$/[\ion{N}{2}] complex simultaneously, tying together the Gaussian standard deviations, and fixing the [\ion{N}{2}] $\lambda6548$ / [\ion{N}{2}] $\lambda6583$ ratio to the expected value of $1 / 2.95$ \cite[][]{1989agna.book.....O}. The [\ion{N}{2}] lines are barely resolved from the H$\alpha$ line in our spectra, so we perform an alternative, single Gaussian fit to just the H$\alpha$ line in every case. If this fit gives a smaller $\chi^2$ value than the three-component fit, we assign the resulting Gaussian parameters to the H$\alpha$ line, and report no [\ion{N}{2}] measurements in these instances. Despite not having sufficient spectral resolution to properly resolve the [\ion{N}{2}] components, in instances where these lines are strong, the resulting line profile has definite asymmetry, which motivates us to decompose the line profile. Uncertainties on the fitted Gaussian parameters are estimated by generating 100 Monte Carlo realisations of the data. For each realisation, we add a number to every pixel flux, randomly generated from a Gaussian distribution of values with standard deviation equal to its $1\sigma$ uncertainty. Each of these 100 realisations are then fit using the same procedure as in the nominal case, and the standard deviation over the resulting best-fit parameter values are taken as the $1\sigma$ uncertainty on the measurement. 

	To identify bad measurements in our galaxy spectra, we have devised a flagging scheme as follows:
	\begin{itemize}
		\item Flag (0): No warnings.
		\item Flag (1): Measurement may be affected by the OH forest between 8600 and 8700\AA.
		\item Flag (2): Line was fit with the maximum/minimum allowed Gaussian standard deviation.
		\item Flag (3): Line coincides with a region above a user-specified sky spectrum threshold.
		\item Flag (4): Line may be affected by the $\textrm{O}_2$ telluric absorption at $\sim 7600$ \AA.
		\item Flag (5): Bad continuum reduced $\chi^2$ $(> 10)$.
		\item Flag (6): No spectral coverage.
	\end{itemize}
	The quality of the sky subtraction in our spectra makes the line measurements reasonably robust to flag (1). Flag (2) is implemented to identify potential broad-line \ac{agn} and contaminating sky lines. Flag (3) is mainly implemented to eliminate contaminating zero-orders. We found that the \ac{vimos} \esorex\ pipeline reduction software often incorrectly identifies zero orders offset from the galaxy spectrum of interest and tries to correct for them, leaving deep, artificial absorption features in the extracted 1D galaxy spectra. Nevertheless, these appear as broad spikes in the extracted 1D sky spectra, and can be identified by adopting a threshold sky value. Flag (4) is implemented because most of our spectra are not corrected for the $\textrm{O}_2$ telluric, and even in spectra that are corrected for this contaminating feature, the correction is highly uncertain due to the narrow `picket-fence' nature of the absorption. Flag (5) is implemented to identify line measurements that are marred by bad continuum estimation. We allow for relatively high reduced $\chi^2$ values in view of occasional absorption that raises the value of this statistic even for reasonable continuum estimations. Flag (6) is implemented to identify lines not measured due to insufficient spectral coverage. We enforce coverage across the entire region defined by each of the line indices in \Cref{tab:spectral_index}. Line measurements that do not raise any of the aforementioned flags are assigned flag (0), to indicate that there are no warnings.

	In all of our spectra, we reject measurements that raise flags (2), (4), (5) and (6). For \ac{vimos} spectra reduced with \esorex, we additionally reject measurements that raised flag (3). In practice, this flag is reserved for those spectra only.

	\subsection{Star-formation activity}
	\label{sec:sf_activity}
	For the \ac{vvds} and \ac{vipers} surveys, and our \ac{vimos}, \ac{gmos} and \ac{cfht} data, we use the spectral line measurements described in the previous section to split the sample of galaxies in terms of their star formation activity. For \ac{gama}, we also do the same using a very similar set of line measurements provided by the \ac{gama} survey team \cite[see][for a description]{2013MNRAS.430.2047H}. We aim simply to define galaxies as `star-forming', `non star-forming', `\ac{agn} dominated', or `unclassified'. Although we could have calculated star formation rates for many of our galaxies using standard procedures \cite[e.g.][]{1998ARA&A..36..189K,2006ApJ...642..775M}, estimating $K$-corrections at redshifts $\gtrsim 0.5$ becomes increasingly uncertain, and in general we lack a homogeneous set of multi-band photometric measurements across our sample to allow for a consistent approach. In any case, simply splitting our sample purely on the basis of spectral line measurements suffices for our requirements. We adopt a very similar prescription to that outlined in \cite{2004MNRAS.351.1151B}, whose classification scheme was applied to the \ac{sdss} galaxies in our sample.

	\begin{figure}
		\centering
		\includegraphics[width=8.4cm]{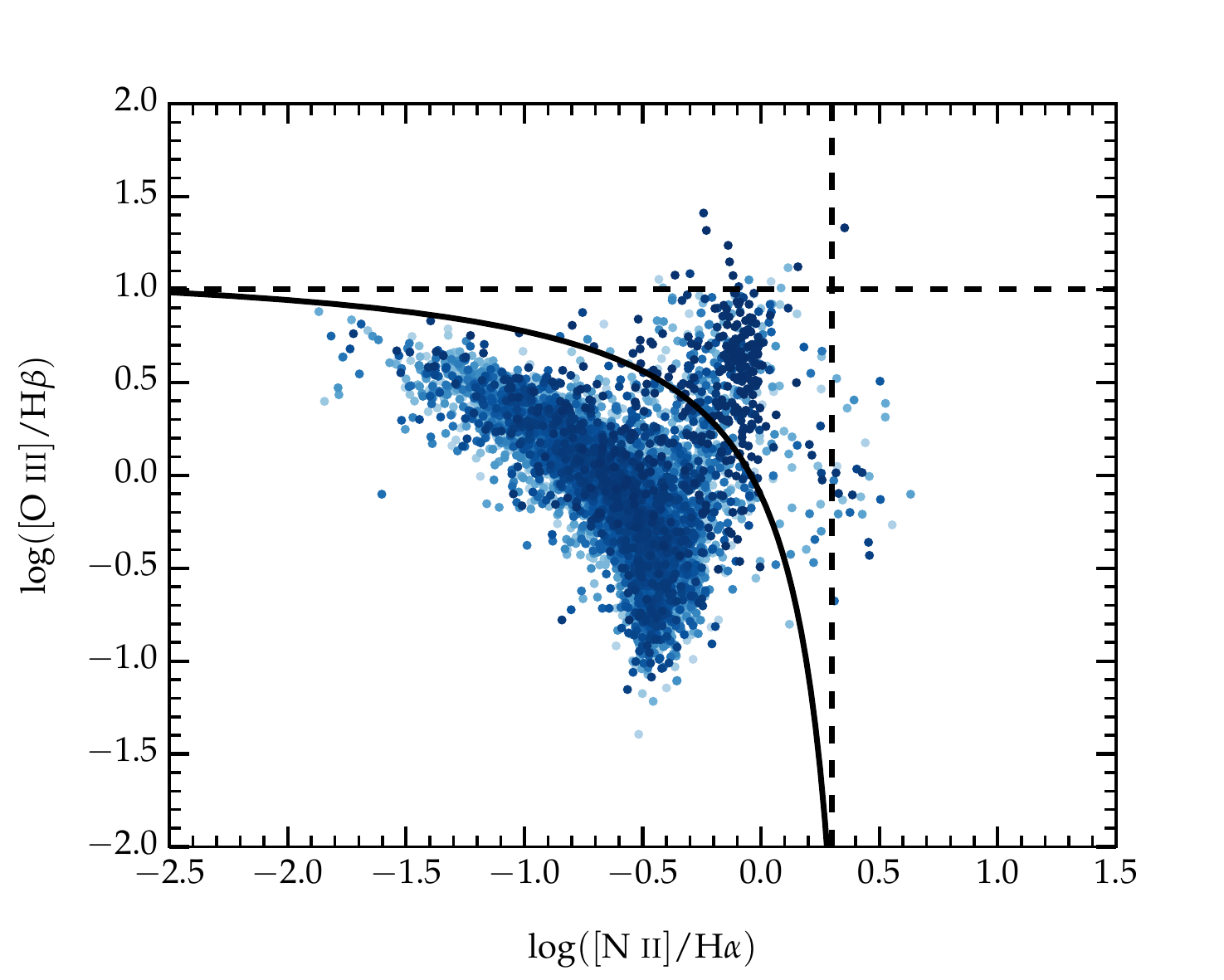}
		\caption[BPT diagram]{The distribution of galaxies in our sample on a BPT line-ratio diagram. The solid line indicates the discrimination line from \cite{2001ApJ...556..121K}, separating star-forming galaxies from \ac{agn}. The dashed lines indicate the single-line ratio diagnostics we also employ to identify \ac{agn} when two line ratios are not available.}
	\label{fig:bpt}
	\end{figure}

	First, we attempt to identify galaxies whose spectra are dominated by an \ac{agn} component. A number of broad-line \ac{agn} are already identified on the basis of their redshift determination via cross-correlation with \ac{agn} templates \cite[][]{2014MNRAS.437.2017T}. For the remaining galaxies, we perform a classification on the basis of a Baldwin, Phillips \& Terlevich (\citeyear{1981PASP...93....5B}, hereafter BPT) diagram, shown in \Cref{fig:bpt}. \cite{2001ApJ...556..121K}  performed \ac{sed} and photoionization modelling to find a theoretical discriminating line between star-forming galaxies and \ac{agn} on a BPT diagram of $\log([$N$\;$\textsc{ii}$] / \textrm{H}\alpha)$ versus $\log([$O$\;$\textsc{iii}$] / \textrm{H}\beta)$. This discriminating line is shown in \Cref{fig:bpt}, along with the subset of our galaxies that have $\textrm{SNR} > 3$ in each of the measured H$\alpha$, H$\beta$, [\ion{N}{2}] and [\ion{O}{3}] line fluxes, and no severe warning flags associated with these (we accept flags 0 and 1 in the scheme devised above). High values of $\log([$N$\;$\textsc{ii}$] / \textrm{H}\alpha)$ and $\log([$O$\;$\textsc{iii}$] / \textrm{H}\beta)$ are driven by a hard \ac{euv} spectrum attributable to \ac{agn} activity in the host galaxies, pushing these galaxies to the top-right corner of the diagram. Although we can make an \ac{agn} classification on the basis of this diagram regardless of \ac{snr}, we use this technique only for $\textrm{SNR} > 3$, since below this, an increasing fraction of galaxies have measured line fluxes that are negative, and the non-symmetric distribution of galaxies on this diagram leads to classification biases. Typically, only a very small subset of our galaxies for which we performed spectral line measurements have all the required lines measured above our \ac{snr} criterion, but we are able to expand the classification by using only single line ratios. These are indicated by the dashed lines in \Cref{fig:bpt}, and they correspond to $\log([$N$\;$\textsc{ii}$] / \textrm{H}\alpha) > 0.3$ and $\log([$O$\;$\textsc{iii}$] / \textrm{H}\beta) > 1$. Clearly these classifiers are less effective than that using both line ratios. However, this does allow us to classify \ac{agn} according to their line ratios over a larger number of galaxies. We find that only around 5\% of our galaxies are classified as \ac{agn}, but we cannot rule out a small additional population that could not be identified in the manner just described.

	After identifying \ac{agn}, we assume that the rest of the galaxies are regular star-forming or non star-forming galaxies. We identify star-forming galaxies as those that show measured fluxes that are positive and with $\textrm{SNR} > 2$ in any one of the H$\alpha$, H$\beta$, or [\ion{O}{2}] lines. Those that do not meet this criterion are identified as non star-forming galaxies. Those galaxies that do not have good measurements of any of the H$\alpha$, H$\beta$, or [\ion{O}{2}] emission lines (due to bad flags and/or lack of spectral coverage) are marked as `unclassified'. These galaxies nevertheless have redshift measurements from e.g. the \ion{Ca}{2} H and K, CH molecule G-band, Mg $\lambda5175$ and Na $\lambda5894$ lines.

	For all other galaxies in our sample, we obtain their spectral classifications from the literature, typically found from principle component analyses \cite[see][for details]{2009ApJ...701.1219C,2011ApJS..193...28P,2013MNRAS.434.1765J,2014MNRAS.437.2017T}. Overall, we find that $\sim 55$\% of the galaxies in our sample are classified as star-forming, $\sim 35$\% are classified as non star-forming, $\sim 5$\% are classified as \ac{agn} and $\sim 5$\% are unclassified.

\section{Gas and galaxies in the EAGLE hydrodynamical simulation}
\label{sec:eagle}
In the following sections, we present the methods used to extract a comparison data set from the \ac{eagle} project, which is a suite of hydrodynamical simulations that follow the formation and evolution of galaxies and supermassive black holes in volumes representative of a $\Lambda$CDM Universe. We begin by briefly describing the pertinent aspects of the simulation, and discuss its key advantages and limitations. We then follow with a detailed description of the processes involved in generating mock catalogues of galaxies and absorption systems, designed to mimic as closely as possible the observations.

	\subsection{The \textsc{Eagle} simulations}
	\label{sec:eagle_description}
	The \ac{eagle} project \cite[][]{2015MNRAS.446..521S,2015MNRAS.450.1937C} is a suite of cosmological hydrodynamical simulations representative of a $\Lambda$CDM Universe. The simulations were run with the \ac{sph} code \gadget3 in cubic volumes 12.5, 25, 50 and 100 comoving Mpc on a side.\footnote{There is also a set of high-resolution `zoom' simulations \cite[see][for details]{2015MNRAS.448.2941S}.} State-of-the-art numerical techniques and subgrid models are used to capture various physical processes important to galaxy formation and evolution. These include radiative gas cooling, star formation, mass loss from stars, metal enrichment, energy feedback from star formation and \ac{agn} and gas accretion onto, and mergers of, supermassive black holes. The efficiency of stellar feedback and the mass accretion onto black holes is calibrated to match the present-day stellar mass function of galaxies (subject to the additional constraint that the galaxies sizes need to be reasonable), and the efficiency of \ac{agn} feedback is calibrated to match the observed relation between stellar mass and black hole mass. Calibrations such as these are necessary, since the underlying physics behind galaxy feedback is neither well understood, nor well constrained observationally, and the resolution of the simulations is insufficient for ab initio predictions of the feedback efficiency.
	
	In this work, we use only the largest simulation volume (100 Mpc$^3$), referred to as L100N1504, containing $1504^3$ \ac{sph} particles. For this, the initial baryonic particle mass is $m_{\textrm{g}} = 1.81 \times 10^6~M_{\odot}$, the dark matter particle mass is $m_{\textrm{dm}} = 9.70 \times 10^6~M_{\odot}$, and the comoving, Plummer-equivalent gravitational softening length is 2.66 kpc. At this resolution, \ac{eagle} is marginally sufficient to resolve the Jeans scales in the warm \ac{ism}. We shall refer to three particle types in \ac{eagle}: dark matter particles, star particles and gas particles. Dark matter particles are evolved using the N-body part of the code, which simulates just the gravitational interactions between particles, while gas particles are also subject to hydrodynamical forces. Gas particles above a metallicity-dependent density threshold are converted to star particles stochastically. More details can be found in \cite{2015MNRAS.446..521S}.

	We will test the feedback prescriptions in \ac{eagle} by examining its predictions for the distribution and dynamics of \ion{O}{6} absorbers around galaxies, which we can then compare to our observational sample. A test such as this has considerable diagnostic power, since the simulation was not calibrated to match observations such as these. Even though we cannot hope to learn much about the detailed physics governing supernova and \ac{agn} feedback, a simulation that matches these observations should nevertheless provide important insights on the gas flows around galaxies, which are responsible for driving the evolution of key galaxy properties, such as their star formation rates, and the mass-metallicity relationship. To perform this test, we need to generate mock catalogues of galaxies and absorbers from the simulation, which is the subject of the following sections.

	\subsection{Intergalactic gas}
	\label{sec:eagle_gas}
	We characterise the \ac{igm} in \ac{eagle} in a very similar manner to observations by drawing synthetic \ac{qso} sight-lines through the simulation volume. We follow the procedure outlined in \cite{1998MNRAS.301..478T} (see their appendix A4), using a modified version of the artificial transmission spectra code, \specwizard. This works as follows. For a given $(x, y, z)$ coordinate and orientation in the simulation volume at a given redshift snapshot, specifying a one-dimensional sight-line, \specwizard\ first extracts all \ac{sph} gas particles that intersect that sight-line. The sight-line is then divided into an arbitrarily high number of bins of width $\Delta$ in real space. These bins are labelled from zero to $\dot{a}L$ in velocity space, each having a width of 1 \kms, where $a(z)$ is the dimensionless scale-factor and $L$ is the box size in comoving coordinates. For each bin, the code then calculates the local physical density, $\rho_X$, and temperature, $T_X$, for an ionic species $X$, weighted by the \ac{sph} smoothing kernel and abundance of species $X$, assuming ionization equilibrium in the presence of a \cite{2001cghr.confE..64H} \ac{uv} background radiation field. Then, for a given atomic transition, $i$, of species $X$, assuming only thermal line broadening, a bin $k$, corresponding to a velocity $v(k)$, will suffer absorption due to material in bin $j$, at velocity $v(j)$, by an amount $\mathrm{e}^{-\tau(k)}$, where
	\begin{equation}
		\tau(k) = \sigma_{X_i} \frac{1}{\sqrt{\pi}} \frac{c}{V_X(j)}\rho_X(j)a\Delta\exp\left[-\left(\frac{v(k) - v(j)}{V_X(j)}\right)^2\right],
	\end{equation}
	and
	\begin{equation}
		V_X^2(j) = \frac{2kT_X(j)}{m_X}
	\end{equation}
	\cite[][]{1998MNRAS.301..478T}. Here, $\sigma_{X_i}$ is the absorption cross-section of the transition, $c$ is the speed of light and $V_X(j)$ is the Doppler width of species $X$ with mass $m_X$. For the vast majority of the absorption along these sight-lines, the physical densities are small enough that a purely thermally broadened line-profile is a good approximation to the real one.
	
	To create mock catalogues of \ion{O}{6} absorbers, we use the method above to calculate the optical depth, $\tau(v)$, in \ion{O}{6} $\lambda1031$ along 25 000 randomly-drawn sight-lines parallel to the $z$-axis (our pseudo-redshift axis) through each of 7 different redshift snapshots over the range $0.1 \lesssim z \lesssim 0.7$ (the dominant range covered by our observational sample). We then take peaks in the $\tau$ distribution above a threshold value of 0.0005 (arbitrary) along each sight-line to correspond to the optical depth at the absorption line centres, and calculate the absorbing column density assuming a Doppler broadening parameter equivalent to $V_X$ in the equations above. Each absorber is then assigned the $(x, y)$ coordinate of the sight-line it was extracted from, and the velocity, $v$, at which it was extracted, which we convert to a position $r_z$ along the $z$-axis via $r_z = v(1 + z) / H(z)$, where $z$ is the redshift of the simulation snapshot, and $H(z)$ is the value of the Hubble parameter at that redshift.
	
	\begin{figure}
		\centering
		\includegraphics[width=8.4cm]{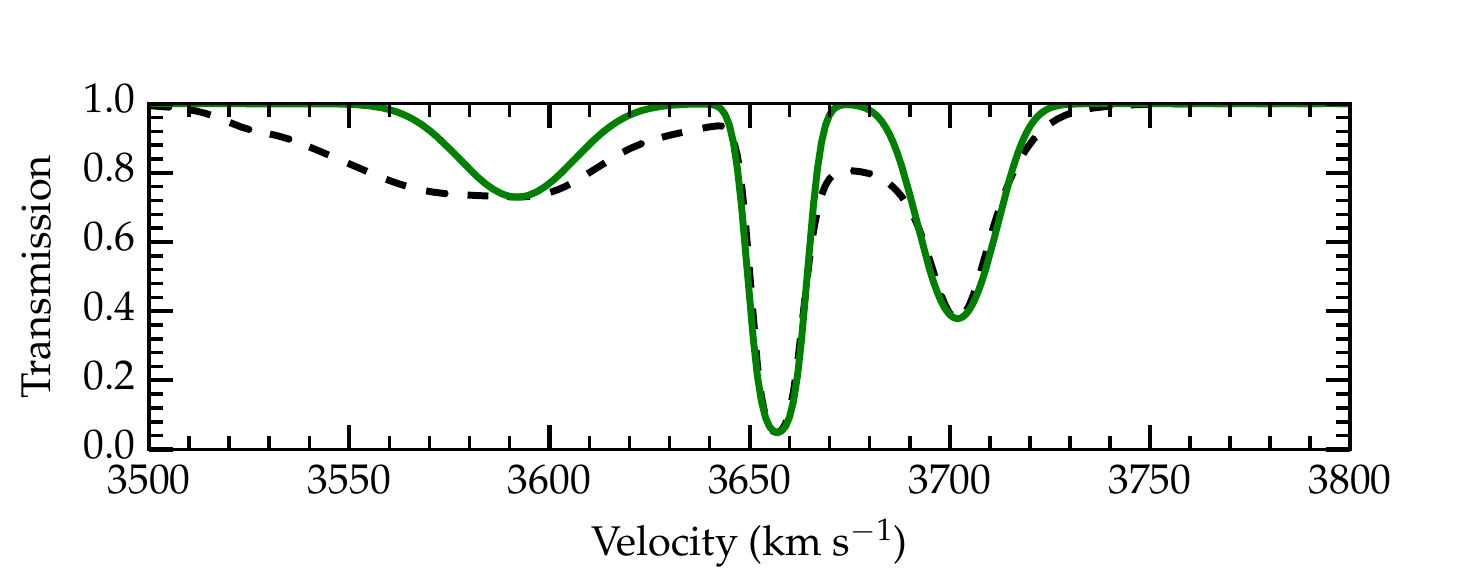}
		\caption[Voigt profiles derived from \textsc{Eagle} spectra]{Voigt profiles derived from peaks in the optical depth distribution (plotted in green) compared to the $\mathrm{e}^{-\tau}$ spectrum. The predicted Voigt profiles show simpler velocity structure than is apparent in the spectrum, highlighting an important caveat to our approach.}
	\label{fig:eagle_absorption}	
	\end{figure}
	
	It is important to note that the above procedure differs from the observational one, whereby Voigt profiles are fit to the absorption features in transmission spectra ($\mathrm{e}^{-\tau}$) in order to extract column densities and Doppler broadening parameters. This approach enables de-blending of multiple-component absorption, and takes into account the broadening of the lines due to the instrumental profile, enabling accurate recovery of their column densities. The technique was originally devised under the premise that the intervening absorption lines in \ac{qso} spectra arose from discrete absorbing clouds, but this picture was challenged early by the smoothly distributed \ac{igm} captured in cosmological hydrodynamical simulations \cite[e.g.][]{1994ApJ...437L...9C,1996ApJ...457L..51H,1996ApJ...471..582M,1998MNRAS.301..478T,1999ApJ...511..521D}. Absorbers that have a large spatial extent take part in the Hubble expansion, which leads to a line profile that deviates from a Voigt profile. Voigt profile fitting the transmission spectra will therefore glean slightly different results to simply taking peaks in the $\tau$ distribution. In particular, we might expect a larger number of absorbers, and some differences in the derived Doppler broadening parameters and column densities. As a check, we derived Voigt profiles for lines recovered in some of the sight-lines extracted from the simulation using our $\tau$ peak method, and plotted these on top of the transmission spectrum. An example is shown in \Cref{fig:eagle_absorption}. The transmission spectrum is shown as the black dashed line, and the predicted Voigt profiles are shown in green. More structure is apparent in the real spectrum, which would yield a larger number of Voigt components in Voigt profile fitting, and column densities and Doppler broadening parameters that differ slightly from those we have recovered via the $\tau$ peak method. This an obvious caveat to our approach, which we bear in mind when interpreting our results later on.
	
	\begin{figure}
		\centering
		\includegraphics[width=8.4cm]{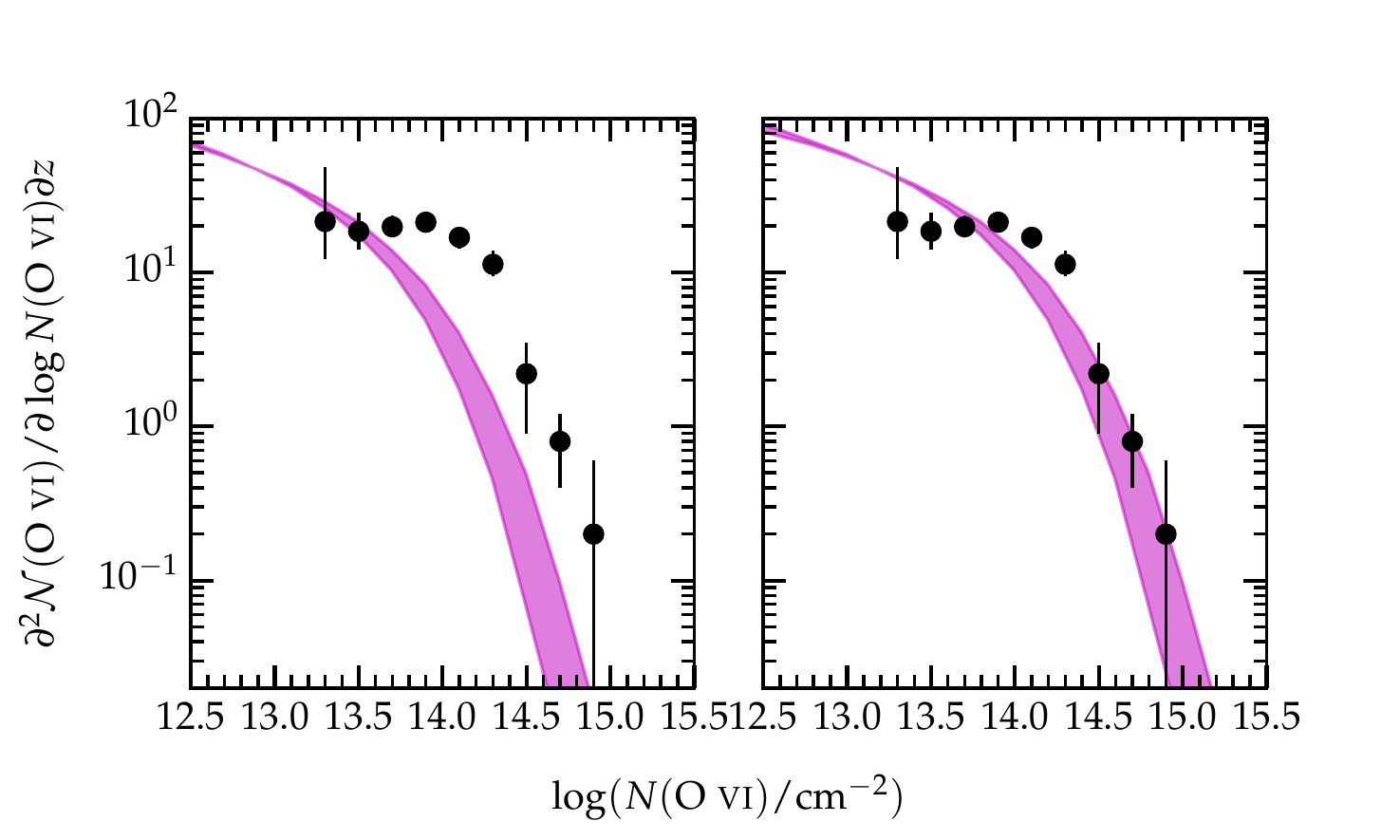}
		\caption[The column density distribution function of $\textrm{O}\;\textsc{vi}$ absorbers compared to \textsc{Eagle} predictions]{The column density distribution function of $\textrm{O}\;\textsc{vi}$ absorbers from \cite{2014arXiv1402.2655D} (data points) compared to predictions from \ac{eagle} snapshots over the redshift range $0.1 \lesssim z \lesssim 0.7$ (shaded magenta region). The shape of the column density distribution function predicted from \ac{eagle} is not far from that in the data (left panel), but a shift of 0.3 dex in column density (right panel) is necessary to obtain the required number of high column density absorbers from the simulation (see the text for justification).}
	\label{fig:o6_fn}	
	\end{figure}
	
	Having created a mock catalogue of absorbers from the \ac{eagle} simulation, we now examine whether the global statistics of our mock population matches the observed one. To do so, we investigate the column density distribution function of \ion{O}{6} absorbers. This is defined as the number, $\mathcal{N}$, of absorption lines per unit column density, $\textrm{d}N_{{\textrm{O}\;\textsc{vi}}}$, per unit redshift, $\textrm{d}z$:
	\begin{equation}
		f(N_{\textrm{O}\;\textsc{vi}}) = \frac{\textrm{d}^2\mathcal{N}}{\textrm{d}N_{\textrm{O}\;\textsc{vi}}\textrm{d}z}.
	\end{equation}
	It is also sometimes defined in terms of the absorption distance $\textrm{d}X$, which is related to $\textrm{d}z$ via $\textrm{d}X/\textrm{d}z = H_0(1 + z)^2 / H(z)$, where $H(z)$ is the Hubble parameter. We compare to the measurement presented in \cite{2014arXiv1402.2655D} (data points in \Cref{fig:o6_fn}), whose data form the majority of our \ion{O}{6} sample, although we remind the reader that this measurement was recently updated in \cite{2016ApJ...817..111D}. To calculate this for the \ion{O}{6} absorbers extracted from \ac{eagle}, we adopt the same bin size as was used for the data, $\Delta (\log N_{{\textrm{O}\;\textsc{vi}}}) = 0.2$, and note that we have $n_{\textrm{LOS}} = 25~000$ sight-lines through the $L = 100$ Mpc simulation volume, then calculate $\Delta z = n_{\textrm{LOS}}LH(z) / c$, where $c$ is the speed of light. We do this for each of the redshift snapshots in the range $0.1 \lesssim z \lesssim 0.7$, and plot the resulting curves as the shaded region in the left-hand panel of \Cref{fig:o6_fn}. We note that a similar calculation has been performed by projecting the simulation volume along one axis, and taking the summed \ion{O}{6} column densities in each two-dimensional grid cell, yielding similar results \cite[see][for details]{2015MNRAS.446..521S}. The shape of the column density distribution function from \ac{eagle} is not far from that of the observed one, although the `knee' seen in the observed distribution is not so well pronounced in the simulation. We note also that the distribution of simulated \ion{O}{6} absorbers falls off at lower column densities than in the data. Uncertainties in the shape and normalisation of the \ac{uv} background may be enough to account for this difference, along with uncertainties in the metal yields from star-formation \cite[][]{2015MNRAS.446..521S}. Due to these differences, it is necessary to shift the simulated \ion{O}{6} absorption column densities by 0.3 dex towards higher values in order to draw a sub-sample of \ion{O}{6} absorbers from the simulation that is matched to the selection functions in the data (right-hand panel in \Cref{fig:o6_fn}). We therefore proceed from this point on with simulated \ion{O}{6} absorbers whose column densities are rescaled by an extra 0.3 dex in column density.
	
\begin{figure}
		\centering
		\includegraphics[width=8.4cm]{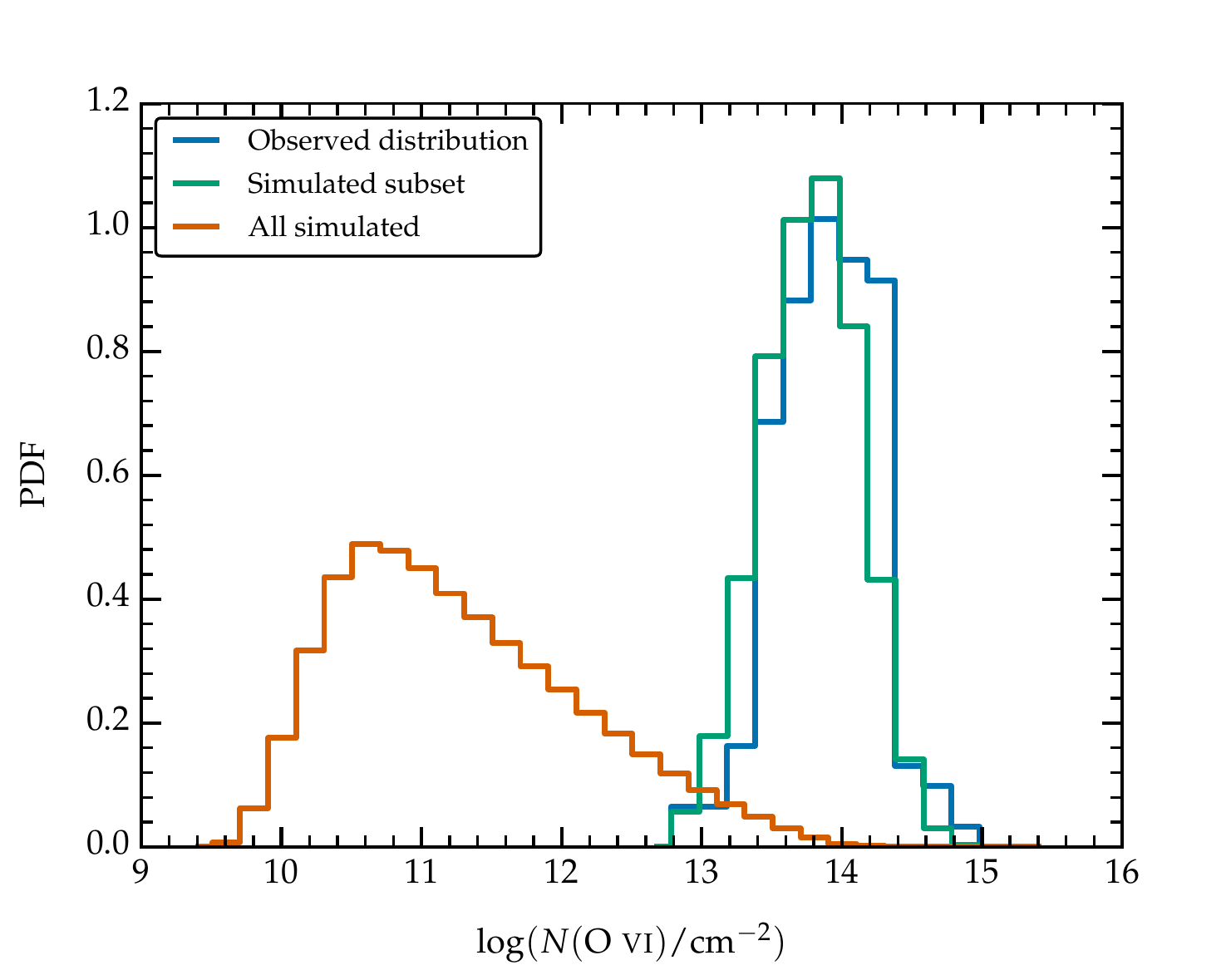}
		\caption[Probability density functions of $\textrm{O}\;\textsc{vi}$ column densities in the observations and the simulations]{Probability density functions for the observed $\textrm{O}\;\textsc{vi}$ column densities (blue), the simulated ones (red) and the simulated subset (green). See the text for a full description.}
	\label{fig:eagle_o6_subset}
	\end{figure}

	To create a matched sample of \ion{O}{6} absorption systems from the \ac{eagle} simulation, we begin by optimally binning the column density histogram of the observed sample using the method presented in \cite{2006physics...5197K} (see \Cref{sec:random_catalogues} for a description and motivation), then smooth the histogram with a Gaussian having standard deviation equal to the bin size. We then interpolate a cubic spline over the smoothed histogram, and normalise by the area underneath the curve to produce a probability density function (PDF) for the observed column densities. We also do the same on the column density distribution of simulated samples, which exhibits a turn over at very low column densities ($N({{\textrm{O}\;\textsc{vi}}}) < 10^{11}~\textrm{cm}^{-2}$) due to the arbitrarily imposed optical depth cut. We then randomly draw a 10\% subset of the simulated \ion{O}{6} absorbers from the probability density distribution of the data, inverse weighted by that of the simulation. We illustrate this in \Cref{fig:eagle_o6_subset}. The observed probability density function is shown in blue, the simulated one in red, and that of the 10\% subset in green. As is clearly evident from the figure, we are able to extract a sub-sample of \ion{O}{6} absorbers from the \ac{eagle} simulation that reproduces well the selection bias towards high column densities in the data.
	
	\subsection{Galaxies}
	\label{sec:eagle_galaxies}
	Galaxies in \ac{eagle} are defined by the minimum of the gravitational potential in `subhaloes' identified by the \textsc{subfind} algorithm \cite[][]{2001MNRAS.328..726S,2009MNRAS.399..497D}. A `subhalo' is defined in this algorithm as a gravitationally bound, locally over-dense group of particles, identified in the following fashion. First, dark matter haloes are found by implementing the friends-of-friends \cite[FoF;][]{1985ApJ...292..371D} algorithm on the dark matter particles. Gas and star particles in the simulation are then assigned to the same (if any) FoF halo as their nearest dark matter particles. Next, substructure candidates are identified by finding overdense regions within each FoF halo that are bounded by saddle points in the density distribution. Finally, particles not gravitationally bound to the substructure are discarded, and the resulting substructures are termed subhaloes. For \ac{eagle} it was also necessary to merge subhaloes separated by less than the minimum of 3 kpc and the stellar half-mass radius, to remove a small number of very low-mass subhaloes whose mass was dominated by a single particle \cite[][]{2015MNRAS.446..521S}.
	
	The stellar mass of each galaxy in \ac{eagle} is defined to be the sum of the masses of all the star particles that belong to the corresponding subhalo, and that are within a 3D aperture of radius 30 kpc. The choice of aperture was motivated by the fact that this gives a nearly identical galaxy stellar mass function as applying the Petrosian apertures used in the observations \cite[][]{2015MNRAS.446..521S}. All galaxy properties in \ac{eagle}, including the star formation rates, are evaluated in this aperture. For the purpose of matching to the observational sample, we limit the \ac{eagle} sample to galaxies with stellar masses $\log(M_{\star} / M_{\odot}) > 8$. We perform this mass cut for two reasons: (i) to eliminate galaxies consisting of $\lesssim 100$ star particles, as these galaxies will have star-formation rates that are not robust; and (ii) to reflect our observed sample, for which stellar masses $\log(M_{\star} / M_{\odot}) < 8$ are comparatively rare due to the limiting magnitudes of the surveys \cite[see, for example,][]{2009MNRAS.398.2177L,2012MNRAS.421..621B}.
	
	To split the \ac{eagle} sample of galaxies in terms of their star-formation activity, we adopt the same scheme as in \cite{2015MNRAS.446..521S}, whereby galaxies with specific star-formation rates $\dot{M}_{\star} / M_{\star} > 0.1~\textrm{Gyr}^{-1}$ are defined to be star-forming galaxies, and all others are defined to be non-star-forming galaxies. We do not attempt to find a cut in star-formation rate (or specific star-formation rate) that will match the relative proportions of star-forming and non star-forming galaxies in the observational sample, since the observational definition is subject to signal-to-noise constraints not present in the simulation. We leave a more precise comparison for future work.
	
	Finally, we define the $z$-axis of the simulation volume to be the pseudo-redshift axis, and replace the $z$-coordinate of each galaxy, which we shall denote $r_z$, with $r_{z,\textrm{pec}} = r_z + v_{\textrm{pec}}(1 + z) / H(z)$, where $v_{\textrm{pec}}$ is the peculiar velocity of the galaxy, and $H(z)$ the Hubble parameter for a redshift $z$ corresponding to the snapshot redshift. We note that the simulation volume is periodic, and we preserve this feature when modifying the $z$-coordinates of the galaxies. This then mimics the distortions introduced by peculiar velocities along the \ac{los} for galaxy surveys constructed in redshift space.

\section{IGM/Galaxy clustering analysis}
\label{sec:correlation_analysis}
We now make use of our samples of \ion{O}{6} absorbers and galaxies, together with the data sets drawn from the \ac{eagle} cosmological hydrodynamical simulation, to examine their two-point correlation functions. We limit the \ac{qso} sight-line fields listed in \Cref{tab:fields} to just those where we found \ion{O}{6} absorption in at least one \ac{qso} spectrum with confidence label `a' or `b'. We also exclude \ion{O}{6} absorbers within 5000 \kms\ of the \ac{qso} redshift, on the conservative assessment that these may be intrinsic to the \ac{qso} host galaxy. Galaxies are limited to those with redshift confidence labels `a' or `b'. The resulting sample for this study then consists of 27 fields and 32 \ac{qso} sight-lines, which contain 51 296 galaxies and 140 \ion{O}{6} absorbers respectively. Our full sample of galaxies and \acp{qso} will be used to investigate the \ion{H}{1}-galaxy cross-correlation function in future work. This will serve as an extension to the work presented in \cite{2014MNRAS.437.2017T}.

	\subsection{The two-point correlation function}
	\label{sec:correlation_function_description}
	To address the statistical connection between the metal-enriched \ac{igm} (traced by \ion{O}{6} absorbers) and galaxies, we focus on a two-point correlation analysis. The advantages of doing so are two-fold: (i) we do not rely on associating a particular intergalactic absorber with a particular galaxy (or set of galaxies), which in many instances is ambiguous; and (ii) we are robust to galaxy/absorber completeness variations in our survey, since we are measuring a clustering excess as a function of scale relative to a random expectation that takes into account the survey selection function. For the latter, it is important to note that a deficit of one population relative to another at a given scale due to a selection bias \emph{will} bias the measurement at those scales if it is not properly modelled and corrected for.

	The two-point correlation function, $\xi(r)$, is defined as the probability \emph{excess} of finding two points separated by a distance $r$ with respect to the random expectation.\footnote{Assumes isotropy, since $r$ in this case is a scalar quantity.} We shall use the correlation function between pairs of galaxies, $\xi_{\textrm{gg}}$ (the galaxy auto-correlation function), between pairs of absorbers, $\xi_{\textrm{aa}}$ (the absorber auto-correlation function) and between absorber-galaxy pairs, $\xi_{\textrm{ag}}$ (the absorber-galaxy cross-correlation function) to gain insights on the relationship between them.

	For each field, we assign a central coordinate in right-ascension and declination $(\alpha_0, \delta_0)$. Then, for every object in that field with a spectroscopic redshift $z$, we calculate its position in Cartesian $(x, y, z)$ coordinates as follows:
	\begin{align}
		x &\equiv r(z) \cos(\Delta\delta) \cos(\Delta\alpha), \nonumber \\
		y &\equiv r(z) \cos(\Delta\delta) \sin(\Delta\alpha), \nonumber \\
		z &\equiv r(z) \sin(\Delta\delta),
	\end{align}
	where $r(z)$ is the comoving distance, $\Delta\delta \equiv (\delta - \delta_0)$ and $\Delta\alpha \equiv (\alpha - \alpha_0)\cos(\delta_0)$. All of our fields are away from the celestial poles and have small angular coverage, making this transformation accurate. The $x$ coordinate is parallel to the \ac{los}, while the $y$ and $z$ coordinates are perpendicular (transverse) to it, so given that peculiar velocities contribute to the redshifts of objects in addition to cosmological expansion, our coordinate system is subject to distortions, often termed `redshift-space distortions' \cite[e.g.][]{1987MNRAS.227....1K}. We therefore measure correlation functions parallel and transverse to the \ac{los} independently, i.e. we measure the two-dimensional two-point correlation function $\xi(r_{\perp}, r_{\parallel})$, where for a given pair of objects denoted $i$ and $j$, we have
	\begin{align}
		r_{\perp}     &\equiv \sqrt{|y_i - y_j|^2 + |z_i - z_j|^2}, \nonumber \\
		r_{\parallel} &\equiv |x_i - x_j|.
	\end{align}
	Deviations from an isotropic signal in these coordinates can then be attributed to peculiar velocities along the \ac{los} and/or large-scale bulk motions between objects in our sample.

	We use the \cite{1993ApJ...412...64L} (LS) estimator to calculate the auto-correlation functions:
	\begin{equation}
		\xi(r_{\perp}, r_{\parallel}) = \frac{DD / n_{\textrm{DD}} - 2DR / n_{\textrm{DR}}}{RR / n_{\textrm{RR}}} + 1,
	\label{eq:correlation_function}
	\end{equation}
	where DD, DR and RR are the number of data-data, data-random and random-random pairs respectively at a given $r_{\perp}$ and $r_{\parallel}$ (or at a given $s$ for the redshift-space correlation function), and the values of $n$ correspond to their normalisation factors:
	\begin{align}
		n_{\textrm{DD}} &= N(N - 1) / 2, \nonumber \\
		n_{\textrm{DR}} &= \vartheta N^2, \nonumber \\
		n_{\textrm{RR}} &= \vartheta N(\vartheta N - 1) / 2,
	\end{align}
	where N is the total number of real objects, and $\vartheta N$ is the total number of random ones. We write the normalisation factors in this way to highlight that the random samples always have an integer number $\vartheta$ times as many objects as the real ones. For the absorber-galaxy cross-correlation function, we adopt the following, generalised form of the LS estimator:
	\begin{equation}
		\xi_{\textrm{ag}}(r_{\perp}, r_{\parallel}) = \frac{D_{\textrm{a}}D_{\textrm{g}} / n_{\textrm{ag}}^{\textrm{DD}} - D_{\textrm{a}}R_{\textrm{g}} / n_{\textrm{ag}}^{\textrm{DR}} - R_{\textrm{a}}D_{\textrm{g}} / n_{\textrm{ag}}^{\textrm{RD}}}{R_{\textrm{a}}R_{\textrm{g}} / n_{\textrm{ag}}^{\textrm{RR}}} + 1
	\label{eq:correlation_function_general}
	\end{equation}
	\cite[e.g.][]{2003ApJ...584...45A,2014MNRAS.437.2017T}, where $D_{\textrm{a}}D_{\textrm{g}}$, $D_{\textrm{a}}R_{\textrm{g}}$, $R_{\textrm{a}}D_{\textrm{g}}$ and $R_{\textrm{a}}R_{\textrm{g}}$ are the data-data, data-random, random-data and random-random absorber-galaxy pairs respectively, and their normalisation factors, $n$, are
	\begin{align}
		n_{\textrm{ag}}^{\textrm{DD}} &= N_{\textrm{a}}N_{\textrm{g}}, \nonumber \\
		n_{\textrm{ag}}^{\textrm{DR}} &= \vartheta_{\textrm{g}}N_{\textrm{a}}N_{\textrm{g}}, \nonumber \\
		n_{\textrm{ag}}^{\textrm{RD}} &= \vartheta_{\textrm{a}}N_{\textrm{a}}N_{\textrm{g}}, \nonumber \\
		n_{\textrm{ag}}^{\textrm{RR}} &= \vartheta_{\textrm{a}}\vartheta_{\textrm{g}}N_{\textrm{a}}N_{\textrm{g}},
	\end{align}
	where $N_{\textrm{a}}$ and $N_{\textrm{g}}$ are the total number of absorbers and galaxies respectively, and $\vartheta_{\textrm{a}}N_{\textrm{a}}$ and $\vartheta_{\textrm{g}}N_{\textrm{g}}$ are the total number of random absorbers and random galaxies respectively. \cite{1993ApJ...412...64L} have shown that these estimators minimise the variance in the correlation function, and so are preferable to 	other proposed estimators.

	A useful quantity, which we also compute, is the projection of the two-dimensional two-point correlation function along the \ac{los}:
	\begin{equation}
		\Xi(r_{\perp}) = 2 \int_0^{\infty} \textrm{d}r_{\parallel}~\xi(r_{\perp}, r_{\parallel}).
	\end{equation}
	In reality, one only integrates $\xi(r_{\perp}, r_{\parallel})$ up to a finite $r_{\parallel}$ where the correlation function ceases to be well measured, or where it is consistent with zero. The advantage of calculating this quantity is that it integrates over correlations smeared along the \ac{los} due to peculiar motions, and is insensitive to redshift-space distortions on small transverse scales where bulk flows are not important \cite[][]{1983ApJ...267..465D}. We can therefore find a relation between the `real-space' correlation function (free of distortions), $\xi(r)$, and $\Xi(r_{\perp})$ as:
	\begin{align}
		\Xi(r_{\perp}) &= 2 \int_0^{\infty} \textrm{d}r_{\parallel}~\xi(r) \nonumber \\
	     		   &= 2 \int_{r_{\perp}}^{\infty} \textrm{d}r~\xi(r)\frac{r}{\sqrt{r^2 - r_{\perp}^2}},
	\label{eq:projected_correlaton_function}
	\end{align}
	which gives $\xi(r)$ as the inverse Abel transform
	\begin{equation}
		\xi(r) = -\frac{1}{\pi} \int_r^{\infty} \frac{\textrm{d}\Xi(r_{\perp})}{\textrm{d}r_{\perp}} \frac{\textrm{d}r_{\perp}}{\sqrt{r_{\perp}^2 - r^2}}.
	\end{equation}
	\cite{1983ApJ...267..465D} showed that when $\xi(r)$ is described by a power law of the form
	\begin{equation}
		\xi(r) = \left(\frac{r}{r_0}\right)^{-\gamma},
	\end{equation}
	\cref{eq:projected_correlaton_function} yields
	\begin{equation}
		\Xi(r_{\perp}) = A(r_0, \gamma) r_{\perp}^{1 - \gamma},
	\label{eq:xcorr_power_law}
	\end{equation}
	where $A(r_0, \gamma) = r_0^{\gamma}\Gamma(1/2)\Gamma[(\gamma - 1)/2]/\Gamma(\gamma/2)$, and $\Gamma$ is the Gamma function. Fitting a power law form to $\Xi(r_{\perp})$ therefore allows determination of $r_0$ and $\gamma$, and hence $\xi(r)$, for $\gamma > 1$. Here $r_0$ is usually referred to as the `correlation length', and $\gamma$ is the slope of the correlation function.

	Given the volume-limited nature of any survey, all estimators are biased towards correlation amplitudes that are lower than the real ones. This arises because the mean density of objects is estimated from the survey itself, and is a well-known bias commonly referred to as the `integral constraint' \cite[][]{1977ApJ...217..385G}. \cite{1993ApJ...412...64L} showed that the $\xi$ measured using \cref{eq:correlation_function} or \cref{eq:correlation_function_general} and the real one, $\xi_{\textrm{real}}$, are related as
	\begin{equation}
		1 + \xi = \frac{1 + \xi_{\textrm{real}}}{1 + \xi_{V}},
	\end{equation}
	where $\xi_V$ is the (scalar) integral constraint, defined as
	\begin{equation}
		\xi_V \equiv \int_V \textrm{d}^2V~G(r)\xi_{\textrm{real}}(r)
	\end{equation}
	Here, $G(r)$ is the normalised geometric window function, which gives the probability of having two volume elements separated by a distance $r$ for a given survey geometry. For a large enough random catalogue, this is accurately approximated as $G(r) \approx \textrm{RR} / n_{\textrm{RR}}$.

	Although we cannot know $\xi_{\textrm{real}}$ a priori, we can still estimate the integral constraint by obtaining $r_0$ and $\gamma$ from a power-law fit to $\Xi(r_{\perp})$, and taking the mean of the random-random pair counts over all $r_{\parallel}$ bins at each $r_{\perp}$ bin to give a proxy for $\xi_V$ as
	\begin{equation}
		\tilde{\xi}_V = \sum_{r_{\perp}} \langle\textrm{RR}/n_{\textrm{RR}}\rangle A(r_0, \gamma)r_{\perp}^{1 - \gamma}.
	\end{equation}
	We can then make a (small) correction to our measured correlation function, $\xi'$, to obtain $\xi$ as follows:
	\begin{equation}
		\xi = \xi' + \tilde{\xi}_V(1 + \xi').
	\end{equation}
	All of the correlation function measurements that follow have this correction applied.

	To interpret our correlation function measurements, we follow \cite{2003ApJ...584...45A} and \cite{2014MNRAS.437.2017T} in using the Cauchy-Schwarz inequality:
	\begin{equation}
		\xi_{\textrm{ag}}^2 \leq \xi_{\textrm{gg}}\xi_{\textrm{aa}}.
	\label{eq:cs_inequality}
	\end{equation}
	The equality can only hold at any given scale when the density fluctuations that give rise to absorbers and galaxies are linearly dependent. In other words, both populations must trace the same underlying distribution of matter with a linear bias (independent of the scale) to achieve $\xi_{\textrm{ag}}^2 = \xi_{\textrm{gg}}\xi_{\textrm{aa}}$ in general.

	We estimate the uncertainty in our correlation function measurements using the bootstrap method, which in our experience provides the most conservative measure of the uncertainty \cite[see][for a discussion]{2014MNRAS.437.2017T}. We do this by creating $N_{\textrm{bs}} = 1000$ sets of 27 fields, randomly chosen (with replacement) from our set of 27 fields, and compute the uncertainty as
	\begin{equation}
		\Delta^2(\xi) = \frac{1}{N_{\textrm{bs}}}\sum_i^{N_{\textrm{bs}}}(\xi_i - \bar{\xi})^2,
	\end{equation}
	where $\xi_i$ is the correlation function measured from the $i$th random set of fields, and $\bar{\xi}$ is the mean of these measurements. Performing bootstrap realisations over fields ensures that we capture the sample variance, as well as the statistical uncertainty in the measurement.

	The bootstrap uncertainty estimation is clearly appropriate for our observational sample, for which we have a large number of independent fields, but is not so easily applied to the simulated sample we have assembled, which is drawn from a single cubic volume, 100 comoving Mpc on a side (see \Cref{sec:eagle} for details). To apply this uncertainty estimator, we would have to break the simulation down into sub-volumes, which limits the scales on which we can measure the correlation functions. For the simulated samples, we therefore quantify the statistical uncertainty on the measurement via the approximate estimator presented in \cite{1993ApJ...412...64L}:
	\begin{equation}
		\Delta_{\textrm{LS}}^2(\xi) \approx \frac{(1 + \xi)^2}{n_{\textrm{DD}}(RR / n_{\textrm{RR}})} \approx \frac{(1 + \xi)^3}{DD}.
	\end{equation}
	Note that this is greater than the commonly used Poissonian estimator, $\Delta^2_{\textrm{DD}}(\xi) = (1 + \xi) / \textrm{DD}$, by a factor of $\sim(1 + \xi)^2$, since it takes into account correlations introduced by non-independent cross-pairs. In practice, we are able to achieve negligibly small statistical uncertainties on the measurements from the simulated data, but we note that the sample variance (which may be much larger than the statistical one) is not taken into account.

	\subsection{Random samples}
	\label{sec:random_catalogues}
	As is clear from the previous section, the construction of random samples is a crucial part of any correlation function analysis. These random samples need to capture the selection functions in the data, so as not to bias the measurement. We present in the following sections a detailed description of the method for generating random samples of observed and simulated \ion{O}{6} absorbers and galaxies.
	
		\subsubsection{Random galaxy catalogues}
		For our observational sample, we create random galaxies for each field and survey independently. This means that individual fields containing galaxies from multiple surveys have separate random catalogues constructed for each of those surveys that are then combined at the end to form the random sample for that field. We do this since the different surveys in our sample have different selection functions (see \Cref{tab:galaxy_sample}), and it is easier to model these separately, rather than attempting to model the combined selection functions. We note that the T14-Q0107 survey listed in \Cref{tab:galaxy_sample} has a complex selection function, since it combines 4 sub-surveys constructed with different instruments, each with their own specific selection biases. Ideally, we would further split this survey down into its constituent parts, but the number of galaxies attributable to each of the different instruments is too small to reliably model their individual selection functions.
	
		Our process for creating random galaxy catalogues for the observational sample expands upon that described in \cite{2014MNRAS.437.2017T}, and works as follows. For a given galaxy, in a given field, and from a given survey, we create $\vartheta_{\textrm{g}} = 10$ random ones, varying the redshift of the galaxy, but preserving its position on the sky and all of its other properties. The random redshifts are drawn from a probability density function that is modelled on the observed redshift histogram for galaxies with matching properties in the survey from which the real redshift was obtained. We take into account the observed magnitude of the galaxy (in the broadband filter that defines the magnitude limit for the survey), and whether it is a star forming galaxy, a non star-forming galaxy, or neither of these. For example, if a galaxy in a particular field is drawn from the \ac{sdss}, and is a star-forming galaxy with an $r$-band magnitude $r = 16$, we randomly draw $\vartheta_{\textrm{g}} = 10$ redshifts from a probability density function that is modelled on the redshift histogram in \ac{sdss} for star-forming galaxies with that magnitude. In this way, our random samples reflect the individual survey sensitivity functions for galaxies of a given magnitude and spectral type, and also the evolution in the star-forming fraction. For galaxies that have no classification of spectral type, or are classified as \ac{agn}, we use the redshift histogram of \emph{all} galaxies from the same survey, and with the same magnitude as that galaxy. Our approach guarantees that we take into account the survey incompleteness in the construction of the random catalogues.
	
		The probability density distributions described above are constructed in the following way. First, we optimally bin redshift distributions using the algorithm presented by \cite{2006physics...5197K}. This is a maximum likelihood method for determining the optimum number of bins needed to both capture the dominant features in the data and minimise the number of random sampling fluctuations. We create histograms in this way for star-forming and non star-forming galaxies separately, in magnitude bins of size 1, shifted by 0.5 magnitudes, over the range $r = 13$ to $r = 25$. We iteratively increase the magnitude bin sizes at the bright and faint ends of the magnitude distributions to ensure that there are a minimum of 20 galaxies in each redshift histogram. We then smooth the histograms with a Gaussian smoothing kernel having a standard deviation equal to the bin size, to remove spikes in the redshift distributions attributable to large-scale structure (galaxy clusters, filaments, sheets and voids). The probability density distributions are then obtained by interpolating a cubic spline over the smoothed histograms, and by normalising to the area underneath the resulting curve.
	
		\begin{figure}
			\centering
			\includegraphics[width=8.4cm]{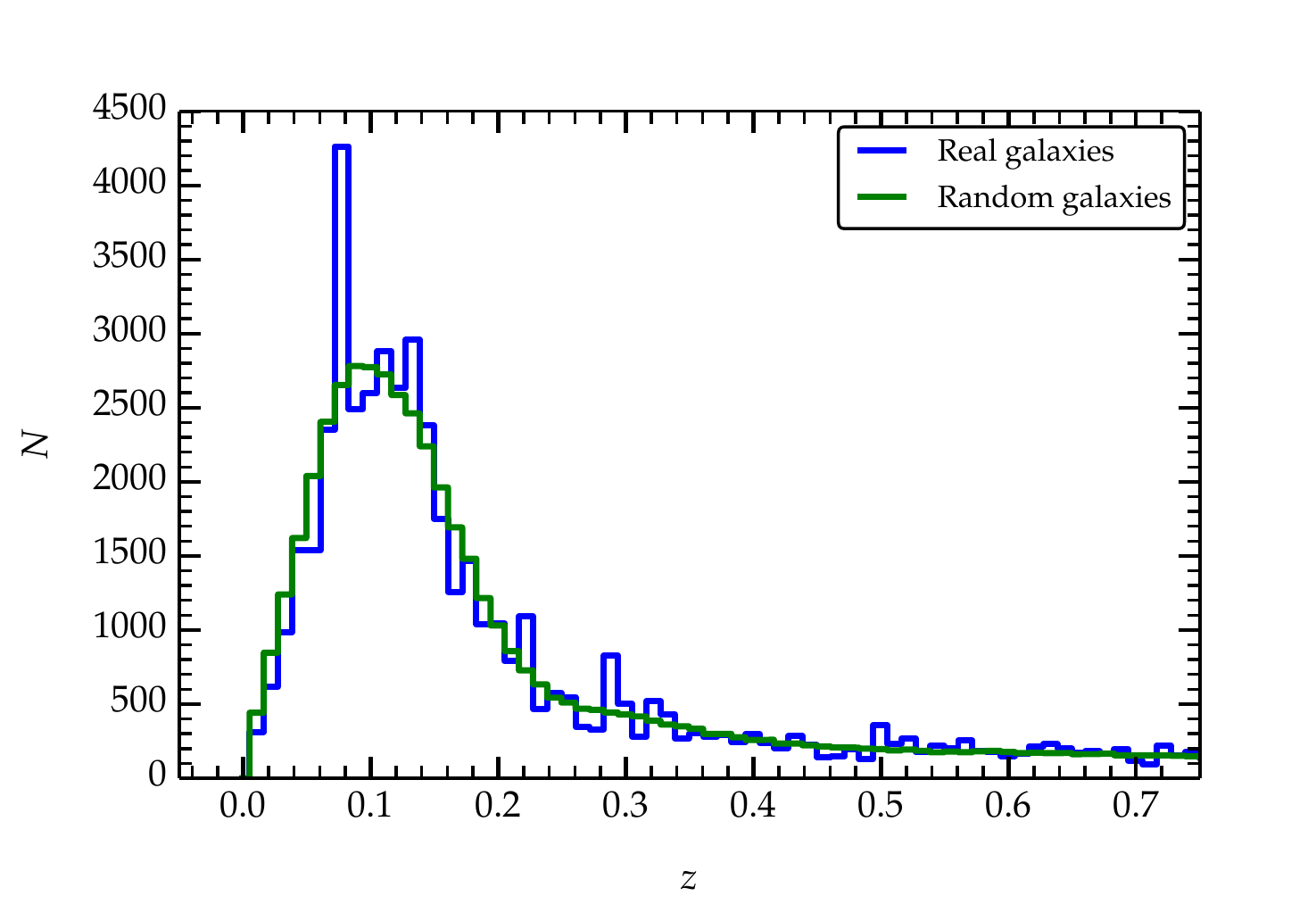}
			\caption[The real and random galaxy redshift distributions]{The real galaxy redshift distribution for our total sample (blue), and the random galaxy redshift distribution divided by $\vartheta_{\textrm{g}}$ (green). The smooth shape of the random distribution, following the broad shape of the real distribution, suggests that our approach to generating random catalogues of galaxies is robust.}
		\label{fig:random_galaxies}
		\end{figure}
	
		In \Cref{fig:random_galaxies}, we plot a histogram showing the redshift distribution of our total galaxy sample in blue, and the random sample divided by $\vartheta_{\textrm{g}}$ in green. The green histogram shows a smooth distribution, reflective of the overall selection function for our sample. Spikes are apparent in the real histogram attributable to large-scale structure. The smooth redshift distribution apparent in our random sample, following the overall shape of the real histogram, suggests that our approach is robust.
	
		Random galaxies for our sample extracted from \ac{eagle} simulation volume are distributed uniformly for each of the redshift snapshots. We created $\vartheta_{\textrm{g}} = 10$ times as many random galaxies as there are real ones.
		
		\subsubsection{Random absorber catalogues}
		For our observed \ion{O}{6} sample, which derives entirely from \ac{cos} \ac{fuv} spectra, we created $\vartheta_{\textrm{a}} = 1000$ random absorbers for every real one, varying the redshift, but preserving all other parameters. In this way, we randomise absorbers only along the \ac{qso} sight-lines, not transverse to them, so as to preserve the geometry of our survey. Random redshifts were chosen on the basis of an equivalent width threshold. For every real \ion{O}{6} absorber with Doppler broadening parameter $b$, and equivalent width $W$, we calculated the minimum equivalent width, $W_{3\sigma}$, at which the weaker transition in the doublet (\ion{O}{6} $\lambda1037$) for that absorber could still be observed above the required $3\sigma$ significance threshold as a function of wavelength in the spectrum from which that absorber was obtained. The significance of absorption features in \ac{cos} spectra cannot be estimated in the usual way due to the non-random noise properties of the instrument \cite[][]{2012PASP..124..830K}. Nevertheless, \cite{2012PASP..124..830K} provide a formalism for doing this, which we adopt (see their equations (4)--(5), (7) and (9)--(10)). We then transformed wavelength coordinates to redshift coordinates to obtain $W_{3\sigma}$ as a function of $z$, and distributed random absorbers in $z$ where the condition $W > W_{3\sigma}(z)$ was satisfied. We enforced a maximum redshift equivalent to a $-5000$ \kms\ offset from the \ac{qso} redshift, as was done in the data.
	
		\begin{figure}
			\centering
			\includegraphics[width=8.4cm]{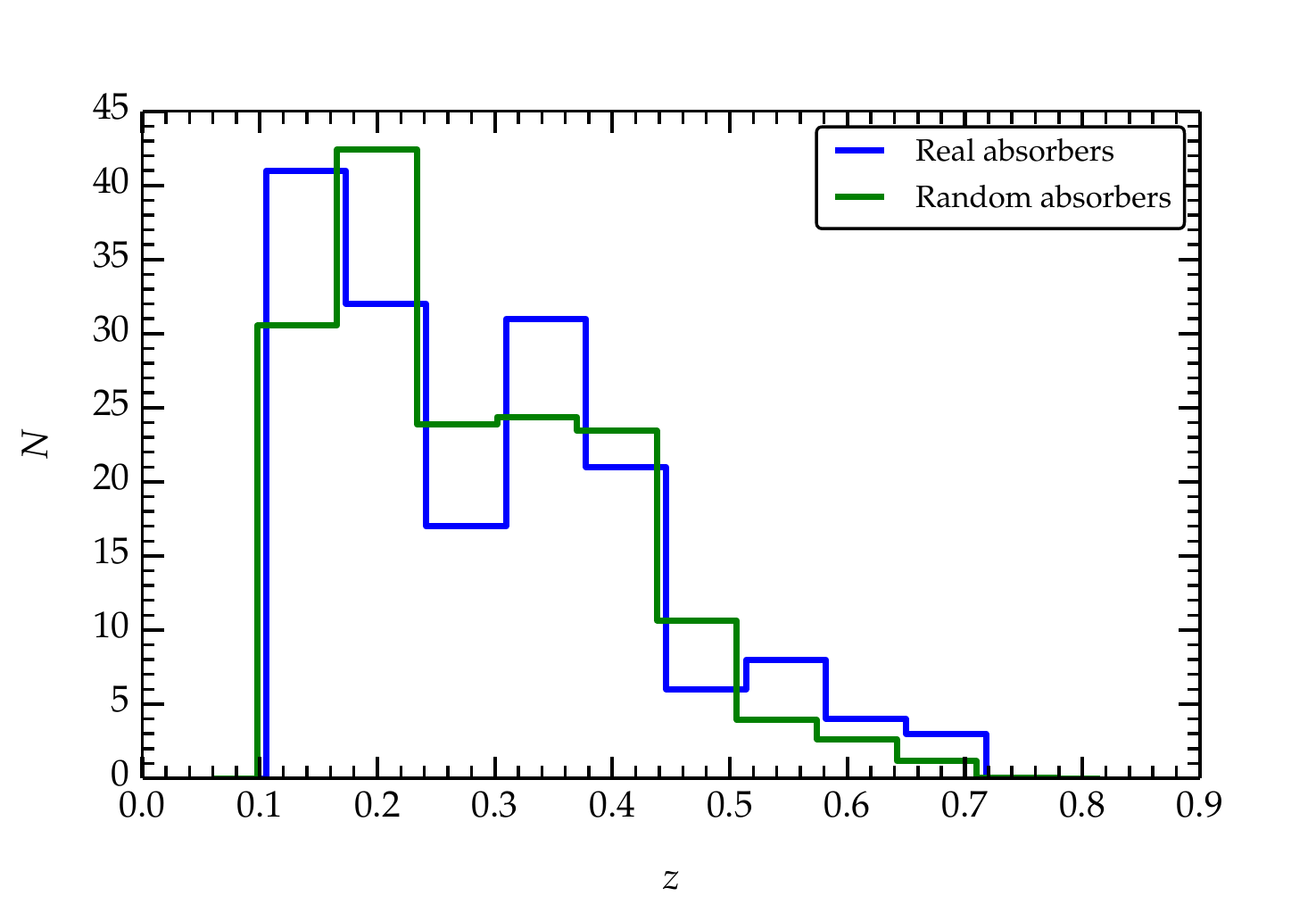}
			\caption[The real and random $\textrm{O}\;\textsc{vi}$ redshift distributions]{The real $\textrm{O}\;\textsc{vi}$ redshift distribution for our total sample (blue), and the random $\textrm{O}\;\textsc{vi}$ redshift distribution divided by $\vartheta_{\textrm{a}}$ (green). The random redshift distribution follows the same selection bias as the data, which suggests our procedure for generating random absorber catalogues is robust.}
		\label{fig:random_absorbers}
		\end{figure}
	
		In \Cref{fig:random_absorbers}, we show the redshift distribution of \ion{O}{6} absorbers for our full sample in blue in comparison to the random sample divided by $\vartheta_{\textrm{a}}$ in green. A larger number of \ion{O}{6} absorbers are found at lower redshifts, where they are detected in the wavelength range covered by the \ac{cos} G130M grating. \ac{cos} has greater sensitivity over this wavelength range, and the spectra from this grating therefore typically have a higher \ac{snr}. The shape of the random distribution follows this closely, which indicates our procedure is robust.
	
		For our simulated set of \ion{O}{6} absorbers extracted from the \ac{eagle} simulation, we randomise absorber velocities uniformly along the sight-lines from which they are extracted. We created $\vartheta_{\textrm{a}} = 10$ times as many random absorbers as there are real ones.

\section{Results}
\label{sec:results}
In this section, we present the results of our two-point correlation analysis following the mathematical formalism outlined in \Cref{sec:correlation_function_description}. The results that follow were computed using the random samples described in the previous section. We present results for our full sample of \ion{O}{6} absorbers and galaxies, and also for the subsamples containing only star-forming and non star-forming galaxies.

	\subsection{2D two-point correlation functions}
	\label{sec:correlation_functions}
\begin{figure*}
		\centering
		\includegraphics[width=14.3cm]{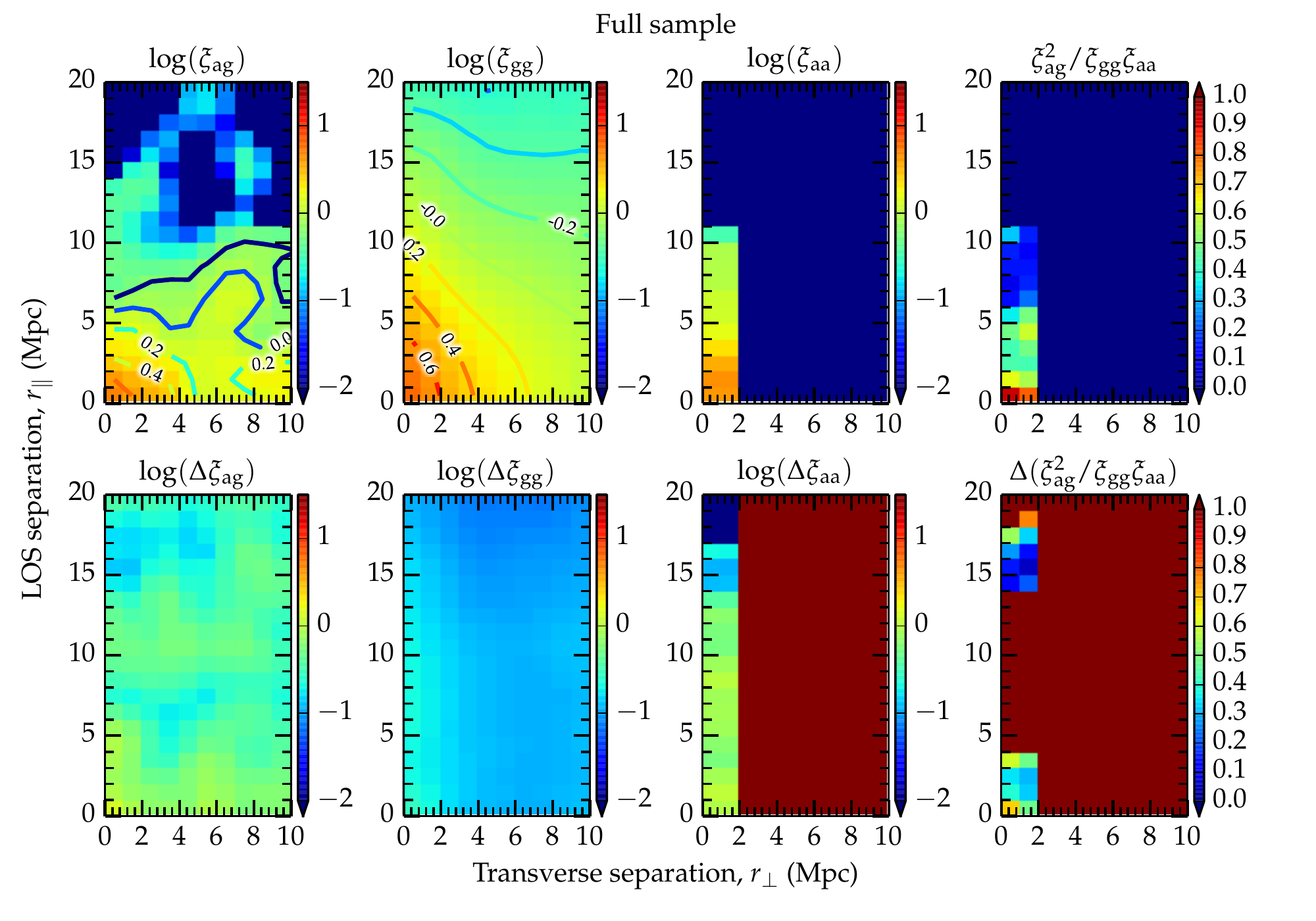}
		\caption[Two-dimensional correlation functions for galaxies and $\textrm{O}\;\textsc{vi}$ absorbers - full sample]{Two-dimensional correlation functions for galaxies and $\textrm{O}\;\textsc{vi}$ absorption systems (top panels) and their respective uncertainties (bottom panels), as a function of comoving separation parallel ($r_{\parallel}$) and perpendicular ($r_{\perp}$) to the LOS. From left to right: the galaxy-$\textrm{O}\;\textsc{vi}$ cross correlation function ($\xi_{\textrm{ag}}$), the galaxy auto-correlation function ($\xi_{\textrm{gg}}$), the $\textrm{O}\;\textsc{vi}$ auto-correlation function ($\xi_{\textrm{aa}}$), and the ratio $\xi_{\textrm{ag}}^2 / \xi_{\textrm{gg}}\xi_{\textrm{aa}}$. Note that our data are not suitable for measuring $\xi_{\textrm{aa}}$ and $\xi_{\textrm{ag}}^2 / \xi_{\textrm{gg}}\xi_{\textrm{aa}}$ on transverse scales $> 2$ Mpc. The correlation functions are calculated using a bin size of 1 Mpc, and the pair-counts are smoothed with a Gaussian kernel of standard deviation 1 Mpc in both directions.}
	\label{fig:o6_real_2d}
	\end{figure*}

	In \Cref{fig:o6_real_2d} we show the two-dimensional correlation functions (top panels) and their uncertainties (bottom panels) for our full sample of \ion{O}{6} absorbers and galaxies. The results are shown in bins of 1 Mpc (comoving) and are derived from pair counts smoothed with a Gaussian kernel having a standard deviation of 1 Mpc. The use of a smoothing kernel strikes a compromise between lowering the shot noise in the measurement, whilst keeping a relatively small bin size. From left to right, the panels show the \ion{O}{6}-galaxy cross-correlation function ($\xi_{\textrm{ag}}$), the galaxy auto-correlation function ($\xi_{\textrm{gg}}$), the \ion{O}{6} auto-correlation function ($\xi_{\textrm{aa}}$) and the ratio $\xi_{\textrm{ag}}^2 / \xi_{\textrm{gg}}\xi_{\textrm{aa}}$. Throughout this section, it is important to bear in mind that correlations along the \ac{los} in $\xi_{\textrm{aa}}$ are subject to the often somewhat subjective decomposition of \ion{O}{6} absorption complexes into multiple absorption `components' (see \Cref{sec:vp_fitting} for a description of our line-fitting approach).
	
	\begin{figure*}
		\centering
		\includegraphics[width=14.3cm]{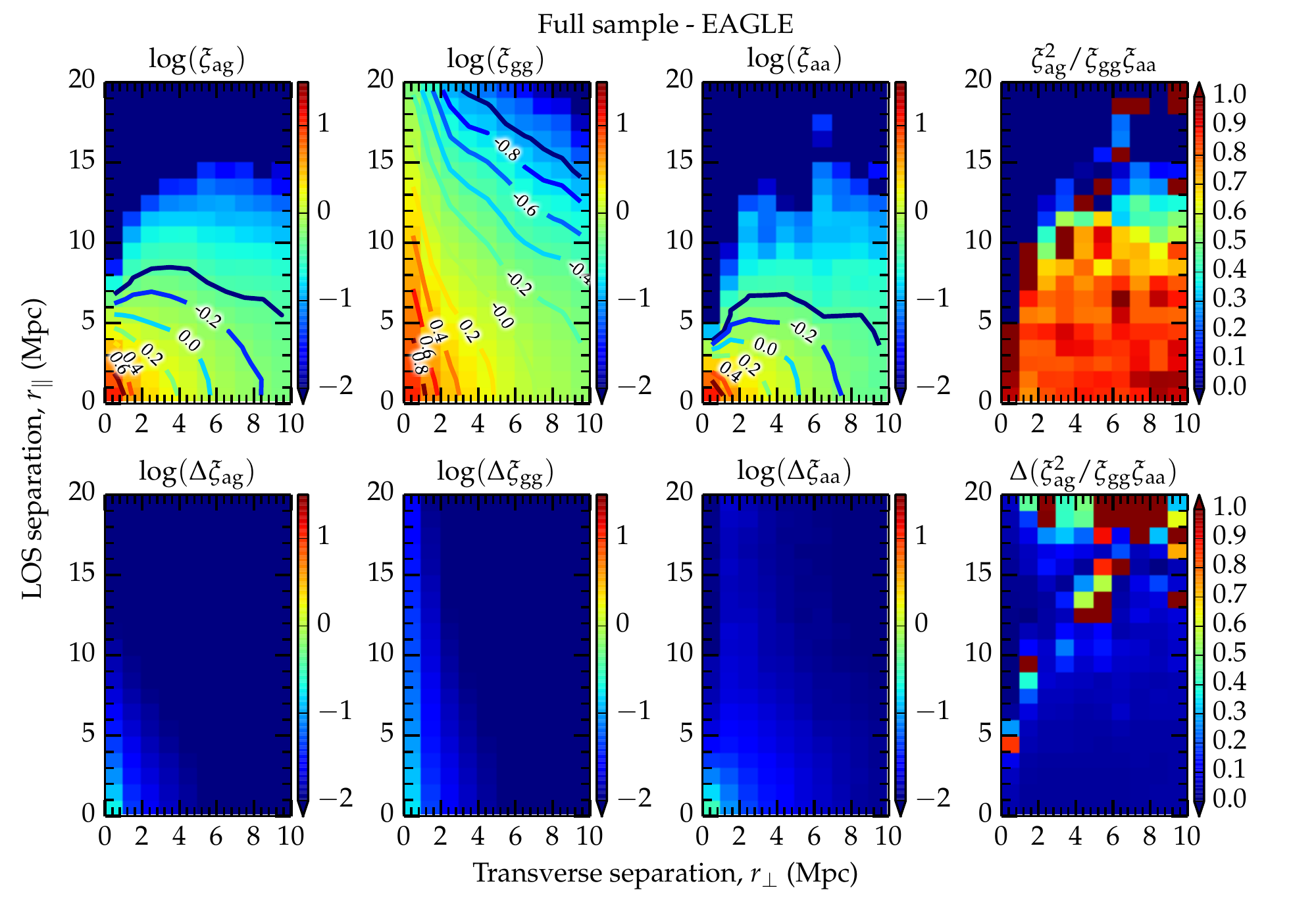}
		\caption[Two-dimensional correlation functions for galaxies and $\textrm{O}\;\textsc{vi}$ absorbers - \textsc{Eagle} full sample]{Same as \Cref{fig:o6_real_2d}, but for the simulated samples extracted from the \textsc{Eagle} simulation. Note that for the simulated samples we do not apply a Gaussian smoothing kernel.}
	\label{fig:o6_eagle_2d}
	\end{figure*}
	
	On inspection of the figure, we see that on small scales, the amplitudes of $\xi_{\textrm{ag}}$, $\xi_{\textrm{gg}}$ and $\xi_{\textrm{aa}}$ are comparable within the uncertainties. Note that we are unable to probe $\xi_{\textrm{aa}}$ on transverse scales $> 2$ Mpc with our data. The ratio $\xi_{\textrm{ag}}^2 / \xi_{\textrm{gg}}\xi_{\textrm{aa}}$ is close to 1 on scales $\lesssim 1$ Mpc, which suggests that \ion{O}{6} absorbers and galaxies are in close correspondence with one another on these scales. Our data do not have sufficient statistical power to quantify the presence of anisotropies in $\xi_{\textrm{ag}}$ or $\xi_{\textrm{aa}}$, but for $\xi_{\textrm{ag}}$ we can nevertheless examine the `isocorrelation' contours shown in these plots. We see a reasonably isotropic signal in $\xi_{\textrm{ag}}$, at least on the small scales where our measurement is stronger, which indicates that \ion{O}{6} absorbers show little velocity dispersion with respect to the galaxies. There is a hint of some compression in the signal along the \ac{los} on large scales, indicative of large-scale bulk motions \cite[e.g.][]{1987MNRAS.227....1K}, but not at a statistically significant level. Deviations from an isotropic signal are present in $\xi_{\textrm{gg}}$, which we expect for a galaxy sample of the size presented here, where a significant fraction of the galaxies reside in groups and clusters, with velocity dispersions of several 100 \kms. We note that the deviation from isotropy in $\xi_{\textrm{gg}}$ complicates the comparison of clustering amplitudes between $\xi_{\textrm{ag}}$ and $\xi_{\textrm{gg}}$ (and our inferences based on $\xi_{\textrm{ag}}^2 / \xi_{\textrm{gg}}\xi_{\textrm{aa}}$ above), since the effect of the redshift-space distortions is to `smear' the total correlation amplitude at a given $r_{\perp}$ over a range in $r_{\parallel}$. The total correlation amplitudes at a given $r_{\perp}$ are therefore not necessarily as comparable as they appear in \Cref{fig:o6_real_2d}. We investigate this further in the next section.
	
	In \Cref{fig:o6_eagle_2d} we present the same calculation as was illustrated in \Cref{fig:o6_real_2d}, but this time for the samples extracted from the \ac{eagle} simulation volume. For this, and all the comparisons that follow, we present results from the $z = 0.271$ \ac{eagle} snapshot, which is roughly the median redshift of our \ion{O}{6} sample. Note that for the \ac{eagle} calculations we do not apply a Gaussian smoothing kernel. Much like in the real case, the amplitudes of $\xi_{\textrm{ag}}$, $\xi_{\textrm{gg}}$ and $\xi_{\textrm{aa}}$ are all very similar at small scales within the uncertainties. We note that the correlation amplitudes in \ac{eagle} are somewhat higher than in the data at these scales, although the Gaussian smoothing kernel employed in the latter does act to lower the correlation amplitude at small separations where they are intrinsically peaked. Inspection of an unsmoothed version of \Cref{fig:o6_real_2d} reveals that the correlation amplitudes on the smallest scales are in fact comparable to those in \ac{eagle} within the uncertainties. Also, much like in the data, there is very little anisotropy on small scales in $\xi_{\textrm{ag}}$. Even without model-fitting, it is clear that the `anisotropy ratio' along the \ac{los} on $< 4$ Mpc scales is no more than 2:1, which limits the velocity dispersion of \ion{O}{6} around galaxies to $\lesssim 100$ \kms. A highly isotropic signal is seen in $\xi_{\textrm{aa}}$ on $< 4$ Mpc scales as well, which suggests that the \ion{O}{6} absorbers are virtually static with respect to one another on these scales. Intriguingly, the hint of a compression along the \ac{los} in $\xi_{\textrm{ag}}$, seen in the real data on large scales, appears in the simulated sample with high significance. The same is seen in $\xi_{\textrm{aa}}$. This then points to a picture in which \ion{O}{6} absorbers show bulk motions towards both galaxies and themselves on $\sim 5 - 10$ Mpc scales. Given the low significance of this result in the real data, we caution that this finding is far from conclusive. We note that there are some differences in $\xi_{\textrm{gg}}$ between the data and the simulation. In particular, there is a larger anisotropy in the signal along the \ac{los} in \ac{eagle} compared to the data. However, we note that a good agreement between observations and simulations is not necessarily expected for $\xi_{\textrm{gg}}$, since it depends on the fraction of passive galaxies. Observational studies are biased against selecting low-mass passive galaxies due to the difficulty in assigning a redshift, whereas galaxies of this type will always be present in the simulated samples. Since the galaxy auto-correlation functions are not the primary focus of this study, and we shall leave a more detailed comparison to future work.
	
	\begin{figure*}
		\centering
		\includegraphics[width=14.3cm]{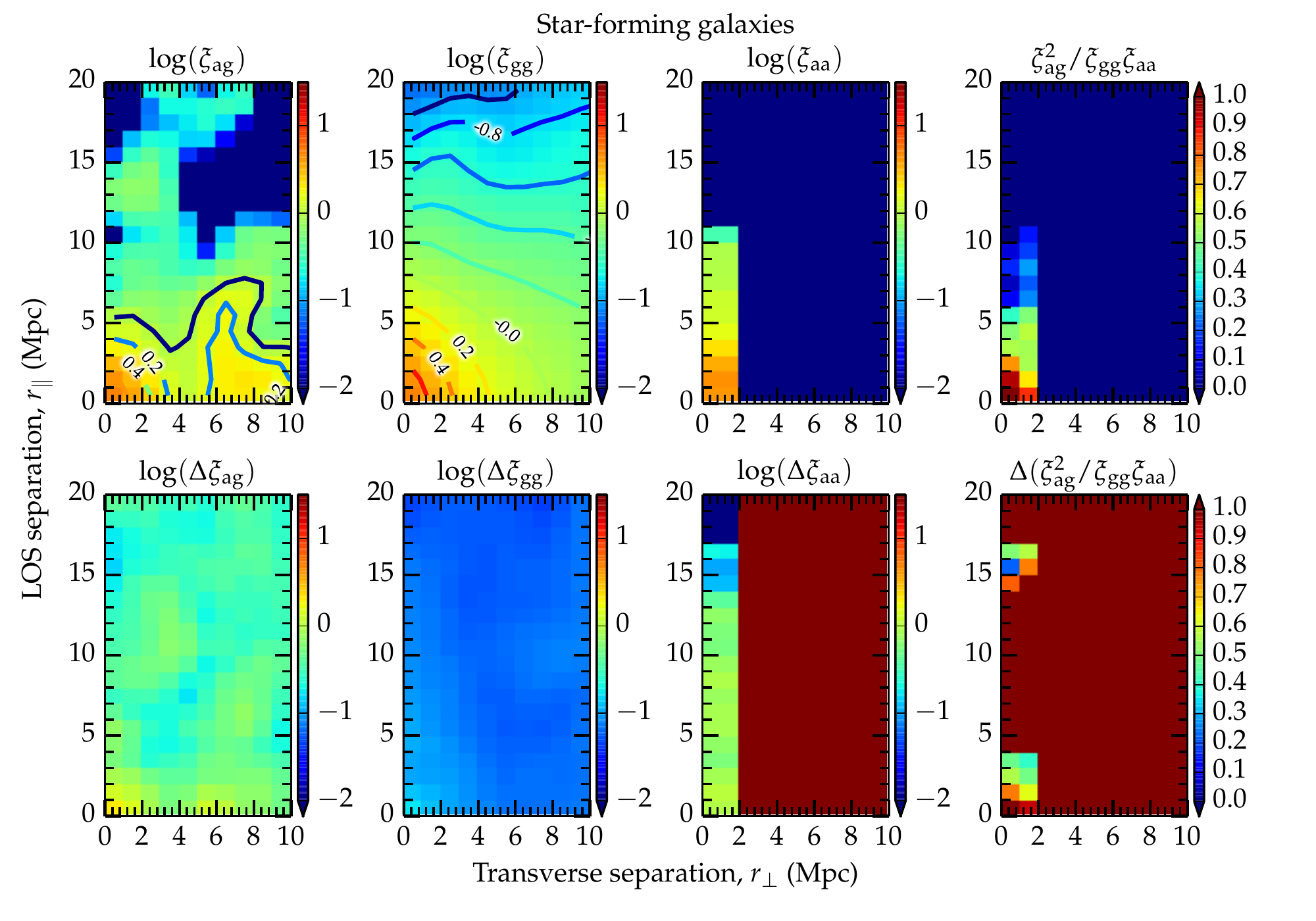}
		\caption[Two-dimensional correlation functions for galaxies and $\textrm{O}\;\textsc{vi}$ absorbers - star forming galaxies only]{Same as \Cref{fig:o6_real_2d}, but for star-forming galaxies only.}
	\label{fig:o6_real_2d-sf}
	\end{figure*}
	
	\begin{figure*}
		\centering
		\includegraphics[width=14.3cm]{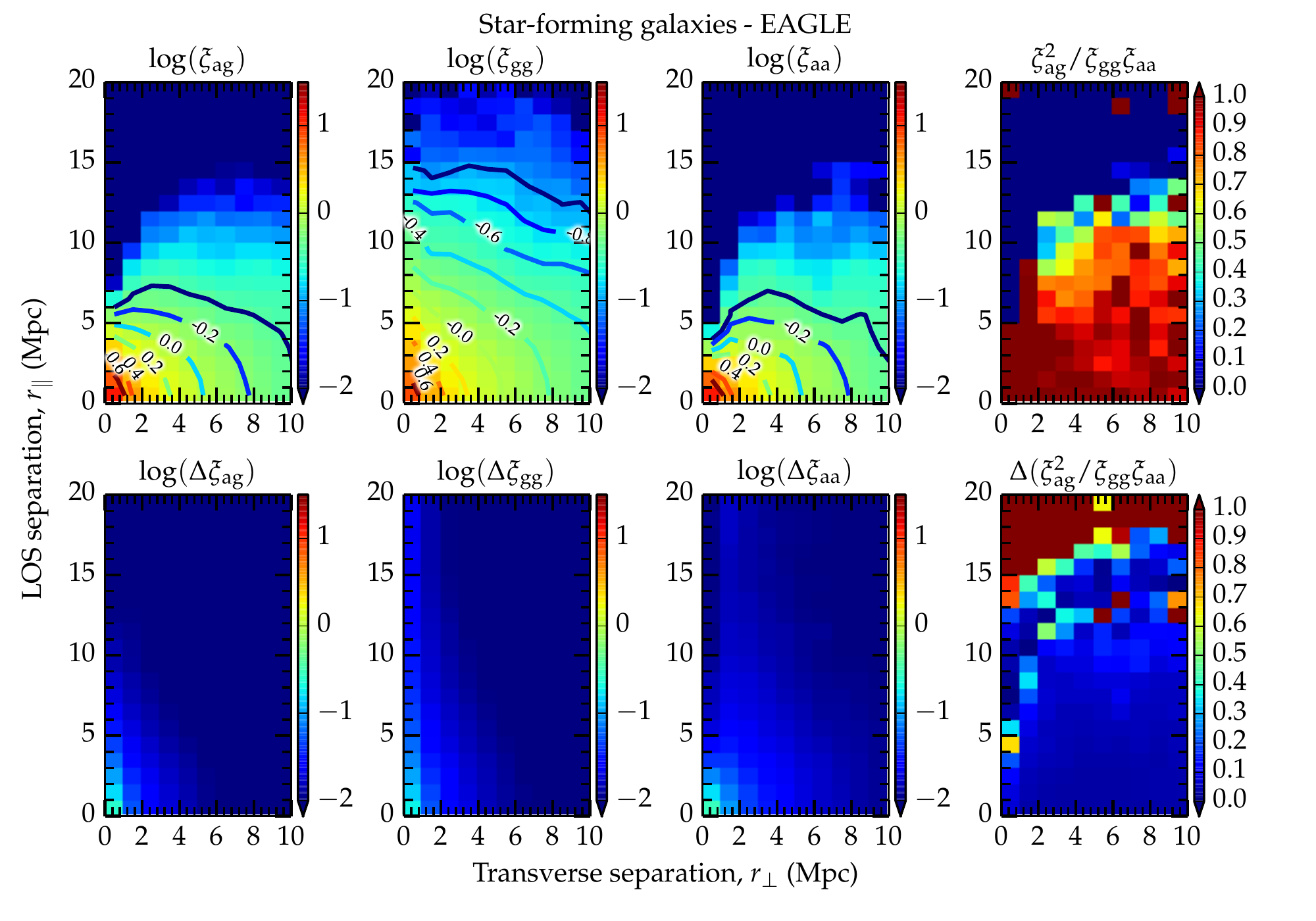}
		\caption[Two-dimensional correlation functions for galaxies and $\textrm{O}\;\textsc{vi}$ absorbers - \textsc{Eagle} star forming galaxies only]{Same as \Cref{fig:o6_real_2d-sf}, but for the simulated samples extracted from the \textsc{Eagle} simulation.}
	\label{fig:o6_eagle_2d-sf}
	\end{figure*}
	
	\begin{figure*}
		\centering
		\includegraphics[width=14.3cm]{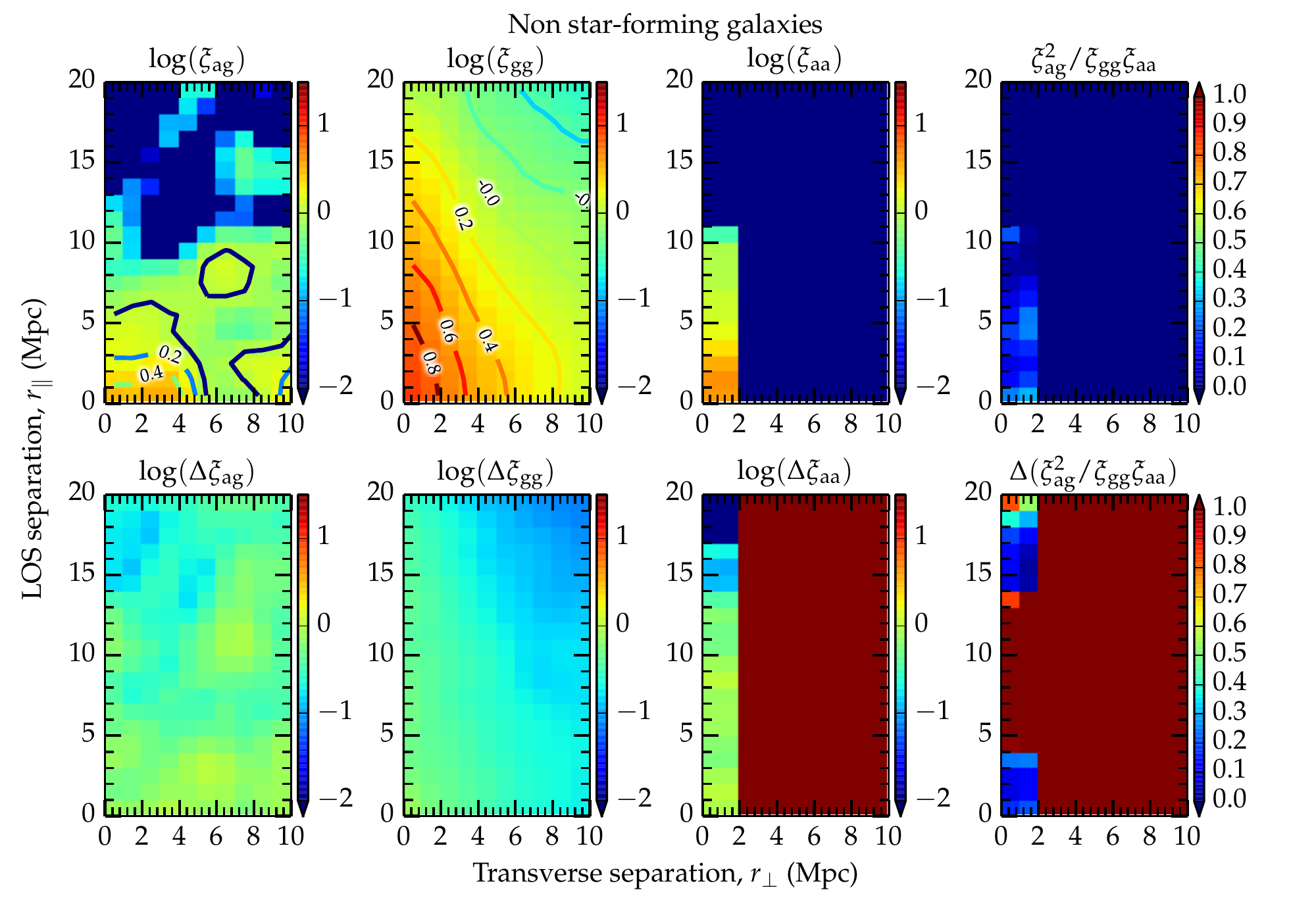}
		\caption[Two-dimensional correlation functions for galaxies and $\textrm{O}\;\textsc{vi}$ absorbers - non star-forming galaxies only]{Same as \Cref{fig:o6_real_2d}, but for non star-forming galaxies only.}
	\label{fig:o6_real_2d-nsf}
	\end{figure*}
	
	\begin{figure*}
		\centering
		\includegraphics[width=14.3cm]{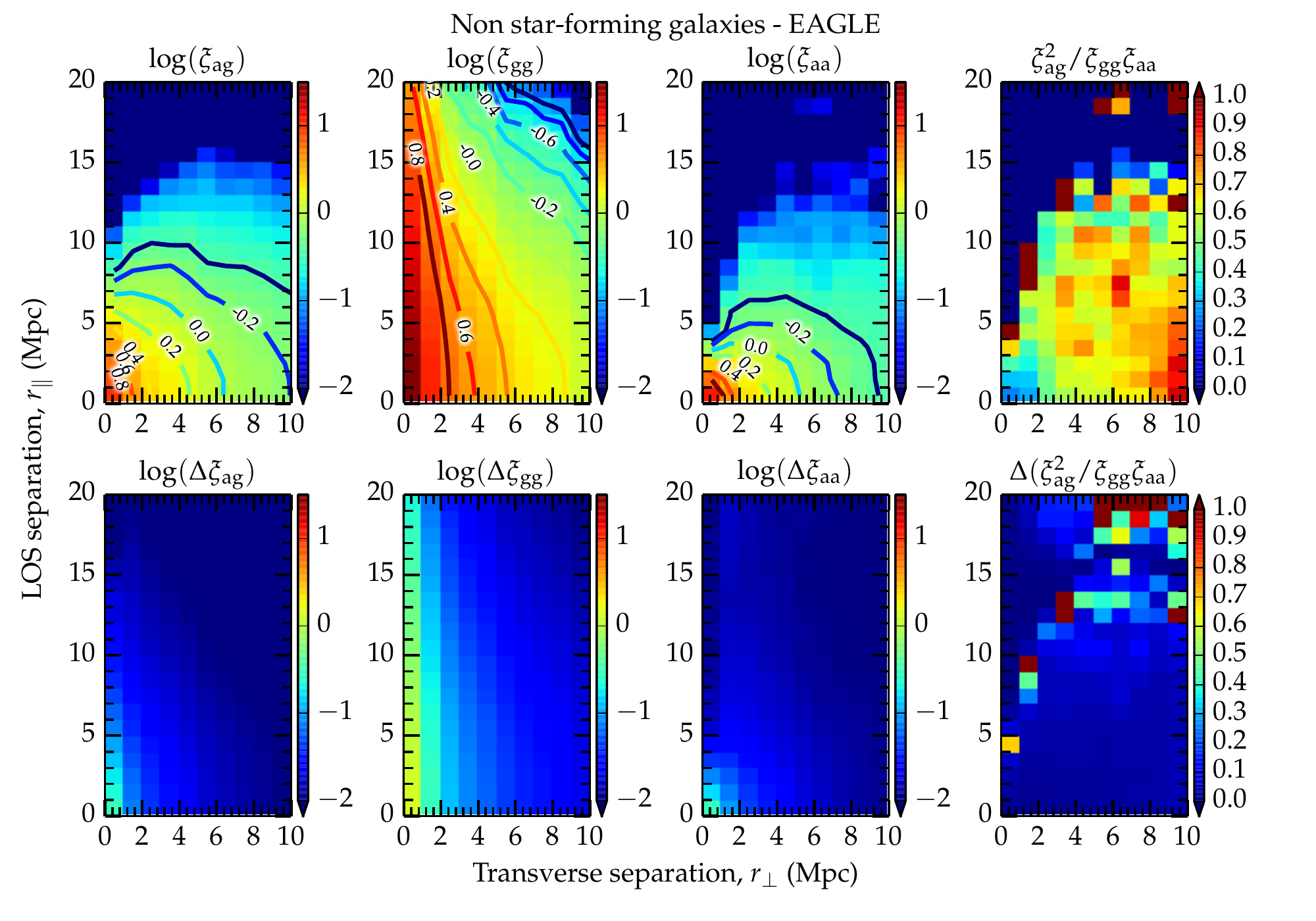}
		\caption[Two-dimensional correlation functions for galaxies and $\textrm{O}\;\textsc{vi}$ absorbers - \textsc{Eagle} non star-forming galaxies only]{Same as \Cref{fig:o6_real_2d-nsf}, but for the simulated samples extracted from the \textsc{Eagle} simulation.}
	\label{fig:o6_eagle_2d-nsf}
	\end{figure*}
	
	\begin{figure}
		\centering
		\includegraphics[width=8.4cm]{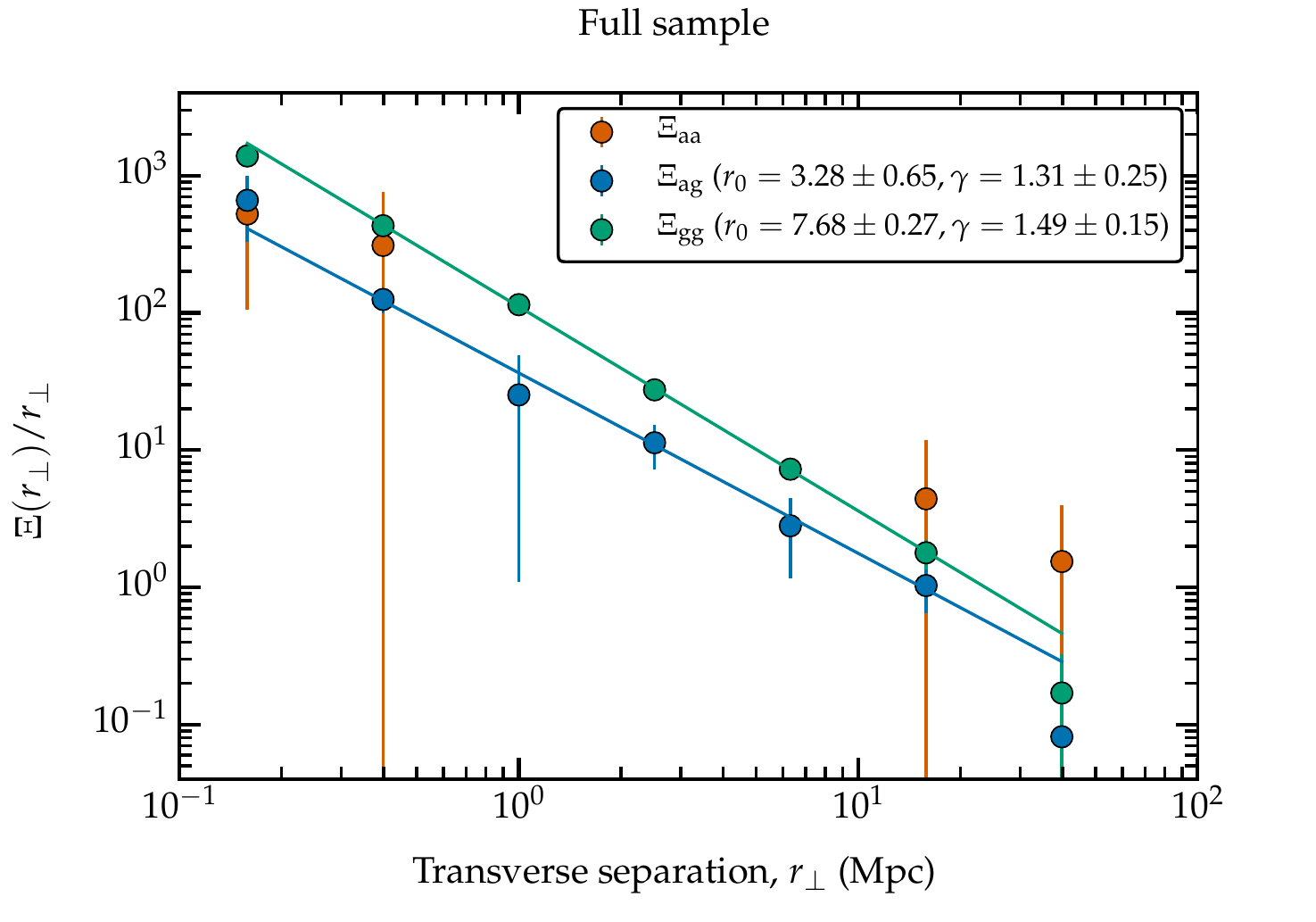}
		\caption[Correlation functions projected along the line-of-sight - full sample]{Correlation functions projected along the line-of-sight and divided by $r_{\perp}$, $\Xi(r_{\perp}) / r_{\perp}$, for our full sample of $\textrm{O}\;\textsc{vi}$ absorbers and galaxies. Our measurement for the projected $\textrm{O}\;\textsc{vi}$-galaxy cross-correlation function, $\Xi_{\textrm{ag}}$, is shown by the blue data points. Green data points show the measurement for the projected galaxy auto-correlation function, $\Xi_{\textrm{gg}}$, and red data points show the measurement for the projected $\textrm{O}\;\textsc{vi}$ auto-correlation function, $\Xi_{\textrm{aa}}$. The lines correspond to the best power-law fits (\cref{eq:xcorr_power_law}) to $\Xi_{\textrm{ag}}$, $\Xi_{\textrm{gg}}$ and $\Xi_{\textrm{aa}}$ in blue, green and red respectively. Note that the data points and their uncertainties are correlated. Uncertainties smaller than the symbols are not shown.}
	\label{fig:o6_real_1d}
	\end{figure}
	
	Next we examine the 2D two-point correlation functions for \ion{O}{6} absorbers and star-forming galaxies only. For the real data, this calculation is shown in \Cref{fig:o6_real_2d-sf}. Again, we see that $\xi_{\textrm{ag}}$, $\xi_{\textrm{gg}}$ and $\xi_{\textrm{aa}}$ are all comparable within the uncertainties. This leads to a ratio $\xi_{\textrm{ag}}^2 / \xi_{\textrm{gg}}\xi_{\textrm{aa}}$ that is consistent with 1 on small scales. We see no significant anisotropies in $\xi_{\textrm{ag}}$ either, which suggests that on small scales, \ion{O}{6} absorbers and star-forming galaxies trace the same underlying distribution of matter, and show very little velocity dispersion with respect to one another (certainly $\lesssim 100$ \kms). Redshift-space distortions in $\xi_{\textrm{gg}}$ are less evident for star-forming galaxies, and are qualitatively similar to those in $\xi_{\textrm{ag}}$ and $\xi_{\textrm{aa}}$, which makes our findings based on $\xi_{\textrm{ag}}^2 / \xi_{\textrm{gg}}\xi_{\textrm{aa}}$ somewhat more robust than those from the full sample.
	
	In \Cref{fig:o6_eagle_2d-sf} we show the same calculation for the \ac{eagle} simulation, again with no Gaussian smoothing kernel applied. In agreement with the real data, $\xi_{\textrm{ag}}$, $\xi_{\textrm{gg}}$ and $\xi_{\textrm{aa}}$ are all consistent within the uncertainties, and the ratio $\xi_{\textrm{ag}}^2 / \xi_{\textrm{gg}}\xi_{\textrm{aa}}$ is consistent with 1. We see the same compression in $\xi_{\textrm{ag}}$ and $\xi_{\textrm{aa}}$ along the line of sight on large scales as was seen for the full sample. However, this is not detected in the real data with significance. A small anisotropy is seen in $\xi_{\textrm{ag}}$, which amounts to a velocity dispersion between \ion{O}{6} absorbers and star-forming galaxies of no more than $\sim 100$ \kms.
	
	Finally, we examine the 2D two-point correlation functions for \ion{O}{6} absorbers and non star-forming galaxies only. We show this calculation for the real data in \Cref{fig:o6_real_2d-nsf}. Now we see that the correlation amplitudes of $\xi_{\textrm{ag}}$ and $\xi_{\textrm{aa}}$ are both comparable within the uncertainties, but the amplitude of $\xi_{\textrm{gg}}$ is significantly higher than both of these on small scales. As a result, we see the ratio $\xi_{\textrm{ag}}^2 / \xi_{\textrm{gg}}\xi_{\textrm{aa}}$ is nearly consistent with zero. This suggests that \ion{O}{6} absorbers and non star-forming galaxies trace the underlying distribution of matter differently, although in this case we note that the interpretation of $\xi_{\textrm{ag}}^2 / \xi_{\textrm{gg}}\xi_{\textrm{aa}}$ is complicated by redshift-space distortions that are highly-prominent in $\xi_{\textrm{gg}}$. Again, we see no evidence for any anisotropy in $\xi_{\textrm{ag}}$ along the \ac{los}, which implies that \ion{O}{6} absorbers show very little velocity dispersion with respect to non star-forming galaxies.
	
	The corresponding calculation for the \ac{eagle} simulation is shown in \Cref{fig:o6_eagle_2d-nsf}. We see a very similar situation to the data, whereby $\xi_{\textrm{ag}}$ and $\xi_{\textrm{aa}}$ are both similar within the uncertainties, but the amplitude of $\xi_{\textrm{gg}}$ is significantly higher than both of these. This leads to a ratio $\xi_{\textrm{ag}}^2 / \xi_{\textrm{gg}}\xi_{\textrm{aa}}$ that is nearly zero on the smallest scales. However, we note that in both the simulation and the data, the clustering amplitudes of \ion{O}{6} absorbers around non star-forming galaxies are highly comparable to those around star-forming galaxies. This suggests that the likelihood of finding an \ion{O}{6} absorber close to a non-star forming galaxy should be similar to the likelihood of finding an \ion{O}{6} absorber close to a star-forming galaxy over the scales probed. As is consistently seen in the simulation, there does exist a small anisotropy in $\xi_{\textrm{ag}}$ along the \ac{los}, but only at the level whereby the velocity dispersion between \ion{O}{6} absorbers and non star-forming galaxies is $\lesssim 100$ \kms, which is broadly consistent with the real data within the uncertainties.

	\subsection{Correlation functions projected along the line-of-sight}
	\label{sec:projected_correlation_functions}
	Comparisons of the clustering amplitudes between $\xi_{\textrm{ag}}$, $\xi_{\textrm{gg}}$ and $\xi_{\textrm{aa}}$ for the 2D two-point correlation functions in the previous section are complicated by the redshift-space distortions that lead to anisotropies in the signal. To better-compare the clustering amplitudes as a function of scale, we now examine the correlation functions that are projected along the \ac{los}, as in \cref{eq:projected_correlaton_function}.
	
\begin{figure}
		\centering
		\includegraphics[width=8.4cm]{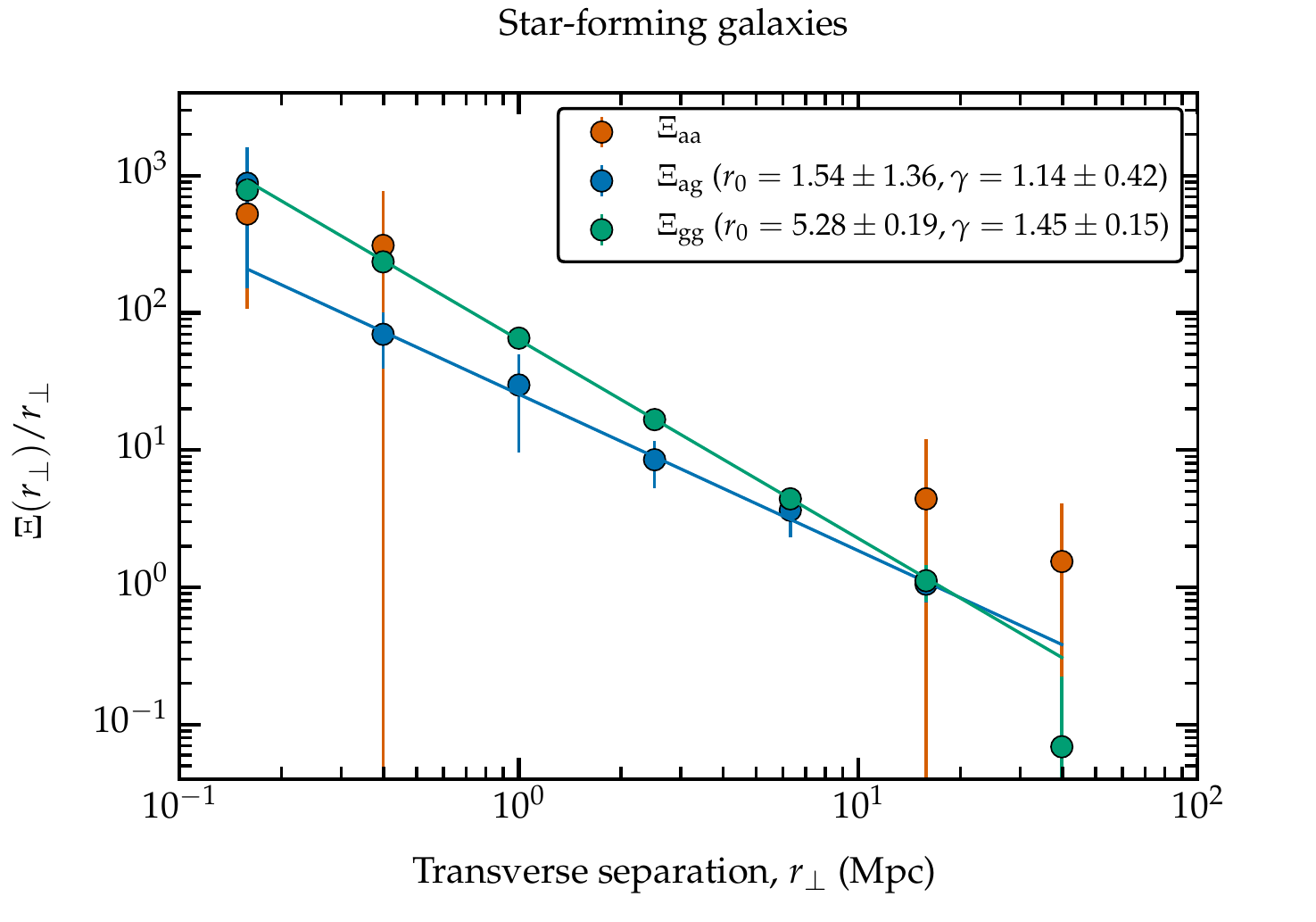}
		\caption[Correlation functions projected along the line-of-sight - star-forming galaxies only]{Same as \Cref{fig:o6_real_1d}, but for star-forming galaxies only.}
	\label{fig:o6_real_1d-sf}
	\end{figure}

\begin{figure}
		\centering
		\includegraphics[width=8.4cm]{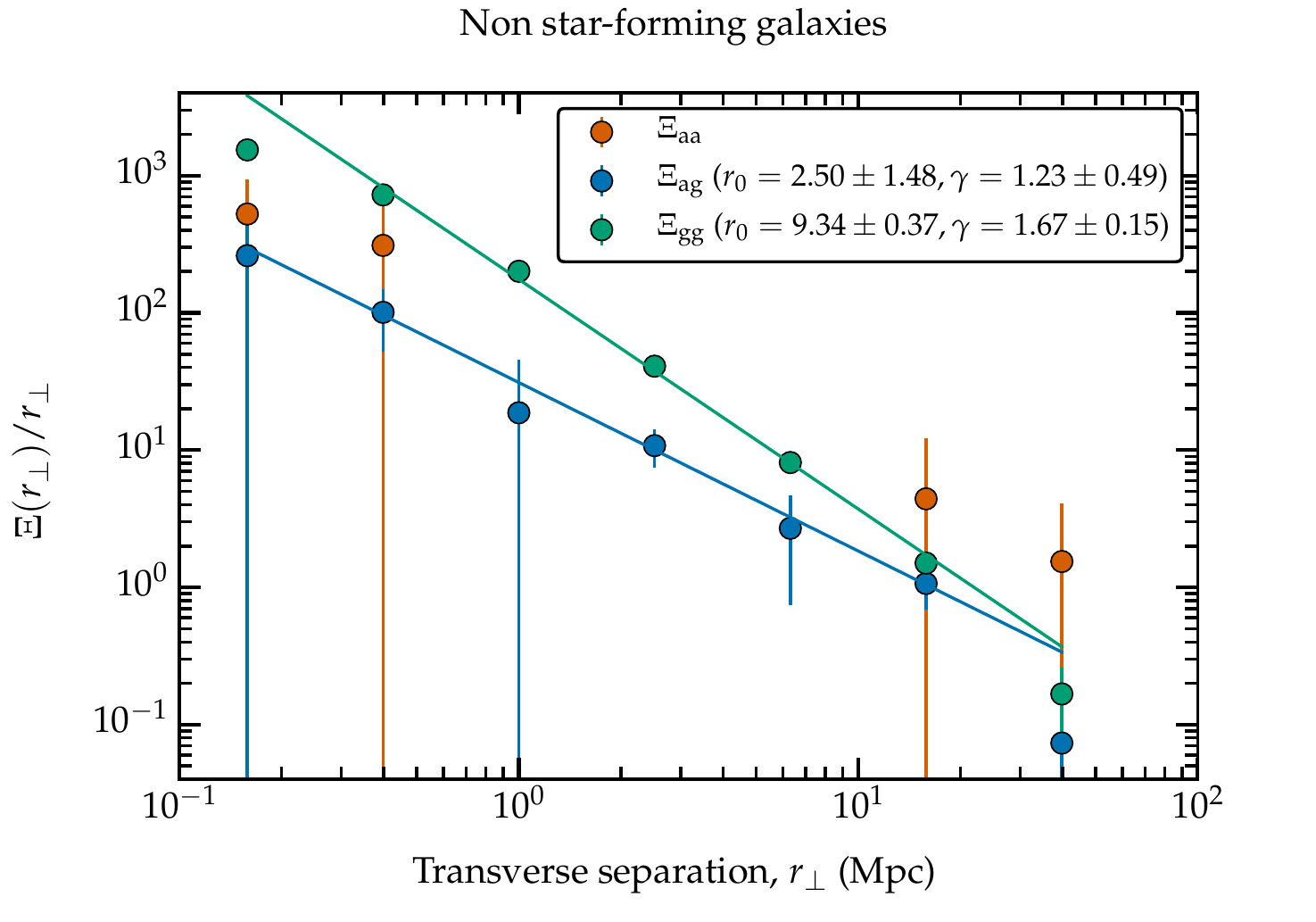}
		\caption[Correlation functions projected along the line-of-sight - non star-forming galaxies only]{Same as \Cref{fig:o6_real_1d}, but for non star-forming galaxies only.}
	\label{fig:o6_real_1d-nsf}
	\end{figure}

	In \Cref{fig:o6_real_1d,fig:o6_real_1d-sf,fig:o6_real_1d-nsf}, we show the correlation functions of \ion{O}{6} absorbers and galaxies projected along the \ac{los} and divided by $r_{\perp}$, $\Xi(r_{\perp}) / r_{\perp}$, for our full sample, and for star-forming and non star-forming galaxies only. We show the projected \ion{O}{6}-galaxy cross-correlation function, $\Xi_{\textrm{ag}}$, in blue data points, the projected galaxy auto-correlation function, $\Xi_{\textrm{gg}}$, in green data points and the projected \ion{O}{6} auto-correlation function, $\Xi_{\textrm{aa}}$, in red data points. For $\Xi_{\textrm{gg}}$ we integrate to $r_{\parallel} = 45$ Mpc, and for $\Xi_{\textrm{ag}}$ and $\Xi_{\textrm{aa}}$ we integrate to $r_{\parallel} = 13$ Mpc. These integration limits are the minimum for which the data points had converged to stable values, indicating that we are fully integrating the reliably measured signal, and minimising the addition of shot noise. We note that the data points are correlated, and that uncertainties smaller than the data points are not shown. The error bars show $1\sigma$ bootstrap uncertainties that include both the variance and covariance in the measurement. The lines show the best-fitting power-laws to the data (\cref{eq:xcorr_power_law}) using the same colour scheme. We only fit these power-laws to the data at $r_{\perp} < 10$ Mpc, as deviations from a power-law are typically seen at $r_{\perp} > 10$ Mpc. We do not attempt to fit a power-law slope to $\Xi_{\textrm{aa}}$, as these data show very low statistical significance and are likely not robust. It is important to note that while the measurement of $\xi_{\textrm{aa}}$ in the 2D two-point correlation function is dominated by \ion{O}{6} pairs along the \ac{los}, these pairs do not contribute to the measurements in \Cref{fig:o6_real_1d,fig:o6_real_1d-sf,fig:o6_real_1d-nsf}, which instead come from transverse pairs in the very few closely-separated \ac{qso} sight-line pairs (and triplets) in our sample. The best-fit parameters for all of the projected correlation functions are summarised in \Cref{tab:xcorr_params}.
	
	Starting with \Cref{fig:o6_real_1d}, which shows projected correlation functions for our full sample, we note that a power-law is a good description of the data for $\Xi_{\textrm{ag}}$ and $\Xi_{\textrm{gg}}$ at $r_{\perp} < 10$ Mpc. We find that $\xi_{\textrm{ag}}(r)$ has a correlation length of $r_0^{\textrm{ag}} = 3.28\pm0.65$ Mpc, and a slope of $\gamma^{\textrm{ag}} = 1.31\pm0.24$, whereas $\xi_{\textrm{gg}}(r)$ has a correlation length of $r_0^{\textrm{gg}} = 7.68\pm0.27$ Mpc, and a slope of $\gamma^{\textrm{gg}}=1.49\pm0.15$. We therefore find that the correlation length of \ion{O}{6} absorbers around galaxies is less than the correlation length of galaxies with themselves, and this finding is significant at a $>4\sigma$ confidence level. The data also suggests that the slope of $\Xi_{\textrm{ag}}$ is shallower than that of $\Xi_{\textrm{gg}}$, although they are consistent with one another within the $1\sigma$ uncertainties. The amplitude of $\Xi_{\textrm{aa}}$, and the apparent slope, is consistent with both $\Xi_{\textrm{ag}}$ and $\Xi_{\textrm{gg}}$. The differing slopes between $\Xi_{\textrm{ag}}$ and $\Xi_{\textrm{gg}}$, combined with the difference in their correlation lengths, suggests that \ion{O}{6} absorbers and galaxies may not linearly trace the same underlying distribution of matter in general, however this is by no means definitive given the uncertainties on the data, in particular for $\Xi_{\textrm{aa}}$.
	
	\begin{table*}
	\setlength{\tabcolsep}{12pt}
	\caption{Best fit projected correlation function parameters.}
	\begin{tabular}{l c c c c}
		\hline
  		Sample & $r_0^{\textrm{ag}}$ (Mpc) & $\gamma^{\textrm{ag}}$ & $r_0^{\textrm{gg}}$ (Mpc) & $\gamma^{\textrm{gg}}$ \\
  		\hline
  		Full sample               & $3.28 \pm 0.65$ & $1.31 \pm 0.25$ & $7.68 \pm 0.27$ & $1.49 \pm 0.15$ \\
  		Star-forming galaxies     & $1.54 \pm 1.36$ & $1.14 \pm 0.42$ & $5.28 \pm 0.19$ & $1.45 \pm 0.15$ \\
  		Non star-forming galaxies & $2.50 \pm 1.48$ & $1.23 \pm 0.49$ & $9.34 \pm 0.37$ & $1.67 \pm 0.15$ \\
 		\hline
	\end{tabular}
	\label{tab:xcorr_params}
	\end{table*}
	
	As we did for the 2D two-point correlation functions, we now examine the correlation functions with star-forming and non star-forming galaxies separately. We show the projected correlation functions with star-forming galaxies only in \Cref{fig:o6_real_1d-sf}. Here we find that $\xi_{\textrm{ag}}(r)$ has a correlation length of $r_0^{\textrm{ag}} = 1.55\pm0.35$ Mpc, and a slope of $\gamma^{\textrm{ag}} = 1.14\pm0.41$, whereas $\xi_{\textrm{gg}}(r)$ has a correlation length of $r_0^{\textrm{gg}} = 5.28\pm0.21$ Mpc, and a slope of $\gamma^{\textrm{gg}}=1.45\pm0.15$. The correlation length of \ion{O}{6} absorbers around star-forming galaxies is therefore still less than the correlation length of star-forming galaxies with themselves, although at a lower significance level ($>2\sigma$), and there remains marginal evidence for a difference in slope between $\Xi_{\textrm{ag}}$ and $\Xi_{\textrm{gg}}$, although this result is not significant ($<1\sigma$). We therefore find indications that \ion{O}{6} absorbers and star-forming galaxies may not linearly trace the same underlying distribution of matter, similar to the full sample, even though the clustering amplitudes in the cross-correlation function do indicate that \ion{O}{6} absorbers have a strong association with star-forming galaxies in general.
	
	Finally, in \Cref{fig:o6_real_1d-nsf}, we show the projected correlation functions with non star-forming galaxies only. Here we find that $\xi_{\textrm{ag}}(r)$ has a correlation length of $r_0^{\textrm{ag}} = 2.63\pm1.56$ Mpc, and a slope of $\gamma^{\textrm{ag}} = 1.22\pm0.51$, whereas $\xi_{\textrm{gg}}(r)$ has a correlation length of $r_0^{\textrm{gg}} = 9.35\pm0.38$ Mpc, and a slope of $\gamma^{\textrm{gg}}=1.67\pm0.15$. We therefore find that the correlation length of \ion{O}{6} absorbers around non star-forming galaxies is less than the correlation length of non star-forming galaxies with themselves at a $>3 \sigma$ significance level. The slopes of the correlation functions are still consistent with each other within the uncertainties, but the data nevertheless suggests that the slope in $\Xi_{\textrm{gg}}$ is steeper than that of $\Xi_{\textrm{ag}}$. This supports indications from \Cref{fig:o6_real_2d-nsf} that \ion{O}{6} absorbers and non star-forming galaxies do not linearly trace the same underlying distribution of matter.
	
	It is well known that non star-forming galaxies are more biased tracers of matter than are non star-forming galaxies, as is clearly seen from the difference in the correlation lengths and slopes of their auto-correlation functions. However, the most striking result is the similarity in the correlation lengths and slopes of the cross-correlation functions of \ion{O}{6} absorbers with star-forming and non star-forming galaxies. These slopes and amplitudes are entirely consistent with one another within the uncertainties, and this demonstrates that the likelihood of finding \ion{O}{6} absorbers around star-forming galaxies is similar to the likelihood of finding \ion{O}{6} absorbers around non star-forming galaxies, at least over the scales probed by our data (to separations as close as $\sim100$ kpc).
	
	\begin{figure}
		\centering
		\includegraphics[width=8.4cm]{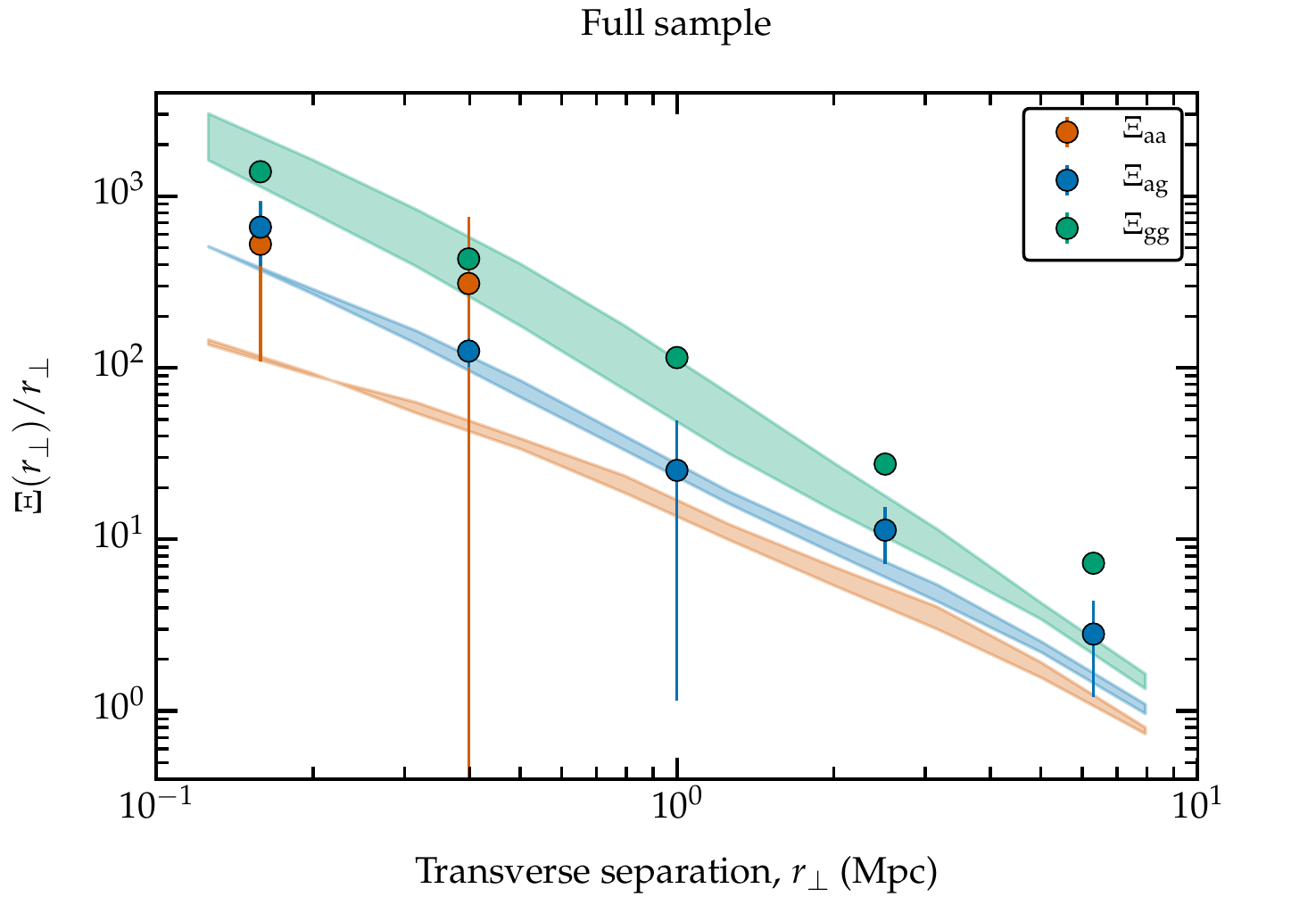}
		\caption[Correlation functions projected along the line-of-sight for galaxies and $\textrm{O}\;\textsc{vi}$ absorbers - full sample - \textsc{Eagle} comparison]{Same as \Cref{fig:o6_real_1d}, but now comparing the data to the predicted slopes and amplitudes of the correlation functions from the \ac{eagle} simulation. The shaded regions represent the predictions over the redshift range $0.1 < z < 0.7$, and adopt the same colour scheme as the data.}
	\label{fig:o6_eagle_1d}
	\end{figure}
	
	We now compare the correlation functions projected along the \ac{los} to predictions from the \ac{eagle} simulation. In \Cref{fig:o6_eagle_1d}, we show the same measurements as in \Cref{fig:o6_real_1d} for our full sample, with the predictions from \ac{eagle} over the redshift range $0.1 < z < 0.7$ shown as shaded regions with the same colour scheme as the data. We see a reasonable agreement in $\Xi_{\textrm{ag}}$ between the real data and that from the simulation given the uncertainties. The simulation also produces predictions for $\Xi_{\textrm{aa}}$ that are consistent with the data. We also find a reasonable agreement in $\Xi_{\textrm{gg}}$ at $<$ Mpc scales, but this agreement ceases at larger scales due to a discernible difference in the slopes of the correlation functions. As highlighted previously, it is likely that differences between the observed and simulated galaxy auto-correlation functions can be explained in part due to the bias against low-mass non star-forming galaxies in the observational sample, owing to the difficulty in assigning redshifts to these galaxies. The relative slopes between $\Xi_{\textrm{gg}}$ and $\Xi_{\textrm{ag}}$ predicted by the \ac{eagle} simulation reveal the same trend suggested by the data, whereby the slope of $\Xi_{\textrm{ag}}$ is shallower than that of $\Xi_{\textrm{gg}}$. It can also be seen that the slope of $\Xi_{\textrm{aa}}$ is shallower still. This indicates that \ion{O}{6} absorbers and galaxies in the \ac{eagle} simulation do not linearly trace the same underlying distribution of matter.

\begin{figure}
		\centering
		\includegraphics[width=8.4cm]{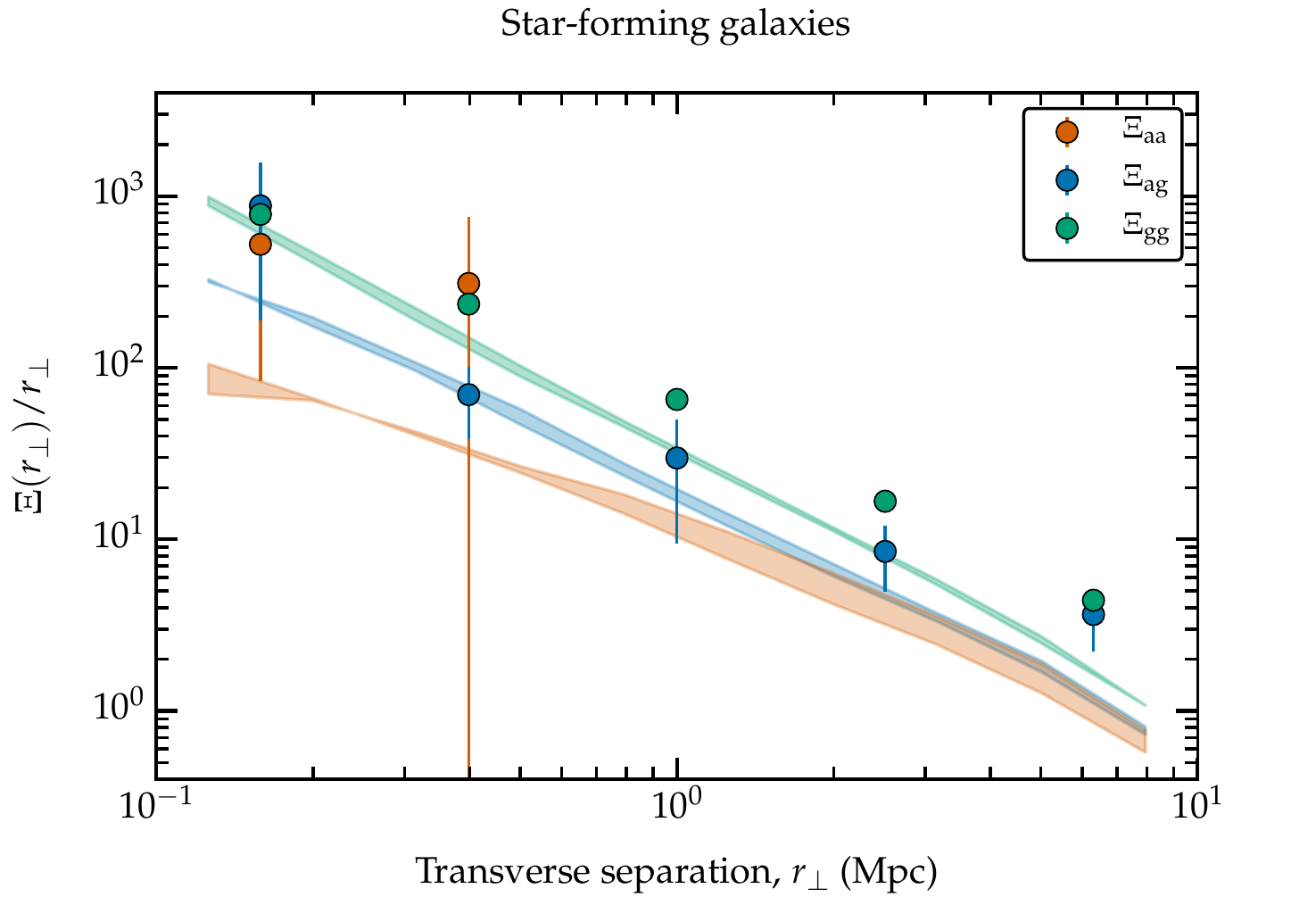}
		\caption[Correlation functions projected along the line-of-sight for galaxies and $\textrm{O}\;\textsc{vi}$ absorbers - star-forming galaxies only - \textsc{Eagle} comparison]{Same as \Cref{fig:o6_eagle_1d}, but for star-forming galaxies only.}
	\label{fig:o6_eagle_1d-sf}
	\end{figure}

\begin{figure}
		\centering
		\includegraphics[width=8.4cm]{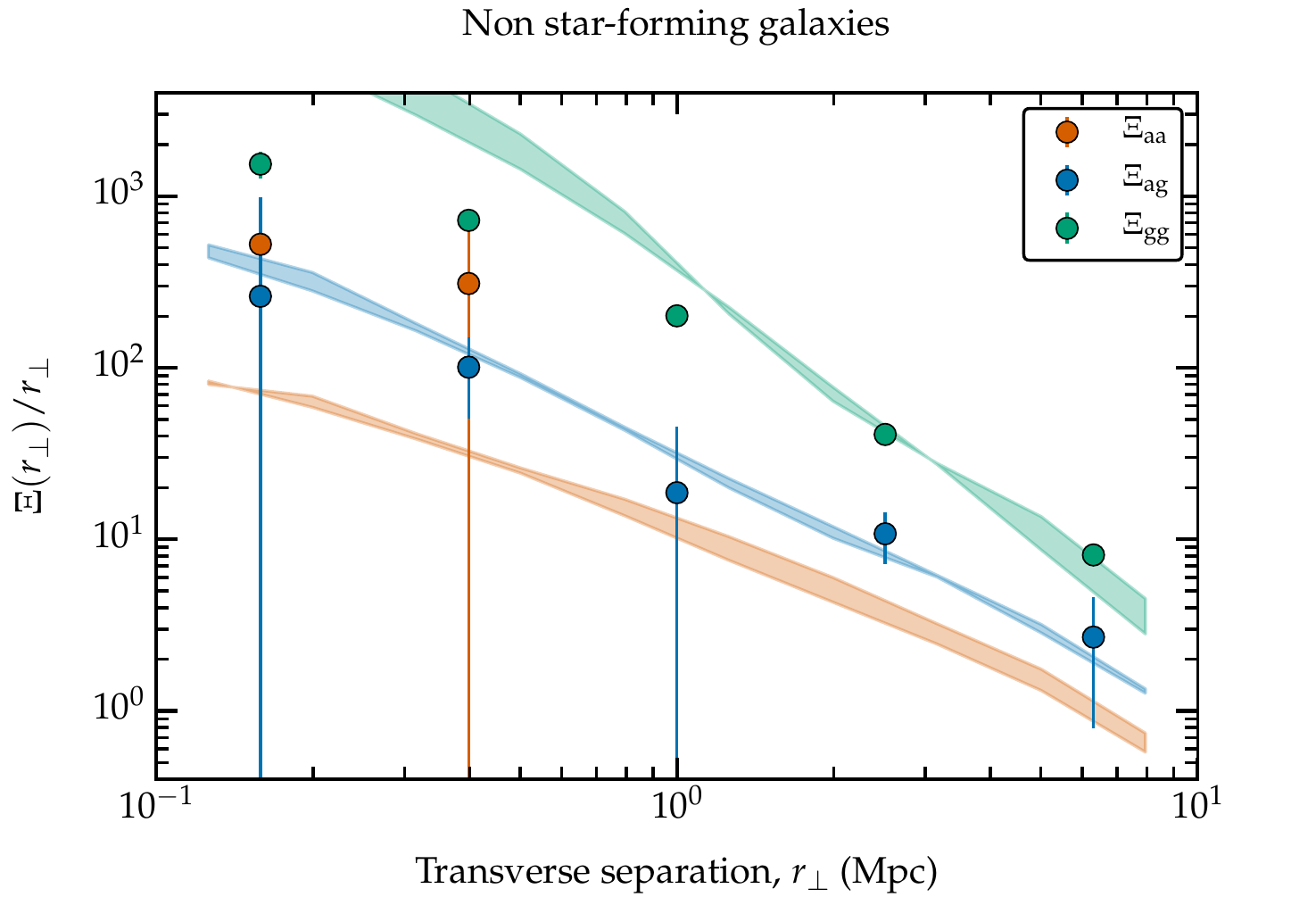}
		\caption[Correlation functions projected along the line-of-sight for galaxies and $\textrm{O}\;\textsc{vi}$ absorbers - non star-forming galaxies only - \textsc{Eagle} comparison]{Same as \Cref{fig:o6_eagle_1d}, but for non star-forming galaxies only.}
	\label{fig:o6_eagle_1d-nsf}
	\end{figure}

	In \Cref{fig:o6_eagle_1d-sf}, we show the same comparison but for star-forming galaxies only. There is again a reasonable agreement between the predicted and measured $\Xi_{\textrm{ag}}$ within the uncertainties, albeit with a discernible difference in the slopes of the correlation functions. A good agreement is seen in \Cref{fig:o6_eagle_1d-nsf} for the slope and amplitude of the cross-correlation function of \ion{O}{6} absorbers with non star-forming galaxies. We also see that the amplitudes in $\Xi_{\textrm{ag}}$ are in general lower than those in $\Xi_{\textrm{gg}}$, and that this is more pronounced for non star-forming galaxies, for which $\Xi_{\textrm{gg}}$ is a more biased tracer of the underlying distribution of matter. Furthermore, the predicted relative slopes in $\Xi_{\textrm{ag}}$ and $\Xi_{\textrm{gg}}$ show the same trends suggested by the data, whereby $\Xi_{\textrm{ag}}$ is shallower than $\Xi_{\textrm{gg}}$ for both star-forming and non star-forming galaxies, with this effect being more pronounced in the case of the latter. The slope in $\Xi_{\textrm{aa}}$ is also consistently shallower than both of these. We can therefore infer in the simulation that \ion{O}{6} absorbers and galaxies do not linearly trace the same underlying distribution of matter, and that this is most markedly the case for non star-forming galaxies. In addition, the likelihood of finding \ion{O}{6} absorbers close to star-forming galaxies is similar to the likelihood of finding \ion{O}{6} absorbers close to non star-forming galaxies over the scales probed. These inferences are all entirely consistent with those that can be drawn from the data.
	
	Finally, it is important to note that all of our comparisons with the \ac{eagle} simulation are subject to the caveats highlighted in \Cref{sec:eagle_gas}, whereby the statistical properties of the simulated and observed absorbers differ. As a result, the simulated absorbers required a systematic $+0.3$ dex shift in their column densities as a matter of procedure. Nevertheless, bearing this caveat in mind, and despite the disparity in their statistical properties, the clustering and dynamical properties of absorbers with galaxies in the \ac{eagle} simulation show encouraging qualitative agreement with the data.

\section{Discussion}
\label{sec:discussion}
In light of these new results on the cross-correlation functions of \ion{O}{6} absorbers and galaxies, we now explore some possible interpretations, and make comparisons to similar studies in the literature.

	\subsection{Comparison with previous results}
	\label{sec:results_comparison}
	\cite{2009ApJ...701.1219C} performed a very similar study to the one presented here, in which they measured the two-point cross-correlation function of \ion{O}{6} absorbers and galaxies projected along the \ac{los} at $z < 1$. We note that all of the galaxy data used in that study form a small subset of the galaxy data used here, whilst their absorption-line data from the \ac{stis} and the \ac{fuse} have been updated to that available from \ac{cos}. We find that our results contrast with these earlier results in two primary aspects: (i) \cite{2009ApJ...701.1219C} find that the clustering amplitudes of \ion{O}{6} absorbers around star-forming galaxies are comparable to those of star-forming galaxies with themselves, whilst we find that in general they are smaller; and (ii) \cite{2009ApJ...701.1219C} find that the clustering amplitudes of \ion{O}{6} absorbers around star-forming galaxies are weaker than those around non star-forming galaxies, whilst we find that they are comparable. To explain these differences, it is important to note the sample sizes. \cite{2009ApJ...701.1219C} used a sample of 13 \ion{O}{6} absorbers and 670 galaxies, which is substantially smaller than the sample assembled for this study. The quoted uncertainties on their measurements are also Poissonian, which underestimates the true uncertainties (see \Cref{sec:correlation_functions}). We therefore believe that our measurements are more statistically robust, and that with more conservative estimates on the uncertainties, the results of these two studies may in fact be consistent.
	
	\cite{2014MNRAS.445..794T} examined the distribution of \ion{O}{6}, \ion{H}{1}, and other metal ions around star-forming galaxies at $z \approx 2.4$. Their study uses a different technique to ours, in which they measure the median optical depth of \ion{O}{6} in spectral pixels binned in terms of their transverse and \ac{los} separation from galaxies. Their study is optimised to investigate the association between absorbers and star-forming high-redshift galaxies on small ($< 2$ Mpc) scales, whereas ours is optimised to investigate the association between absorbers and galaxies at $z < 1$ on predominantly large scales, spanning the range $0.1 \lesssim r_{\perp} \lesssim 10$ Mpc. Given the difference in approach between these studies, our comparison is restricted to being purely qualitative. Nevertheless, we note that the distribution of \ion{O}{6} around galaxies presented in \cite{2014MNRAS.445..794T} reveals stronger anisotropies along the \ac{los} than are evident in our study. It is important to note that these anisotropies are revealed on scales of a few hundred (proper) kpc, which are substantially smaller than the $\sim$ Mpc scales (comoving) that are considered here. It will be interesting to investigate the evolution in the distribution and dynamics of \ion{O}{6} around galaxies from high redshifts, around the peak in star-formation activity at $z \sim 2$, to the present day, but there are no straight-forward comparisons at present.
	
	\cite{2011ApJ...740...91P} investigated the incidence of \ion{H}{1} and \ion{O}{6} absorbers around galaxies at $z \lesssim 0.5$. Note that we make use of a subset of their galaxy redshift data in this study \cite[][]{2011ApJS..193...28P}. By examining the incidence rate of \ion{O}{6} absorbers as a function of galaxy impact parameter, they infer that the covering fraction of \ion{O}{6} around sub-$L^{\star}$ galaxies is nearly unity to impact parameters of 300 kpc. Comparing to the total incidence rate per absorption path length, they conclude that the majority of \ion{O}{6} absorbers to an equivalent width limit $W > 0.03$ \AA\ in \ion{O}{6} $\lambda1031$ arise within 300 kpc of sub-$L^{\star}$ galaxies in the `extended \ac{cgm}', and that they rarely originate in the intergalactic \ac{whim} predicted by cosmological hydrodynamical simulations \cite[e.g.][]{1999ApJ...514....1C,2001ApJ...552..473D}. The strategy of \cite{2011ApJ...740...91P} is (i) to identify galaxies at small ($< 300$ kpc) impact parameters from the \ac{qso} sight-lines in their sample observed with the \ac{ghrs}, \ac{stis} and \ac{fuse}, then to search for \ion{O}{6} absorbers close to those sight-lines; and (ii) to search for galaxies close to particular \ion{O}{6} absorbers. Using the latter approach, they find that 5 out of the 30 \ion{O}{6} absorbers do not have a galaxy within 1 Mpc, although this result is sensitive to the completeness limit of the galaxy survey, which will miss low luminosity galaxies. Our results are robust to galaxy completeness, and indicate that the clustering amplitudes of \ion{O}{6} absorbers with galaxies are weaker than those of galaxies with themselves. This may imply that not all \ion{O}{6} absorbers can be found very close to galaxies, or that covering factors of \ion{O}{6} around galaxies on the scales probed ($\gtrsim 100$ kpc) are less than 100\% (or both). The conclusions of \cite{2011ApJ...740...91P} are based on assumed covering factors of \ion{O}{6} around sub-$L^{\star}$ galaxies of close to 100\% to 300 kpc, which highlights a potential source of tension with our results. Similar studies to \cite{2011ApJ...740...91P}, e.g. \cite{2005ApJ...623L..97T} and \cite{2013ApJ...763..148S}, suggest that the covering fractions are significantly below 100\%. We suggest that studies such as these may need to probe galaxies to fainter luminosities in order to reconcile results that may be biased by the galaxy survey completeness limits.
	
	\cite{2011Sci...334..948T} performed another study of the incidence of \ion{O}{6} absorbers around galaxies as part of the COS-halos survey, and found a dichotomy between the incidence rate around star-forming galaxies and non star-forming galaxies. In particular, they find that \ion{O}{6} absorbers are nearly ubiquitous within 150 kpc of star-forming galaxies, but that only a small fraction of non-star forming galaxies show \ion{O}{6} absorption within this distance. This result contrasts with ours, in which we find that the likelihood of finding \ion{O}{6} around star-forming galaxies is similar to the likelihood of finding \ion{O}{6} around non star-forming galaxies. However, it is important to note that our study is optimised for scales $\gtrsim 100$ kpc away from galaxies, and so if this dichotomy only exists on the smallest scales, we would not have been able to detect it. Furthermore, in terms of appropriately calculated virial radii for the COS-halos survey galaxy sample, their \ac{qso} sight-lines intersect numerous star-forming galaxies at impact parameters greater than the virial radius, but no non star-forming galaxies beyond this point \cite[][]{2014ApJ...784..142S}. Our approach is not subject to this particular bias, which may also help explain the differences between these two findings.

	\subsection{Interpretation of the results}
	\label{sec:results_interpretation}
	The principal findings of this study are as follows:
	\begin{enumerate}
		\item \ion{O}{6} absorbers show little velocity dispersion ($\lesssim 100$ \kms) with respect to galaxies on $\sim$ Mpc scales.
		\item The correlation length of the \ion{O}{6}-galaxy cross-correlation function is smaller (at the $4 \sigma$ significance level for our full sample), and the slope potentially shallower, than that of the galaxy auto-correlation function in general, which indicates that \ion{O}{6} absorbers and galaxies may not linearly trace the same underlying distribution of matter.
		\item The \ion{O}{6}-galaxy cross-correlation function split by star-formation activity gives correlation lengths that are entirely consistent within their $1 \sigma$ bootstrap uncertainties. Thus, on the scales probed ($\gtrsim 100$ kpc), the likelihood of finding \ion{O}{6} absorbers around star-forming galaxies is similar to the likelihood of finding \ion{O}{6} absorbers around non star-forming galaxies.
	\end{enumerate}
	Given that the enrichment of the \ac{igm} is attributable to galaxy feedback, it is convenient to think of these results from a galaxy-centric viewpoint, as follows.
	
	In item (ii) from the list above, the lower correlation amplitudes in the \ion{O}{6}-galaxy cross-correlation functions compared to the galaxy auto-correlation functions imply that either \ion{O}{6} absorbers are not ubiquitous near to the galaxies in our sample, or that their distribution around them is patchy, i.e. the covering factor of \ion{O}{6} around galaxies is substantially less than 100\%, as was suggested in the previous section. In reality, both of these inferences could be true. This may be a function of \ion{O}{6} column density and/or Doppler broadening parameter, although we have not attempted to split the \ion{O}{6} absorber sample in this study due to low number statistics. We also note that this conclusion may not necessarily apply to lower column density \ion{O}{6} absorbers and/or fainter galaxies, below the detection limits of the present survey.
	
	Also in item (ii), indications of a shallower slope in the \ion{O}{6}-galaxy cross correlation function compared to the galaxy auto-correlation function could indicate that the distribution of \ion{O}{6} around galaxies is in general more extended than the distribution of galaxies around themselves. From the data, we cannot rule out the possibility that \ion{O}{6} absorbers are primarily attached to galaxies, and that the cross-correlation amplitudes are primarily driven by clustering of the galaxies with themselves. Nevertheless, the inference of an extended distribution of \ion{O}{6} around galaxies is supported by predictions from the \ac{eagle} simulation, which also indicate a shallower slope in the \ion{O}{6}-galaxy cross-correlation function, and an even shallower slope in the \ion{O}{6} auto-correlation function. The difference in the slopes and amplitudes of these correlation functions leads to the possibility that \ion{O}{6} absorbers and the galaxies in our sample do not linearly trace the same underlying distribution of matter, and that some \ion{O}{6} absorbers may be found far from galaxies that are bright enough to be included in our survey. This picture agrees well with the inferences made by \cite{2006ApJ...641..217S,2013ApJ...763..148S}, who find that broad \ion{O}{6} absorbers may trace hot gas ($T \approx 10^6$ K) that extends to large distances ($\sim 400$ -- 600 kpc) around galaxies, distributed in the cosmic web or in supercluster filaments. They argue that for a close correspondence between \ion{O}{6} and galaxies, low luminosity galaxies ($L < 0.1 L^{\star}$) must contribute to the total cross-section to match the observed \ion{O}{6} absorption-line frequency. Our results are entirely compatible with these inferences.
	
	A similar scenario is seen in various cosmological hydrodynamical simulations, where \ion{O}{6} is distributed far from galaxies and often with a relatively flat radial profile \cite[][]{2011ApJ...731...11C,2011ApJ...731....6S,2012MNRAS.420..829O,2012ApJ...759...23S,2013MNRAS.432...89F,2013MNRAS.430.1548H,2014MNRAS.444.1260F}. Nevertheless, obtaining predictions that match observational findings is often met with varied success, and found to be sensitive to the sub-grid physics prescriptions, particularly with reference to the physical and statistical properties of the absorbers. Comparisons made here with the \ac{eagle} simulation largely echo these findings, but do confirm that the simulations are at least broadly capable of reproducing the observed distribution and dynamics of \ion{O}{6} absorbers around galaxies, as inferred from the slope and amplitude of the projected \ion{O}{6}-galaxy cross-correlation function, and the anisotropies in the two-dimensional \ion{O}{6}-galaxy cross-correlation function. These results also raise the tantalising possibility that a fraction of \ion{O}{6} absorbers do in fact arise in the \ac{whim}, outside of galaxy haloes and groups at temperatures of $10^5 < T < 10^7$ K, as predicted by the simulations. Nevertheless, a targeted approach to detecting the \ac{whim} is still needed if we are to confirm these predictions \cite[e.g.][]{2016MNRAS.455.2662T}, and it is not yet clear whether the commonly assumed tracers of the \ac{whim} (\ion{O}{6}, \ion{Ne}{8}, broad \lya) trace the bulk of this hot intergalactic plasma \cite[e.g.][]{2006A&A...445..827R,2007ApJ...658..680L,2010ApJ...710..613D,2011ApJ...730...15N,2011MNRAS.413..190T,2012MNRAS.425.1640T,2013MNRAS.436.2063T}.
	
	In many respects, the level of agreement on the clustering of \ion{O}{6} absorbers around galaxies between the \ac{eagle} simulation and the real data is quite surprising when we consider the potential origins the \ion{O}{6} absorbers, some of which are expected to arise from conductive and turbulent interfaces that are not resolved by the cosmological simulations at present. Either these processes are relatively unimportant for the overall population of \ion{O}{6} absorbers detected by current instrumentation, or they are co-spatial with other production mechanisms (e.g. photoionization, collisional ionization, shocks), the physics of which is captured by the simulations. Given the level of agreement seen here, we also suggest that the shortcomings of our approach to extracting \ion{O}{6} absorbers from the \ac{eagle} simulation, described in \Cref{sec:eagle_gas}, have little overall effect in the measured clustering signal of \ion{O}{6} around galaxies, although it will be important in future work to verify our results with a rigorous Voigt profile fitting procedure.
	
	Item (i) in our list of findings from the present study indicates that we have not found substantial evidence for gas outflows or inflows traced by \ion{O}{6} around galaxies on $\sim$ Mpc scales at low redshifts. This scenario is also consistent with that predicted by the \ac{eagle} simulation, which implements subgrid prescriptions for effective feedback from star formation and \ac{agn} in order to match the present day statistics of the galaxy population. Our constraints on the velocity dispersion of \ion{O}{6} around galaxies are consistent with a scenario in which the majority of the \ion{O}{6} absorbers detected with current instrumentation move with the galaxies, and that those bound to galaxy haloes are not moving with velocities sufficient to escape their local gravitational potential. This is then suggestive of a scenario in which the wider \ac{igm} not bound to individual galaxy haloes may have been enriched early in the history of the Universe \cite[e.g.][]{2010MNRAS.409..132W,2011ApJ...731...11C,2012MNRAS.420..829O}.

	Our other main finding indicates that the presence of \ion{O}{6} absorbers on $\gtrsim 100$ kpc scales around galaxies is not strongly biased towards whether those galaxies are star-forming or not. This situation is clearly echoed in the \ac{eagle} simulation, and indicates that the instantaneous star-formation activity in galaxies bears no relation to the overall distribution of metals around them. This further supports the inference that a significant proportion of the metals in the \ac{igm} have been distributed into the \ac{igm} early, and that the extent of ongoing star-formation has no discernible effect on the metal enrichment of the low-redshift \ac{igm} on $\gtrsim 100$ kpc scales.

\section{Summary \& conclusions}
\label{sec:summary_conclusions}
Our analysis of the two-point cross- and auto-correlation functions of \ion{O}{6} absorbers and galaxies at $z < 1$ resulted in the findings outlined below:
\begin{enumerate}
	\item \ion{O}{6} absorbers show very little velocity dispersion with respect to galaxies on $\sim$ Mpc scales at low redshifts. We estimate that this velocity dispersion amounts to $\lesssim 100$ \kms.
	\item The slope of the \ion{O}{6}-galaxy cross-correlation function is potentially shallower than that of the galaxy auto-correlation function. We therefore find that these populations may not linearly trace the same underlying distribution of matter. In particular, these results indicate that the distribution of \ion{O}{6} around galaxies could be more extended than the distribution of galaxies around themselves. We therefore speculate that a fraction of the \ion{O}{6} absorbers might trace the \ac{whim} predicted by cosmological hydrodynamical simulations.
	\item The clustering amplitudes of \ion{O}{6} absorbers around star-forming galaxies are consistent with those around non star-forming galaxies within the uncertainties. We therefore find that the likelihood of finding \ion{O}{6} absorbers around star forming galaxies is similar to the likelihood of finding \ion{O}{6} absorbers around non star-forming galaxies, at least on scales $\gtrsim 100$ kpc.
	\item The amplitude of the \ion{O}{6}-galaxy cross-correlation is typically lower than that of the galaxy auto-correlation function by factors of a few. This indicates that \ion{O}{6} absorbers are either not ubiquitous to galaxies, that they are predominantly attached to fainter galaxies than those typical in our sample, or that their distribution around them is patchy on scales $\gtrsim 100$ kpc (or any combination of these possibilities), at least for the column densities at which most are currently detected.
	\item We find that predictions from the \ac{eagle} cosmological hydrodynamical simulation, subject to a correction in the predicted \ion{O}{6} column densities, are reasonably consistent with the observational findings outlined above. This suggests that simulations such as these may be regarded as a useful tool for understanding the distribution and dynamics of metal-enriched gas traced by \ion{O}{6} around galaxies on $\gtrsim 100$ kpc scales.
\end{enumerate}

\section*{Acknowledgments}
We would firstly like to thank the referee, whose comments and suggestions improved this paper.

C.W.F would like to acknowledge the support of an STFC studentship (ST/J201013/1). N.H.M.C thanks the Australian Research Council for \textsl{Discovery Project} grant DP130100568 which supported this work. T.T. acknowledges support from the Interuniversity Attraction Poles Programme initiated by the Belgian Science Policy Office ([AP P7/08 CHARM]). The research was supported in part by the European Research Council under the European Union's Seventh Framework Programme (FP7/2007-2013)/ERC grant agreement 278594-GasAroundGalaxies.

We thank the contributors to \textsc{SciPy},\footnote{\url{http://www.scipy.org}} \textsc{Matplotlib}\footnote{\url{http://matplotlib.org}} and the \textsc{Python} programming language,\footnote{\url{http://www.python.org}} the free and open-source community and the NASA Astrophysics Data system\footnote{\url{http://adswww.harvard.edu}} for software and services. This work also made use of \textsc{AstroPy}; a community-developed core Python package for Astronomy \citep{2013A&A...558A..33A}.

The work in this paper was in part based on observations made with the NASA/ESA Hubble Space Telescope under programmes GO 11585 and GO 12264, obtained at the Space Telescope Science Institute, which is operated by the Association of Universities for Research in Astronomy Inc., under NASA contract NAS 5-26555; and on observations collected at the European Southern Observatory, Chile, under programmes 070.A-9007, 086.A-0970 and 087.A-0857. Some of the data presented herein were obtained at the W.M. Keck Observatory, which is operated as a scientific partnership among the California Institute of Technology, the University of California and NASA. The authors wish also to recognise and acknowledge the very significant cultural role and reverence that the summit of Mauna Kea has always had within the indigenous Hawaiian community. This work was partially based on observations obtained at the Gemini Observatory, which is operated by the Association of Universities for Research in Astronomy, Inc., under a cooperative agreement with the NSF on behalf of the Gemini partnership: the NSF (United States), the National Research Council (Canada), CONICYT (Chile), the Australian Research Council (Australia), Minist\'{e}rio da Ci\^{e}ncia, Tecnologia e Inova\c{c}\~{a}o (Brazil) and Ministerio de Ciencia, Technolog\'{i}a e Innovaci\'{o}n Productiva (Argentina).

GAMA is a joint European-Australasian project based around a spectroscopic campaign using the Anglo-Australian Telescope. The GAMA input catalogue is based on data taken from the Sloan Digital Sky Survey and the UKIRT Infrared Deep Sky Survey. Complementary imaging of the GAMA regions is being obtained by a number of independent survey programmes including GALEX MIS, VST KiDS, VISTA VIKING, WISE, Herschel-ATLAS, GMRT and ASKAP providing UV to radio coverage. GAMA is funded by the STFC (UK), the ARC (Australia), the AAO, and the participating institutions. The GAMA website is \url{http://www.gama-survey.org/}.

This work also made use of the DiRAC Data Centric system at Durham University, operated by the Institute for Computational Cosmology on behalf of the STFC DiRAC HPC Facility\footnote{\url{http://www.dirac.ac.uk}}. This equipment was funded by BIS National E-infrastructure capital grant ST/K00042X/1, STFC capital grant ST/H008519/1 and STFC DiRAC Operations grant ST/K003267/1 and Durham University. DiRAC is part of the National E-Infrastructure.

The raw data from \ac{hst}/\ac{cos} may be accessed from the MAST archive.\footnote{\url{http://archive.stsci.edu}} That from the European Southern Observatory may be accessed from the ESO Archive,\footnote{\url{http://archive.eso.org/eso/eso_archive_main.html}} and that from the Gemini Observatory may be accessed from the Gemini Science Archive.\footnote{\url{http://www.cadc-ccda.hia-iha.nrc-cnrc.gc.ca/en/gsa/}} Absorption line data from \cite{2016ApJ...817..111D} can be accessed from \url{http://archive.stsci.edu/prepds/igm/}.

\bibliographystyle{mnras}
\bibliography{bibliography}

\label{lastpage}
\end{document}